\documentclass[runningheads]{llncs}
    \usepackage[square,sort,comma,super,authoryear]{natbib}

\usepackage[fleqn]{amsmath}
\usepackage{amsmath, amssymb}
\usepackage{csquotes}
\usepackage{graphicx}
\usepackage{amsmath}
\usepackage{amsfonts}
\usepackage{dutchcal}
\usepackage{comment}
\usepackage{mathabx}
\usepackage{tabularx}
\usepackage{tabulary}
\usepackage{hhline}
\usepackage{mathrsfs}
\usepackage{setspace}
\usepackage{array}
\usepackage[gen]{eurosym}
\usepackage{comment}
\usepackage[pagewise,mathlines,right]{lineno}\linenumbers

\usepackage{expl3}

\usepackage{xcolor}
\usepackage{hyperref}
\hypersetup{
    colorlinks=false,  
    linkcolor=black,
    linkbordercolor=black,
    urlcolor=black,
    allbordercolors={0 0 0},
    pdftitle={Unification Synthesis},
     pdfborderstyle={/S/U/W .1},     
    pdfborder={0 0 .4},
    breaklinks=true,
    hypertexnames = false
    } 

\DeclareMathOperator{\uand}{\emph{and}\,}
\DeclareMathOperator{\uor}{\emph{or}\,}
\DeclareMathOperator{\unot}{\emph{not}\kern-.0pt}

\DeclareMathOperator{\uf}{\emph{f}\,}
\DeclareMathOperator{\uimplies}{\Rightarrow}
\DeclareMathOperator{\uimpliedby}{\Leftarrow}
\DeclareMathOperator{\uiff}{\iff}
\DeclareMathOperator{\uif}{\emph{if}\,}
\DeclareMathOperator{\uthen}{\emph{then}\,}
\DeclareMathOperator{\uelse}{\emph{else}\,}

\DeclareMathOperator{\unify}{\emph{unify}}
\DeclareMathOperator{\lex}{\emph{lex}}
\DeclareMathOperator{\isatm}{\emph{is-atom}} 
\DeclareMathOperator{\iscnst}{\emph{is-const}} 
\DeclareMathOperator{\isvar}{\emph{is-var}} 
\DeclareMathOperator{\lef}{\emph{left}}
\DeclareMathOperator{\rig}{\emph{right}}

\DeclareMathOperator{\ufirst}{\emph{first}}
\DeclareMathOperator{\usecond}{\emph{second}}
\DeclareMathOperator{\idem}{\emph{idem}}
\DeclareMathOperator{\mgi}{\emph{mgi}}
\DeclareMathOperator{\mgiu}{\emph{mgiu}}
\DeclareMathOperator{\reduce}{\emph{reduce}}

\DeclareMathOperator{\uempty}{\{\}}

\bibpunct[, ]{(}{)}{;}{a}{,}{,}
 \setcitestyle{square}
 
\begin{document}
\nolinenumbers
%

\title{\texorpdfstring{Automating the Derivation of \\ Unification Algorithms}
{Automating the Derivation of Unification Algorithms}
}
\titlerunning{Automating the Derivation of Unification Algorithms}
\subtitle{A Case Study in Deductive Program Synthesis
\thanks{To appear in the Journal of Symbolic Computation.}}
%
\author{Richard Waldinger\thanks{\email{waldinger@ai.sri.com}, \fnmsep \\
 \url{https:www.sri.com/people/richard-j-waldinger} \fnmsep \\ \\ \orcidID{0000-0002-4520-3907}}}

\authorrunning{R. Waldinger}
%
\institute{ 
Artificial Intelligence Center, SRI, Menlo Park, CA 94025, USA
}

 \maketitle              
 
 \begin{center}revised \today \end{center}
\newcommand{\e}[1]{e_{#1}}
\renewcommand{\d}[1]{d_{#1}}
\newcommand{\E}[1]{E_{#1}}
\newcommand{\ttt}[1]{t_{#1}}
\newcommand{\expsub}[2]{{#1}_{#2}}
\newcommand{\x}[1]{x_{#1}}
\newcommand{\y}[1]{y_{#1}}
\newcommand{\thet}[1]{\theta_{#1}}
\newcommand{\Thet}[1]{\Theta_{#1}}

\newcommand{\bb}[1]{\mathbb{#1}}


\newcommand{\var}[1]{#1}
\newcommand{\varsub}[2]{{#1}_{#2}}
\newcommand{\cons}{\, \,\begin{picture}(-1,1)(-1,-3)\circle*{3} \end{picture}\, \, \, \, }

\newcommand{\addto}{+ \,}
\newcommand{\apply}{\blacktriangleleft}

\newcommand{\compose}{ \blackdiamond }
\newcommand{\union}{ \cup }
\newcommand{\fail}{\bot}
\newcommand{\blk}{\ast}
\newcommand{\nil}{\emph{nil}}
\newcommand{\emptysubst}{\{\}}
\newcommand{\moregen}{\succsim_{gen}}
\newcommand{\weakmoregen}{\succsim_{gen^-}}
\newcommand{\issubst}{\emph{is-subst}}
\newcommand{\isprop}{\emph{is-proper}}
\newcommand{\occursin}{\, \varepsilon \,}
\newcommand{\occurseq}{\, \underline{\varepsilon} \,}
\newcommand{\dom}{\emph{dom}}
\newcommand{\range}{\emph{range}}
\newcommand{\misses}{\emph{misses}}

\newcommand{\size}{\emph{size}}
\newcommand{\vars}{\emph{vars}}
\newcommand{\Vars}{\emph{Vars}}
\newcommand{\varssize}{\emph{vars-size}}
\renewcommand{\uiff} {\Leftrightarrow}
\newcommand{\true}{\emph{true}}
\newcommand{\false}{\emph{false}}
\newcommand{\cond}[3]{\emph{if}\,( {#1},\allowbreak \ {#2},\allowbreak \ {#3})}
\newcommand{\repl}[2]{\{{#1}\mapsto{#2}\}}
\newcommand{\ucond}[3]
{\begin{aligned}[t]
&\uif {#1}  \\
&\uthen {#2}\\
&\uelse 
{\begin{aligned}[t]
&\textbf{$\mathbf{\{\unot ({#1})\}}$} \\
&{#3}
\end{aligned}}
\end{aligned}
}
\setlength{\abovedisplayskip}{2pt}
\setlength{\belowdisplayskip}{2pt}
\setlength{\mathindent}{1cm}
\setlength{\tabcolsep}{10pt}
\newcolumntype{T}{|m{0.26\linewidth}|m{0.26\textwidth}||m{0.28\textwidth}|}

\newcolumntype{S}{|m{0.20\linewidth}|m{0.40\textwidth}||m{0.20\textwidth}|}
\newcolumntype{V}{ 
  | >{\raggedright\arraybackslash}X 
  | >{\raggedright\arraybackslash}X 
  || >{\raggedright\arraybackslash}X | }
\newcolumntype{Y}{>{\centering\arraybackslash}X}
\newcolumntype{U}{| Y | Y || Y |}
\newcommand{\SNARK} {\textsc{snark}}
\makeatletter
\newcommand{\vast}{\bBigg@{4}}
\newcommand{\Vast}{\bBigg@{4}}
\makeatother
\date{\today}

\begin{abstract}
The unification algorithm has long been a target for program synthesis research, but a fully automatic derivation remains a research goal. In deductive program synthesis, computer programming is phrased as a task in theorem proving; a declarative specification is expressed in logical form and presented to an automatic theorem prover, and a program meeting the specification is extracted from the proof. The correctness of the program is supported by the proof, which also provides an explanation of how the program works. The proof is conducted in an appropriate axiomatic subject-domain theory, which defines the concepts in the specification and the constructs in the target programming language and provides the background knowledge necessary to connect them. \\

   For the unification proof, we generalize and automate the manual proof presented in \citet{man:wal}. The new program unifies two given symbolic expressions (s-expressions) relative to a given “environment” substitution. The proof establishes the existence of an output substitution that is a most-general idempotent unifier of the given expressions and is an “extension" of the environment substitution. If no such substitution exists and the expressions are not unifiable, the program is to produce a failure indicator $\bot$. \\
     
     Initially the environment substitution is the empty substitution, which makes no replacements at all;  during execution of recursive calls, the environment substitution records the replacements that have been found so far. Our own unification algorithm employs an environment, and such algorithms appear in the literature \citep[e.g.,][]{lug}.  We suspect, in addition to being more efficient, the three-argument algorithm with an environment is easier to synthesize automatically than the two-argument version from the Manna-Waldinger paper.\\

    The proof is conducted relative to an axiomatic theory of expressions and substitutions. The structure of the derived program reflects the proof from which it was extracted. Conditional expressions in the program are obtained from case analysis in the proof; recursion is introduced using well-founded induction, with respect to a well-founded relation axiomatized in the theory. The proof was obtained automatically by the  first-order resolution theorem prover Snark \citep{sti}. The extracted program, as we remarked, improves on Manna and Waldinger's and contains some novel elements.
\end{abstract}

\begin{quote}
\emph{
  In the literature unification is often treated as a simple and straightforward matter, even though it is recognized as a deep and fundamental concept. However when a thorough presentation is attempted, it is then realized that the matter is fairly subtle and treacherous.
  }
  \\---\citep{las:mah:mar}
  
 \end{quote}

\begin{quote}
\emph{
  You have used the unification algorithm to construct the unification algorithm.  You have achieved precisely nothing!
  }
  \\---[Geoff Sutcliffe, personal communication, 2023]
\end{quote}

\begin{quote}
\emph{
I'm sorry to write such a long letter---I didn't have time to make it shorter.
}
\\---[Blaise Pascal, \emph{Lettres Provinciales}, 1657]
\end{quote}

\section*{Introduction}

Program synthesis, the automatic construction of computer programs, was one of the first applications of automated theorem proving and is still one of the most appealing.  Several early researchers  \citep[e.g.,][]{sla, wal:lee, ccg} exhibited the synthesis of simple programs as demonstrations.  The NASA system Amphion  \citep{low} used a theorem prover to construct software for, among other things, the Cassini mission to Saturn.  Yet deductive program synthesis has had little impact on the practice of software engineering, even though it seems as if programming and theorem proving are cognitively similar, and that many of the difficulties software engineers have in program construction are like those mathematicians have in proving theorems. Considering the progress theorem provers have made in recent years, it is natural to wonder how well they can now do in program construction.
     
     In deductive program synthesis, we regard the task of constructing a program as one of proving a theorem.  A given specification describes the purpose of the desired program without giving any indication of the method by which that purpose is to be achieved. For this reason, a correct specification may be easier to construct than a completed program. The specification is phrased as a conjecture, i.e., a potential theorem to be proved, and is submitted to a theorem prover.  For an applicative program, the conjecture typically states the existence of an output entity that, for a given input entity, satisfies the specification.  The proof is conducted with respect to a subject domain theory, an axiomatic theory that defines the concepts of the specification language, the constructs of the target programming language, and the background knowledge necessary to connect them.  The proof is restricted to be sufficiently constructive that, in proving the existence of a desired output, it is forced to indicate a method for finding such an entity.  That method becomes the basis for an algorithm that is extracted from the proof.  

We focus on concepts that play a role in the derivation of unification algorithms.
Roughly speaking, unification is the process of finding a way of making two mathematical expressions identical by replacing variables with other expressions.  Unification was introduced to the theorem-proving community as part of the resolution principle  \citep{rob}.  Robinson credited it to  Herbrand's [\citeyear{her}] thesis, but added to its conceptualization. Unification is a process of central importance not just for theorem proving, but also for such applications as logic programming, type checking, and natural language understanding.  Surprisingly, unification algorithms presented in widely used textbooks by highly qualified authors and incorporated into at least one reasoning system are incorrect  \citep{nor}.  

Unification was an early target in deductive program synthesis research. \citet{man:wal} presented a derivation in which a unification algorithm was extracted from an informal manual proof.
\citet{nardi} showed a manual synthesis of a unification algorithm from a proof within Manna and Waldinger's deductive-tableau formalism, which we also use in this paper. 
\citet{paulson} used an interactive proof checking system to show that Manna and Waldinger's algorithm met its specification, but that argument took 
the algorithm as given, so it is a semi-automatic verification rather than a synthesis. \citet{eriksson} and \citet{armando} produced partial interactive syntheses.  So it seems as if a fully automatic synthesis of an entire unification algorithm is still a research goal.

In this paper, we first provide an introduction to a theory of expressions and substitutions, which serves as the subject domain theory for the derivation.  We outline how to adapt a resolution-style theorem prover for program synthesis, with special emphasis on the use of case analysis to produce conditional expressions and mathematical induction to produce recursion.  We limit our discussion to the use of a first-order-logic theorem prover to produce applicative (side-effect-free) programs. Our illustrative examples will be taken from the proof-based derivation of a unification algorithm.

    Snark was able to produce dozens of unification algorithms for symbolic expressions from the same specification. For this paper we have selected one that was relatively simple and efficient.
    We rephrase Snark's derivation proof for this  algorithm for comprehensibility.  We present the algorithm, which contains some novelties discovered by Snark.  

    Our hope is that this example may serve as an introduction to deductive program synthesis. It is too terse to provide all the necessary logical background---a more leisurely tutorial appears in, say, \citep{man:wal:book}---but a reader with some mathematical training should be able to make sense of this paper without having read the earlier work.
    
 \section*{A Theory of Expressions, Substitutions, and Unifiers}   
Our introduction is brisk and  semi-formal, and we omit the proof of some known results. We do not attempt to survey the entire theory of expressions; for a more general introduction, see \citep{baa:sny}.  Even people familiar with the theory may want to scan the following section, because we have altered some of the concepts to facilitate the automation of the proofs.

\subsection*{Expressions}\hypertarget{par:exp} The expressions in this version of the theory are symbolic expressions modeled on the S-expressions of Lisp  \citep{jmc}.  We distinguish between atoms and nonatomic expressions (conses).  The atoms comprise constants $(a, b, c, \ldots)$, and $\nil$ and variables $(\var{W}, \var{X}, \var{Y}, \var{Z})$.  These are characterized by the relations $\iscnst(e)$ and $\isvar(e)$, respectively.
For the unification study, we also include a special constant $\blk$, called the $\emph{black hole}$; this has the properties of an ordinary constant.

A nonatomic expression $ \e{l} \cons e_{2} $ is the cons ($\cons$) of two subexpressions $e_{1}$ and $e_{r}$.   For instance, the constant $a$ and the variable  $\var{X}$ are atomic expressions, and their cons,  $(a \cons \var{X})$, is a nonatomic expression, and so is $(a \cons (a \cons \var{X}))$. We can think of an expression as a binary tree whose tips are atoms. We may omit the outermost parentheses for brevity.

\subsubsection{Left and Right Functions.} We introduce functions $\lef$ and $\rig$ to decompose a nonatomic expression into its respective components. We characterize atomicity by the relation $\isatm$. We define these constructs to satisfy the \hypertarget{prop:nonatom}{\emph{nonatomic property}},
\begin{align*}
&\unot ({\isatm}(e)) \uimplies \\
&\,e =  {\lef}(e) \cons {\rig}(e),
\end{align*} 
\noindent for all expressions $e$, and the \hypertarget{prop:atom}{\emph{atomic property}},
\begin{align*} 
&{\isatm}(e) \uimplies \\
&\unot (e =  \expsub{e}{l} \cons \expsub{e}{r}),
\end{align*} 
\noindent for all expressions $e$, $\expsub{e}{l}$, and $\expsub{e}{r}$. The functions $\lef$ and $\rig$ correspond to the functions \emph{car} and \emph{cdr} of Lisp. While applying $\lef$ or $\rig$ to an atom is not an error, no axiom specifies what its value is.

 Two nonatomic expressions $d$ and $e$ are equal 
if their corresponding left and right components are equal.  That is, 
\begin{align*}
&{\unot} \,(\isatm(d)) \uand {\unot} \,(\isatm(e)) \uimplies \\
&\lef(d) = \lef(e)). \end{align*} 
It follows that, for any expressions $\e{l}$ and $\e{r}$, 
     $\lef(\e{l} \cons \e{r}) = \e{l}$ and $\rig(\e{l} \cons \e{r}) = \e{r}$. 

\subsubsection{The Occurrence Relations.} For any two expressions $\e{1}$ and $\e{2}$, we write $\e{1}$ \emph{occurs in} $\e{2}$ as  $\e{1} \occurseq \e{2}$ and  $\e{1}$ \emph{occurs properly} in $\e{2}$ as  $\e{1} \occursin \e{2}$. We say that 
$\e{1}$ occurs in $\e{2}$ if $\e{1}$ occurs properly in $\e{2}$ or if they are identical, as expressed in the \hypertarget{prop:occurs-ref}{\emph{reflexive-closure property}}, 
\[ \e{1} \occurseq \e{2} \uiff (\e{1} \occursin \e{2} \uor \e{1} = \e{2}).\]
We shall also call $\occurseq$ and $\occursin$ the \emph{subexpression} and the \emph{proper subexpression} relation, respectively. They satisfy the \hypertarget{prop:occur-atom}{\emph {atomic property}},  that
\begin{align*}
&\isatm(e) \uimplies 
\unot(d \occursin e), 
\end{align*} 
and the \hypertarget{prop:occur-cons}{\emph{cons property}}, that
\begin{align*}
&d  \occursin  (\e{1} \cons \e{2}) \uiff 
(d   \occurseq  \e{1} \uor d  \occurseq  \e{2}),
\end{align*}
for all expressions $d$, $e$,  $e_1$, and $e_2$.
\subsubsection{Size and Vars Functions.} We define the \hypertarget{def:size}{\emph{size}} of an expression to be the number of nonvariable symbols, including conses, in an expression; that is
\[\begin{aligned}
&\size(c) =\,  1 &&\text{if $c$ is a constant} \\
  &  \size(v) =\,  0 &&\text{if $v$ is a variable} \\
  &  \size(\e{1} \cons \e{2}) =\,  1 + \size(\e{1}) + \size(\e{2}) &&\text{if $\e{1}$ and $\e{2}$ are expressions.}
\end{aligned}
\]
The $\size$ function has been defined so that variables are less than constants.  We can also show that 
  \[\e{1} \occursin \e{2} \uimplies \size(\e{1})  < \size(\e{2}),\]
  for any expressions $\e{1}$ and $\e{2},$
  and hence, if $e$ is a nonatomic expression, that
  \[
\size(\lef(e))) < \size(e)  \]
and 
\[\size(\rig(e))) < \size(e).\]
        
We denote by \hypertarget{def:vars}{$\vars(\var{e})$} the set of variables that occur in $\var{e}$, defined by
\[\begin{aligned}
    &\vars(v) =\,  \{v\} &&\text{if $v$ is a variable} \\
    &\vars(c) =\,  \{\} &&\text{if $c$ is a constant} \\
    &\vars(\e{1} \cons \e{2}) =\,  \vars(\e{1}) \cup \vars(\e{2}) &&\text{if $\e{1}$ and $\e{2}$ are expressions.}
    \end{aligned}
\]
Because, for a nonatomic expression $e$, we know $e = \lef(e)\cons \rig(e)$, we have the 
\hypertarget{prop:vars-left-right}{\emph{left-right property} of $\vars$},
\[\vars(e) = vars(\lef(e)) \cup vars(\rig(e)).\]
We allow ourselves to use $\vars$ as an n-ary function, by the \hypertarget{prop-vars-union}{\emph{union property}} of $\vars$, \[\vars(e_1, \ldots , e_n) =
\vars(e_1) \cup \ldots \cup vars(e_n).\]
We also have the \hypertarget{prop:occ-sub}{\emph{occurs-subset property}} of $\vars$, namely
\[\e{1} \occurseq \e{2} \uimplies \vars(\e{1}) \subseteq \vars(\e{2}),\]
for all expressions $\e{1}$ and $\e{2}$. The subset relation here is not always proper; for example, $x \occursin (x \cons x)$  but 
\[
\begin{aligned}
\vars(x) 
     &= 
  \{x\} \\
    &=  \vars(x \cons x). \\
\end{aligned}
\] 

\subsubsection{Pairs, Triples, and Tuples.} Imitating Lisp, we sometimes use the \emph{empty tuple}  $\langle \rangle$ as an alternative notation for $\nil$; the \emph{singleton tuple} $\langle e  \rangle$ as an abbreviation for $\e \cons \nil$; the \emph{pair} $\langle \e{1}, \e{2} \rangle$ as an abbreviation for $\e{1} \cons (\e{2} \cons \nil\,)$; and, in general, the \emph{tuple} $\langle \e{1}, \e{2}, \ldots, \e{n} \rangle$ as an abbreviation for $\e{1} \cons (\e{2} \cons (\ldots (\e{n} \cons \nil\,)\ldots))$. Tuples of length three are called \emph{triples}. 
We can establish the \hypertarget{prop:vars-union}{\emph{vars-union property}} of tuples: 
\[\vars(\langle \e{1}, \e{2}, \ldots, \e{n} \rangle)
= \vars(\e{1}) \cup \vars(\e{2}) \cup \ldots \cup \vars(\e{n}).
\]

When regarding symbolic expressions as tuples, it is conventional to use the symbols $\emph{head}$ and $\emph{tail}$ instead of $\lef$ and $\rig$.  Thus
\[\emph{head}(\langle \e{1}, \e{2}, \ldots, \e{n} \rangle) = \e{1}\]
and 
\[\emph{tail}(\langle \e{1}, \e{2}, \ldots, \e{n} \rangle) = \langle \e{2}, \ldots, \e{n} \rangle.\]
We also use $\ufirst$, $\usecond$, etc.  Thus
\[\begin{split}
&\ufirst(\langle \e{1}, \e{2}, \ldots, \e{n} \rangle) = \e{1},  \\ 
&\usecond(\langle \e{1}, \e{2}, \ldots, \e{n} \rangle) =  \e{2},
\end{split}
\]
etc.

 It is more common to define unification on functional terms, such as $\emph{f}(\e{1},\e{2},\allowbreak \ldots, \e{n})$, but this is more complex and gives us no additional computational power; if we take function symbols to be constants we can encode $\emph{f}\,(\e{1},\e{2}, \ldots, \e{n})$ as 
          $(\emph{f} \cons \langle \e{1}, \e{2}, \ldots, \e{n} \rangle),$ that is 
                    $(\emph{f} \cons (\e{1} \cons (\e{2} \ldots (\e{n} \cons \nil\,)\ldots ))),$
          as in Lisp. 
   
We also apply $\vars$ to sets of expressions, via the \hypertarget{prop:vars-set}{\emph{vars property}} of sets, namely, 
\[\vars(\{ \e{1}, \e{2}, \ldots, \e{n} \})
= \vars(\e{1}) \cup \vars(\e{2}) \cup \ldots \cup \vars(\e{n}).
\]
\subsection*{Substitutions}
A \emph{proper substitution} $\{\varsub{x}{1} \mapsto \e{1}, \varsub{x}{2} \mapsto \e{2}, \ldots, \varsub{x}{n} \mapsto \e{n}\}$ is a function from expressions to expressions that replaces all occurrences of each variable $\varsub{x}{i}$ with the corresponding term $\e{i}$, where we assume that the $\varsub{x}{i}$'s are distinct and that each variable $\varsub{x}{i}$ is distinct from its corresponding term $\e{i}$.  The replacements are viewed as being performed simultaneously.

We denote by $e \apply \theta$ the result of applying the substitution $\theta$ to the expression $e$; we shall also say that $e \apply \theta$ is an \emph{instance} of $e$ under $\theta$. For example, $(\var{X} \cons (a \cons \var{X})) \apply \{\var{X} \mapsto \var{Y},  \var{Y} \mapsto c\}$ is $(\var{Y} \cons (a \cons \var{Y}))$. Both occurrences of $\var{X}$ are replaced and, because the replacements are done simultaneously, the second replacement $\var{Y} \mapsto c$ has no effect.

The \emph{empty substitution} $\{\}$ makes no replacements at all.  Therefore \hypertarget{def:empty-subst}{$e \apply \{\} = e$} for any expression $e$. 

We introduce one \emph{improper substitution}, the \emph{failure} substitution $\fail$.  This substitution maps any expression into the black hole; that is, \hypertarget{def:blackhole}{$e \apply \bot = \blk$}, for any expression $e$. We say that the relation $\isprop$ holds for any proper substitution; that is, \hypertarget{def:isproper}{$\isprop(\theta)\uiff(\theta \neq \bot)$}. The improper substitution does not follow the axioms for proper substitutions.

To apply a proper substitution to a nonatomic expression, apply it to the components; that is, $ (d \cons e) \apply \theta = (d \apply \theta) \cons (e \apply \theta)$. We call this the \hypertarget{prop:cons-dist}{\emph{distributivity property}} for conses. Equivalently, 
we have the \hypertarget{prop:nonatom-apply}{\emph{apply left-right property}}
\begin{align*}
&{\unot} \,(\isatm(e)) \uand \isprop\,(\theta) \uimplies \\
&e \apply \theta = 
(\lef(e) \apply \theta) \cons (\rig(e) \apply \theta). \end{align*}   Consequently, the result of applying a proper substitution to a nonatomic expression is nonatomic.  
We can establish that proper substitutions distribute over tuples and sets;  that is, we have the \hypertarget{prop:distr-tup}{\emph{distributivity property}} for substitutions over tuples,
\[\langle e_1, \ldots, e_n \rangle \apply \theta = \langle e_1\! \apply \theta, \ldots, e_n\! \apply \theta \rangle, \]
and \hypertarget{prop:distr-set}{\emph{distributivity property}} for substitutions over sets,
\[\{ e_1, \ldots, e_n \} \apply \theta = \{e_1\! \apply \theta, \ldots, e_n\! \apply \theta\}
,\]
for all proper substitutions $\theta$ and expressions $e_1, \ldots, e_n$.
Two substitutions are \hypertarget{def:subst-eq}{equal} if they have the same effect on all expressions;  that is, $\thet{1}=\thet{2}$ provided that, for any expression $e$, 
\[e \,\apply\, \thet{1} = e\, \apply\, \thet{2}.\] 
To show equality of two proper substitutions $\thet{1}$ and $\thet{2}$, it suffices to show, by the  \hypertarget{prop:subst-eq-vars}{\emph{variable equality property}}, that they have the same effect on variables; that is, for any variable {$\var{x}$, 
\[\var{x} \apply \thet{1} = \var{x} \apply \thet{2}.\] 
For example, $\{\var{X} \mapsto a,  \var{Y} \mapsto b\}$ and $\{\var{Y} \mapsto b,  \var{X} \mapsto a\}$ are equal substitutions.

Applying a substitution to expressions preserves the  subexpression relations; that is, for any proper substitution $\theta$,
\hypertarget{prop:subst-propsubexp}{\begin{align*}
& \e{1} \occursin \e{2} \, \uimplies\\
& (\e{1} \apply \theta) \occursin (\e{2} \apply \theta)
\end{align*}} 
and, for any substitution $\theta$,
\hypertarget{prop-subst-subexp}
{\begin{align*} 
& \e{1} \occurseq \e{2} \, \uimplies\\
& (\e{1} \apply \theta) \occurseq (\e{2} \apply \theta)
\end{align*}} 
for all expressions $\e{1}$ and $\e{2}$.

We can establish a \hypertarget{prop-distr-vars}{\emph{distributivity property}} of the $vars$ function for expressions, namely 
\[
\vars(e_1,...,e_n)\!\apply\theta = \vars(e_1\!\apply\theta,...,e_n\!\apply\theta) .\]

By an abuse of notation, we call the set of replaced variables $\{\varsub{x}{1}, \ldots, \varsub{x}{n}\}$ of a substitution $\theta$ its \hypertarget{def:dom}{\emph{domain}}, denoted by $\dom(\theta)$;  and the set of variables that occur in the introduced expressions $\e{1}$, $\e{2}$, \ldots, or $\e{n}$ its \hypertarget{def:range}{\emph{range}}, denoted by  ${\range}(\theta)$.  For example, the domain of $\{\var{X} \mapsto (\var{W} \cons a), \var{Y} \mapsto (\var{X} \cons b)\}$ is $\{\var{X}, \var{Y}\}$, and its range is $\{\var{X}, \var{W}\}$. 
We define the variables \hypertarget{def:subst-vars}{$\vars(\theta)$} of a substitution $\theta$ to be those that occur in either its domain or its range; that is,
\[\vars(\theta) = \dom(\theta) \cup \range(\theta),\] and hence
\[\dom(\theta) \subseteq \vars(\theta)\]
and
\[\range(\theta) \subseteq \vars(\theta).\]
We define $dom(\fail) = \range(\fail) = vars(\fail) = \{\}$, where $\{\}$ is the empty set.

A substitution $\theta$ is said to \hypertarget{def:miss}{\emph{miss}} an expression $e$, written $\misses(\theta, e)$, if applying the substitution has no effect on the expression; that is, if $e \apply \theta = e$. This happens if no variable in the domain of $\theta$ occurs in $e$. 

The domain and range of a substitution need not be disjoint. 
We have
 \hypertarget{prop:dom-apply}{$u \in \dom(\theta) \uiff  u \apply \theta \neq u$}  and \hypertarget{prop:range-apply}{$v \in \range(\theta) \uiff (\exists u)[u \in \dom(\theta) \uand v \occurseq (u \apply \theta)  ].$}  If a variable belongs to the range of a substitution, we also say that the substitution \emph{introduces} the variable. 
 

 It can be proved that two substitutions \hypertarget{prop:substs-agree-vars}{agree} on an expression $\var{e}$ if they agree on all the variables of the expression;  that is $\var{e} \apply \thet{1} = \var{e} \apply \thet{2}$ precisely when $ \var{v} \apply \thet{1} = \var{v} \apply \thet{2}$ for all variables $\var{v}$ that occur in $\var{e}.$

\subsubsection{Addition.} The result of \hypertarget{def:subst-add}{adding} two substitutions $\thet{1}$ and $\thet{2}$, denoted by $\thet{1} \addto \thet{2}$, is the substitution that applies the replacements in parallel.  If the same variable occurs in the domains of both substitutions, the replacement in the first substitution is applied; the replacement in the second substitution is ignored.  More precisely, for any variable $\var{x}$,
\[\begin{split}
\var{x} \apply (\thet{1}\addto \thet{2})= \, &\var{x} \apply  \thet{1} &&\text{if } \var{x} \in \dom(\thet{1})  \\
    = \, &\var{x} \apply \thet{2} &&\text{if } \var{x} \notin \dom(\thet{1}).
\end{split} \]   Addition is not in general comutative.

\subsubsection{Replacement.} A substitution that replaces a single variable $x$ with an expression $e$ is itself called a \emph{replacement}, denoted by $\{x \mapsto e\} $.  It satisfies the \hypertarget{prop:repl-ident}{\emph{replacement identity property}}
\[x \apply  \{x \mapsto e\}  = e\]
 and the \hypertarget{prop:repl-nonocc}{\emph{replacement nonoccur property}}
\[d \apply  \{x \mapsto e\}  = d \, \uimpliedby  \,   \unot(x \occurseq d),\]
for all variables $x$ and expressions $d$ and $e$. In other words, a replacement has no effect on an expression if its variable doesn't occur in that expression. We call $x$ the \emph{replaced variable} and $e$ the \emph{introduced expression}.

We have the \hypertarget{prop-repl-dom}{\emph{{domain property}}},  $dom(\{x \mapsto e\}) = \{x\}$, and the \hypertarget{prop-repl-range}{\emph{range property}},  $\range(\{x \mapsto e\}) = vars(e)$, of the replacement function, if $x$ and $e$ are distinct. We take $\{\var{x} \mapsto \var{x}\}$ to be the empty substitution $\emptysubst$. 

\subsubsection{Composition.} Applying  the \emph{composition} $\expsub {\theta}{1} \compose \thet{2}$ of two substitutions $\thet{1}$ and $\thet{2}$ has the same effect as applying them sequentially;  that is, we have a \hypertarget{prop:subst-comp}{\emph{composition property}} \[e \apply (\thet{1}\! \compose \thet{2}) = (e \apply \thet{1}) \apply \thet{2}.\]For example,
$\{\var{X} \mapsto (\var{Y} \cons a)\} \compose \{\var{X} \mapsto b, \var{Y} \mapsto c\} = \{\var{X} \mapsto (c \cons a), \var{Y} \mapsto c\}$.  Here, the replacement $\var{X} \mapsto b$ plays no role in the composition, because $\var{X}$ has already been replaced. 

The empty substitution \{\} acts as an \hypertarget{prop:comp-emp}{identity} under composition;  that is,  $\theta \compose \{\} = \{\} \compose \theta = \theta$. This holds even if $\theta$ is the improper failure substitution $\fail$. The failure substitution acts as an annihilator; that is,  $\theta \compose \fail = \fail \compose \theta = \fail$.

 Composition is \hypertarget{prop:comp-assoc}{associative}; that is, $(\thet{1}\! \compose \thet{2}) \compose \thet{3}  = \thet{1} \compose (\thet{2} \compose \thet{3})$, whether or not the substitutions are proper.   For this reason, we can write the composition of three or more substitutions without parentheses, as in $\thet{1}\! \compose \thet{2} \compose \thet{3}$.

 The domain and range of the composition of two substitutions $\thet{1}$ and $\thet{2}$ satisfy the inclusions 
 $\dom(\thet{1}\! \compose \thet{2})  \subseteq  \dom(\thet{1}) \cup \dom(\thet{2})$ and  $\range(\thet{1}\! \compose \thet{2})  \subseteq  \range(\thet{1}) \cup \range(\thet{2})$, respectively.  We call these the \hypertarget{prop-domain-comp}{\emph{domain}} and \hypertarget{prop-range-comp}{\emph{range properties of composition}}, respectively.
 
 The inclusions cannot be replaced by equalities. For example, the composition $\{\var{X} \mapsto \var{Y}\} \compose \{\var{Y} \mapsto \var{X}\} = \{\var{Y} \mapsto \var{X}\},$ but
   \begin{align*}
  \dom&(\{\var{X} \mapsto \var{Y}\}) \cup \dom(\{\var{Y} \mapsto \var{X}\}) \\
     & =  \{X\} \cup \{Y\} \\
     & =  \{X, Y\}  
 \end{align*}
and
 \begin{align*}
  \dom&(\{\var{X} \mapsto \var{Y}\} \compose \{\var{Y} \mapsto \var{X}\}) \\
     & = \dom(\{\var{Y} \mapsto \var{X}\}) \\
     & =  \{Y\}. 
   \end{align*}

\subsubsection{Permutation Substitutions and Standardizing Apart.} A \emph{permutation} substitution merely rearranges the variable in its domain. That is, if $\pi$ is a permutation substitution  
 $\{\varsub{x} {1} \mapsto \e{1}, \varsub{x}{2} \mapsto \e{2}, \ldots, \varsub{x}{n} \mapsto \e{n}\},$ the expressions $\{\e{1}, \e{2}, \ldots, \e{n}\}$ are a permutation of the domain variables $\{\varsub{x}{1},\varsub{x}{2}\mathbin{,}  \ldots,  \varsub{x}{n}\}$. 
Every  permutation substitution $\pi$ has an \emph{inverse};  $\pi^{-1}$ is 
 $\{\e{1} \mapsto \varsub{x}{1}, \e{2} \mapsto \varsub{x}{2}, \ldots, \e{n} \mapsto \varsub{x}{n}\}.$ 
 This is a legal substitution because the expressions $\e{1}, \e{2}, \ldots, \e{n}$ are distinct variables. The inverse has the property that $\pi \compose \pi^{-1} = \pi^{-1} \compose \pi = \emptysubst$.  The empty substitution $\emptysubst$ is a permutation.
 
 When we apply a permutation to an expression,  we say that we \emph{rename} its variables.  Two permutation substitutions $\expsub{\pi}{1}$ and $\expsub{\pi}{2}$ \emph{standardize apart} two expressions $\e{1}$ and $\e{2}$ if the results of applying the substitutions to the respective expressions, that is, $\e{1}\apply \expsub{\pi}{1}$ and $\e{2} \apply \expsub{\pi}{2}$, have no variables in common.  It is always possible to standardize apart two expressions.  For instance, suppose  $\varsub{x}{1}$, $\varsub{x}{2}$, $\ldots$ $\varsub{x}{n}$ is a complete list of all the variables that occur in either expression.  Suppose $\varsub{y}{1}$, $\varsub{y}{2}$, $\ldots$ $\varsub{y}{n}$ and $\varsub{z}{1}$, $\varsub{z}{2}$, $\dots$ $\varsub{z}{n}$ are variables that are distinct from each other and from those on the original list of $\varsub{x}{i}$'s.  Then, the two permutation substitutions
 \[\expsub{\pi}{1}: \{\varsub{x}{1} \mapsto \varsub{y}{1}, \varsub{y}{1} \mapsto \varsub{x}{1}, \varsub{x}{2} \mapsto \varsub{y}{2}, \varsub{y}{2} \mapsto \varsub{x}{2}, \ldots, \varsub{x}{n} \mapsto \varsub{y}{n}, \varsub{y}{n} \mapsto \varsub{x}{n}\}\] and 
 \[\expsub{\pi}{2}:\{\varsub{x}{1} \mapsto \varsub{z}{1}, \varsub{z}{1} \mapsto \varsub{x}{1}, \varsub{x}{2} \mapsto \varsub{z}{2}, \varsub{z}{2} \mapsto \varsub{x}{2}, \ldots, \varsub{x}{n} \mapsto \varsub{z}{n}, \varsub{z}{n} \mapsto \varsub{x}{n}\}\]
 will standardize  $\e{1}$ and $\e{2}$ apart.
 All the variables in $\e{1}\apply \expsub{\pi}{1}$ will be $\varsub{y}{i}$'s, and all the variables in $\e{2} \apply \expsub{\pi}{2}$ will be $\varsub{z}{i}$'s. 
 One can even standardize apart two expressions with a single permutation, by renaming all the variables in one of the expressions to variables that don't occur in the other.
 
\subsubsection{Idempotent Substitutions.}  A substitution $\theta$ is \hypertarget{def:idem}{\emph{idempotent}}, written $\idem(\theta)$, if applying it twice (or more) has the same effect as applying it once;  that is, if $\theta \compose \theta = \theta$.  For example, the substitution $\{\var{X} \mapsto \var{Y}\}$ is idempotent, because after it has been applied once, all occurrences of $\var{X}$ have been removed from the expression, so subsequent applications have no effect.  On the other hand, the substitution $\{\var{X} \mapsto (\var{X} {\cons} \var{X})\}$ is not idempotent, because each application doubles the number of occurrences of $\var{X}$.
 
  A substitution is \hypertarget{prop:idem-dom-ran}{idempotent} precisely when no variable appears in both its domain and its range.
  To see this, suppose $\theta$ is such a substitution.  Then, for any expression $e$, the result $e \apply \theta$ of applying $\theta$ to $e$ contains no variables in the domain of $\theta$.  Hence applying $\theta$ a second time has no effect; that is $(e \apply \theta) \apply \theta = e \apply \theta$ and hence $e \apply (\theta \compose \theta) = e \apply \theta$. Since this holds for any expression $e$, we can conclude that it $\theta \compose \theta = \theta.$

 To show the inverse, suppose some variable $\var y$ is in both the domain and range of $\theta$. Then, because  $\var{y}$ is in the domain, $\var{y} \apply \theta \neq \var{y}.$  And because $\var{y}$ is in the range, there is a variable $\var{x}$ such that $\var{y} \occurseq (\var{x} \apply \theta).$  Because $\var{y}$ is in its domain, applying $\theta$ again to $\var{x} \apply \theta$ alters $\var{x} \apply \theta$.  In other words, $(\var{x} \apply \theta) \apply \theta \neq \var{x} \apply \theta$;  that is, $\var{x} \apply (\theta \compose \theta) \neq \var{x} \apply \theta$, and therefore $\theta \compose \theta \neq \theta;$  that is, $\theta$ is not idempotent.

 The empty substitution $\{\}$ is idempotent, but no other permutation substitution is.   The composition of two idempotent unifiers is not necessarily idempotent;  for example, \[\{\var{X} \mapsto \var{Y}\} \compose \{\var{Y} \mapsto (\var{X} {\cons} \var{X})\} = \{\var{X} \mapsto (\var{X} {\cons} \var{X}),\var{Y} \mapsto (\var{X} {\cons} \var{X})\},\] which has $\var{X}$ in both its domain and range.

\subsubsection{The Vars-Range Subset Properties.}
 For any expression $\var{e}$ and proper substitution $\theta$, we can informally establish the \hypertarget{prop:vars-range-subset}{\emph{vars-range subset property}}
\[\vars(\var{e} \apply \theta) \subseteq \vars(\var{e}) \cup \range(\theta).\]
In other words, for a variable to occur in $\var{e} \apply \theta $, it must already occur in $\var{e}$ or be introduced by $\theta.$

In general, the subset relation in the property is not necessarily proper; for instance, if $\theta$ is the empty substitution $\emptysubst$, $\range(\theta)$ is the empty set and both sides of the property reduce to $\vars(\var{e})$.  But if $\theta$ is a proper idempotent substitution that does not miss the expression $\var{e}$, we do have the \hypertarget{prop:vars-range-proper-subset}{\emph{vars-range proper-subset property}}
\[\vars(\var{e} \apply \theta) \cup \range(\theta) \subset \vars(\var{e}) \cup \range(\theta).\]
For, if $\theta$ does not miss $\var{e}$, there is a variable $\var{v}$ that occurs in $\var{e}$ and belongs to $\dom(\theta)$.  Because $\theta$ is idempotent, $\var{v}$ does not belong to $\range(\theta)$ and hence does not occur in $\var{e} \apply \theta$.  Thus $\var{v}$ belongs to the right side of the formula but not the left.

These properties hold as well if $e$ is a tuple or set of expressions, rather than a single expression. 
They turn out to be useful in the derivation of unification algorithms

\subsection*{Unifiers}  A substitution $\theta$ is a unifier of two expressions $\e{1}$ and $\e{2}$ if applying it makes them identical, that is, if $\e{1} \apply \theta = \e{2} \apply \theta$. For example, the substitution  $\{\var{X} \mapsto a, \var{Y} \mapsto b\}$ is a  unifier of the expression $(\var{X} \cons b)$ and $(a \cons \var{Y})$ because applying it to either expression maps it into $(a \cons b)$.
 
 Unifiers are not unique. For example,  $\{\var{X} \mapsto a, \var{Y} \mapsto b, \var{Z} \mapsto c\}$ is also  unifier of the expression $(\var{X} \cons b)$ and $(a \cons \var{Y})$.   The third replacement $\var{Z} \mapsto c$ has no effect on either expression.  
 
 The empty substitution $\emptysubst$ is a unifier of two expressions only if they are already identical.  The improper failure substitution $\fail$, on the other hand,  is a unifier of any two expressions; applying it always yields the black hole \hypertarget{prop:unif-blk}{$\blk$}.
 
 We say that two expressions are \hypertarget{def:unifiable}{\emph{unifiable}} if they have a proper unifier, i.e., a unifier other than $\fail$. For example, the two distinct constants $a$ and $b$ are ununifiable, because no proper substitution can make them identical. A constant and a nonatomic expression can never be unifiable, because the result of applying a proper substitution to a constant is the constant itself, an atom, while the result of applying the same substitution to a nonatomic expression is nonatomic. If one expression $\e{1}$ is a proper subexpression of another $\e{2}$, that is, if $\e{1} \occursin \e{2}$, then
 $(\e{1}\apply \theta) \occursin (\e{2}\apply \theta)$; hence the results cannot be equal and the two given expressions are ununifiable.

If $\theta$ is a unifier of two expressions, the result $\theta \compose \delta$ of composing $\theta$ with any substitution $\delta$ is also a unifier of the two expressions. Hence if two expressions are unifiable, they have an infinite set of unifiers. 

 For a substitution $\theta$ to unify two nonatomic expressions $\e{1}$ and $\e{2}$, it must unify both their left components and their right components; that is, we have the \hypertarget{prop:unify-left-right}{\emph {unify-left-right property}},

    \[\begin{aligned}
  &\e{1} \apply \theta = \e{2} \apply \theta
  \uiff \\
  &\left[{\begin{aligned}
  &\lef(\e{1}) \apply \theta = \lef(\e{2}) \apply \theta \uand \\
  & \rig(\e{1}) \apply \theta = \rig(\e{2}) \apply \theta
  \end{aligned}} \right].\\[10pt]
\end{aligned}\]
\noindent This holds even if $\theta$ is the improper substitution $\fail$.

 We have seen that, if two expressions are unifiable, they have an infinite set of different unifiers.  But we want our unification algorithm to yield a unifier that leaves our variables as unconstrained as possible.
 
 \subsubsection{More Generality.}  A substitution $\thet{1}$ is \hypertarget{def:more-gen}{\emph{strongly more general}} than a substitution $\thet{2}$, written  $\thet{1}$ $ \moregen$ $\thet{2}$, if $\thet{1}\! \compose \thet{2} = \thet{2}$. In this case, we can also say that $\thet{2}$ is an \emph{extension} of $\thet{1}$.  For example, the substitution $\{\var{X} \mapsto \var{Y}\}$ is strongly more general than $\{\var{X} \mapsto A, \var{Y} \mapsto A\}$ because $\{\var{X} \mapsto \var{Y}\} \compose \{\var{X} \mapsto A, \var{Y} \mapsto A\} = \{\var{X} \mapsto A, \var{Y} \mapsto A\}$. 
 
 The empty substitution $\{\}$ is \hypertarget{prop:more-gen-empty}{strongly more general} than any substitution $\theta$, because $\{\} \compose \, \theta = \theta$. However, no other substitution $\theta$ is \hypertarget{not-more-gen-empty}{strongly more general} than the empty substitution, because $\theta \compose $\{\}$= \theta \neq \{\}.$ On the other hand, the failure substitution $\fail$ is \hypertarget{prop:fail-not-more-gen}{not strongly more general} than any proper substitution $\theta$, because $\fail \compose \, \theta = \fail \neq \theta.$ But any substitution $\theta$ is strongly more general than the failure substitution, because $\theta \compose \fail = \fail.$

 The strongly-more-general relation has a \hypertarget{prop:more-gen-comp}{\emph{composition property}}, namely
 \[\thet{0} \moregen \thet{1} \uimplies \thet{0} \moregen \thet{1}\! \compose \thet{2},\]
for all substitutions $\thet{0}$, $\thet{1}$, and $\thet{2}.$  For, if
\[\thet{0} \moregen \thet{1},\]
that is [by the definition of $\moregen$]
\[\thet{0} \compose \thet{1} = \thet{1},\]
then
\[\thet{0} \compose \thet{1}\! \compose \thet{2} = \thet{1}\! \compose \thet{2},\]
that is [by the definition of $\moregen$, again]
\[\thet{0} \moregen \thet{1}\! \compose \thet{2},\]
as we wanted to show.

 If $\theta$ is a unifier of two expressions, then clearly \hypertarget{prop:unif-ext}{any extension} of $\theta$ is also a unifier.  

 \citet{rob} used the term “most-general unifier" for the output of the unification algorithm, but he did not define what it meant for one substitution to be more general than another.   \citet{man:wal} said that  $\thet{1}$ is more general than $\thet{2}$ if $(\exists \delta)[\thet{1} \compose \, \delta = \thet{2}]$ holds; we call this the weakly more-general relation $\thet{1} \weakmoregen \thet{2}$.  Clearly the strongly more-general relation implies this weakly more-general relation.  When $\thet{1}$ is idempotent, the weakly more-general relation also implies the  strongly-more-general relation.  To see this, assume that $\thet{1} \weakmoregen \thet{2}$; that is, for some substitution $\delta$,  $\thet{1} \compose \, \delta = \thet{2}$.  Then we have 
 \begin{align*}
   \thet{1}\! \compose \thet{2}
     & =  \thet{1} \compose \thet{1} \compose \, \delta  && \text{[by assumption]} \\
     & = \thet{1} \compose \, \delta  && \text{[by idempotence]} \\
     & = \thet{2} && \text{[by assumption, again].}
 \end{align*}
 \noindent Hence $\thet{1} \moregen \thet{2}$, as we wanted to show.

 On the other hand, if $\theta$ is the  permutation $\{\var{X} \mapsto \var{Y}, \allowbreak \var{Y} \mapsto \var{X}\}$, which is not idempotent, we have $\theta \compose \theta = \emptysubst$, and thus $\theta \weakmoregen \emptysubst$; that is,  $\theta$
is  weakly more general than 
$\emptysubst$.  But $\theta \compose \emptysubst = \theta \neq \emptysubst$, and hence it is not true that $\theta \moregen \emptysubst$; that is, $\theta$ is not strongly more general than the empty substitution.

Although the stronger relation is more difficult to grasp, using it allows us to simplify the specification and shorten the proof. In this paper, we will use only the strongly-more-general relation, sometimes dropping the adverb “strongly."

Clearly, a substitution $\theta$ is idempotent precisely when it is (strongly) more  general than itself; that is, we have the \hypertarget{prop:idem-more-gen}{\emph{idempotent-more-general property}}, that $\idem(\theta)$ is equivalent to $ \theta \moregen \theta$, because then $\theta \, \compose \,  \theta = \theta$. Thus, while the strongly-more-general relation is not reflexive, it is \hypertarget{prop:more-gen-refl}{reflexive} on idempotent substitutions. 

The more-general relation can also be shown to be \hypertarget{prop:moregen-trans}{transitive}; that is,
\[(\thet{1} \moregen \thet{2} \uand 
\thet{2} \moregen \thet{3}) \uimplies \thet{1} \moregen \thet{3}.\]
For, suppose \[\thet{1} \moregen \thet{2} \uand 
\thet{2} \moregen \thet{3}.\]  
Then, by the definition of $\moregen$, we have \[\thet{1}\! \compose \thet{2} = \thet{2} \uand \thet{2} \compose \thet{3} = \thet{3}.\]  Hence \[\thet{1}\! \compose \thet{2} \compose \thet{3} =  \thet{2} \compose \thet{3}\] [by the supposition that [$\thet{1}\! \compose \thet{2} = \thet{2}$] and thus \[\thet{1} \compose \thet{3} = \thet{3}\] [by the supposition that $\thet{2} \compose \thet{3} = \thet{3}$]. In other words, \[\thet{1} \moregen \thet{3},\] as we wanted to show.

\subsubsection{Relativizing the Definitions.} 
It is remarked that, in a proof by induction, it is sometimes easier to prove a stronger, more general theorem, to have the benefit of a stronger induction hypothesis; this is an instance of what \citet{pol}  termed the \emph{inventor's paradox}, famously exploited by the Boyer-Moore [\citeyear{boy:moo}] theorem prover.  Had we initially specified that we merely wanted to find a unifier of two expressions, we would have been led to require that the unifier be most general.  Then, we would have been led to further require that the unifier be idempotent. We don't know how to automate the generalization of these specifications; we regard this as a \emph{eureka!} step, as in program transformation \citep{bur:dar}. Perhaps a failed inductive proof will give us a hint of how we would like to strengthen the theorem; that is, how would we strengthen the corresponding induction hypothesis to allow the proof to succeed.

We now generalize the specification further. For the current derivation, we have found it easier to construct an algorithm that performs unification with respect to an idempotent \emph{environment substitution}.  That is, we search for a unifier that is an extension of a given idempotent substitution $\thet{0}$. This is a strengthening, because, when the environment $\theta$ is the empty substitution $\emptysubst$, it reduces to the original problem, since any substitution is an extension of the empty substitution.  Relative unification requires a more general specification, but a simpler proof gives us a more efficient resulting program.  Indeed, in the implementation of resolution and other rules,  Snark and other theorem provers perform unification with respect to an environment substitution.

\subsubsection{Most-General Idempotence Relation.}  To specify the relative unification algorithm, it is convenient for us to introduce the \hypertarget{def-mgi}{\emph{most-general idempotence relation}}  ${\mgi} (\thet{0},  \e{1}, \e{2}, \theta)$  to stand for the condition that the substitution $\theta$ 
\hypertarget{def-mgi}{\emph{is most-general idempotent}}  for the expressions
$\e{1}$ and $\e{2}$ with respect to the environment substitution $\thet{0}$, that is
\begin{align*}
 {\mgi}(&\thet{0}, \e{1}, \e{2}, \theta) \uiff \\
&\begin{aligned}
 (\forall \theta') \left[{
 \begin {aligned} &\e{1} \apply \theta' = \e{2} \apply \theta' \, \uand \,  
  \,\thet{0} \moregen \theta' \,\uimplies\\
                    &\theta \moregen \theta'
   \end{aligned}}
   \right]. \end{aligned}
   \end{align*}
The relation does not require $\theta$ to be a unifier of the expressions.

Because the roles of $\e{1}$ and $\e{2}$ are interchangeable, the most-general-idempotent relation has a \hypertarget{prop:sym-mgi}{\emph{symmetry property}} 
\begin{align*}
 {\mgi}(\thet{0}, \e{1}, \e{2}, \theta) \,\, \uiff \,\, 
{\mgi}(\thet{0}, \e{2}, \e{1}, \theta).
   \end{align*}
The relation also has a kind of \hypertarget{prop:reflex-mgi}{\emph{reflexivity property}}; when $\thet{0}$ and $\theta$ are equal, the relation clearly holds, because then $\thet{0} \moregen \theta'$ appears in the antecedent and as the consequent of the implication.  Hence
\begin{align*}
[\thet{0} = \theta] \,\, \uimplies \,\,
 {\mgi}(\thet{0}, \e{1}, \e{2}, \theta).
   \end{align*}
   \noindent In other words, the environment $\thet{0}$ itself is always most-general idempotent for any expressions.

If the two expressions $\e{1}$ and $\e{2}$ are ununifiable, any substitution $\theta$ will satisfy the most-general-idempotent relation.   When the substitution $\theta'$ is the improper substitution $\fail$, the consequent  $\theta \moregen \theta'$ of the above definition  of $\mgi$ is true, because any substitution $\theta$ is strongly more general than the failure substitution.  But if $\theta'$ is proper, because the two expressions are ununifiable, the antecedent of the definition will be false, and the implication itself will be true.  In either case, if $\e{1}$ and $\e{2}$ are ununifiable, any substitution $\theta$ is most-general idempotent. We call this the \hypertarget{prop:unun-mgi}{\emph{ununifiable property} of $\mgi$}.

For instance, if the two expressions are distinct constants, any substitution will be most-general idempotent with respect to any environment. That is,
\begin{align*}
 &{\mgi}(\thet{0}, \e{1}, \e{2}, \theta) \,\, \uimpliedby \\ \,\,
&[\iscnst(\e{1}) \uand \iscnst(\e{2}) \uand \e{1} \neq \e{2} ].
   \end{align*}
Similarly, if one of the expressions is a \hypertarget{prop-mgi-const}{constant} and the other is nonatomic, any substitution will be most-general idempotent; that is, 
\begin{align*}
 &{\mgi}(\thet{0}, \e{1}, \e{2}, \theta) \,\, \uimpliedby \\ \,\,
&[\iscnst(\e{1}) \uand \unot(\isatm(\e{2}))].
   \end{align*}
And if one of the expressions is a 
\hypertarget{prop:mgi-occ}{proper subexpression} of the other, any substitution will be most-general idempotent; that is, 
\begin{align*}
 &{\mgi}(\thet{0}, \e{1}, \e{2}, \theta) \,\, \uimpliedby \,\,
[\e{1} \occursin \e{2}].
   \end{align*}

    While the improper failure substitution $\fail$ unifies any two expressions, it will not be most-general idempotent unless the two expressions are ununifiable with respect to that environment; we call this the \hypertarget{prop:mgi-fail}{\emph{failure property}} of $\mgi$.  For, if $\e{1}$ and $\e{2}$ are unified by a proper substitution $\theta'$ that is an extension of the environment $\thet{0}$, the antecedent is true but
it cannot be true that the consequent $\fail \moregen \theta'$ holds;  the failure substitution is not more general than any proper substitution. In other words, the implication
 {\begin {align*} &\e{1} \apply \theta' = \e{2} \apply \theta' \, \uand \,     \thet{0} \moregen \theta' \,\uimplies\\
                    &\fail \moregen \theta'
   \end{align*}}
must be false.
\paragraph{Replacement Property of Most-General Idempotence.}
 The \hypertarget{prop:repl-mgi}{\emph{$\mgi$ replacement property} } expresses the relationship between replacement and the most-general-idempotent relation:
\[
\begin{aligned}
    &\mgi(\thet{0}, \e{1}, \e{2}, \thet{0} \compose \{\e{1} \mapsto \e{2}\}) \uimpliedby \\
    &\isvar(\e{1})
\end{aligned}
\]
for all substitutions $\thet{0}$ and expressions $\e{1}$ and $\e{2}$.

To prove this, it suffices (by the definition of the relative \hyperlink{def:more-gen}{most-general idempotence relation} $\mgi$) to 
assume that, for some substitution $\theta$, variable $\e{1}$ and expression $\e{2}$, 
\[\hypertarget{ass:theta-un}{\e{1} \apply \theta = \e{2} \apply \theta}\]
and
\[{\thet{0} \moregen \theta},\]
that is [by the \hyperlink{def:more-gen}{definition of the relation $\moregen$}], 
\[\hypertarget{ass:thet0-moregen}{\thet{0} \compose \theta = \theta}.\]
We show that then
\[\thet{0} \compose \{\e{1} \mapsto \e{2}\} \moregen \theta,\]
that is [by definition of $\moregen$, again],
\[\thet{0} \compose \{\e{1} \mapsto \e{2}\} \compose \theta = \theta .\]
It suffices to show [because $\thet{0} \compose \theta = \theta$] that
\[\thet{0} \compose \{\e{1} \mapsto \e{2}\} \compose \theta = \thet{0} \compose \theta, \]
or [by properties of equality] that
\[\{\e{1} \mapsto \e{2}\} \compose \theta = \theta. \]
It further suffices [by the \hyperlink{prop:subst-eq-vars}{\emph{variable equality property}} for substitutions] to show that, for any variable $v$,
\[v \apply \{\e{1} \mapsto \e{2}\} \compose \theta =  v \apply \theta. \] 

    When $v$ is actually $\e{1}, $ this reduces [by the \hyperlink{prop:repl-ident}{\emph{replacement identity property}} and the \hyperlink{prop:subst-comp}{\emph{composition property}} of substitutions] to
\[\e{2} \apply \theta = \e{1} \apply \theta,\]
which we have \hyperlink{ass:theta-un}{assumed}.

When $v$ is distinct from $\e{1}$, this reduces [by the \hyperlink{prop:repl-nonocc}{\emph{replacement nonoccur property}}] to
\[v \apply \theta = v \apply \theta, \] which is true.  This concludes the proof.

\paragraph{Instance Property of Most-General Idempotence.} The most-general idempotence relation has an \hypertarget{prop:mgi-instance}{\emph{instance property}} that is useful in the derivation of unification algorithms, namely, that
\begin{align*}
 {\mgi}(\thet{0}, \e{1}, \e{2}, \theta) \,\, \uiff \,\, 
  {\mgi}(\thet{0}, \e{1}\!\apply\thet{0}, \e{2}\!\apply\thet{0}, \theta),
 \end{align*}
that is, 
 \begin{align*}
 &\begin{aligned}
 (\forall \theta') \left[{
 \begin {aligned} &\e{1} \apply \theta' =
                    \e{2} \apply \theta' \, \uand \,  
  \,\thet{0} \moregen \theta' \,\uimplies\\
                    &\theta \moregen \theta'
   \end{aligned}}
   \right] \end{aligned} \uiff \\
&\begin{aligned}
 (\forall \theta') \left[{
 \begin {aligned} &\e{1}\!\apply\thet{0} \apply \theta' =
                   \e{2}\!\apply\thet{0} \apply \theta' \, \uand \,  
  \,\thet{0} \moregen \theta' \,\uimplies\\
                    &\theta \moregen \theta'
   \end{aligned}}
   \right] \end{aligned}
   \end{align*}
for any expressions $\e{1}$, $\e{2}$, and substitutions $\thet{0}$ and $\theta$. The consequents of the implications on the left and the right side of the equivalence are the same, $\theta \moregen \theta'$. So it suffices to show that the antecedents are equivalent.  If $\thet{0} \moregen \theta'$ is false, both antecedents are false, so we may assume $\thet{0} \moregen \theta'$, that is $\thet{0} \compose \theta' = \theta'$, is true.

The antecedent of the right side,
\[\e{1}  \apply \thet{0} \apply \theta' = \e{2} \apply\thet{0} \apply \theta' \, \uand \,   
  \thet{0} \moregen \theta', \]
  is equivalent to
 \[\e{1}  \apply \thet{0} \compose \theta' = \e{2}  \apply \thet{0} \apply \theta' \, \uand \,   
  \thet{0} \compose \theta' = \theta' \]
[by the \hyperlink{prop:subst-comp}{\emph{composition property}} and the \hyperlink{def:more-gen}{definition of $\moregen$]},  which is equivalent to 
\[\e{1} \apply \theta' = \e{2} \apply \theta' \, \uand \,   
  \thet{0} \compose \theta' = \theta'\]
[because $\thet{0} \compose \theta' = \theta'$],  which is equivalent to
\[\e{1} \apply \theta' = \e{2} \apply \theta' \, \uand \,   
  \thet{0} \moregen \theta'. \]
  But this is the antecedent of the left side, ${\mgi}(\thet{0}, \e{1}, \e{2}, \theta).$
Hence, we have the desired result.

  \paragraph{Most-General Idempotence on Nonatomic Expressions.} We can construct most-general idempotent substitutions for nonatomic expressions from most-general idempotent substitutions for their components. There is a kind of transitivity in the relative most-general-idempotence relation.
  More precisely, for all expressions $\d{l}$, $\d{r}$, $\e{l}$, $\e{r}$ and all substitutions $\thet{0}$, $\thet{1}$, $\thet{2}$, we have the \hypertarget{prop:mgi-trans}{\emph{transitivity property}} of $\mgi$,
   \begin{alignat*}{1}
 &\left[\begin{aligned}
  &\mgi (\thet{0}, \d{l}, \e{l}, \thet{1})
  \, \uand \\
   &\mgi (\thet{1}, \d{r}, \e{r}, \thet{2}) 
   \end{aligned}\right] \, \uimplies\\
   &\mgi(\thet{0}, (\d{l} {\cons} \d{r}), (\e{l} {\cons} \e{r}), \thet{2}). 
  \end{alignat*}

   \noindent To prove this, assume that ${\mgi}(\thet{0}, \d{l}, \e{l}, \thet{1})$ and ${\mgi}(\thet{1}, \d{r}, \e{r}, \thet{2}) $; it suffices to show that ${\mgi}(\thet{0}, (\d{l} {\cons} \d{r}), (\e{l} {\cons} \e{r}), \thet{2})$.  For this purpose, expanding the \hyperlink{def-mgi}{definition of $\mgi$}, we assume  that $\theta'$ is a substitution such that $(\d{l} {\cons} \d{r}) \apply \theta' = (\e{l} {\cons} \e{r}) \apply \theta'$ and $\thet{0} \moregen \theta'$; we attempt to show that $\thet{2} \moregen \theta'$.
   
   Because  $(\d{l} {\cons} \d{r}) \apply \theta' = (\e{l} {\cons} \e{r}) \apply \theta'$, we have [by the \hyperlink{prop:cons-dist}{\emph{distributivity property}} of \emph{cons}] that $\d{l} \apply \theta' = \e{l} \apply \theta'$ and  $\d{r} \apply \theta' = \e{r} \apply \theta'$.
   
   Because ${\mgi}(\thet{0}, \d{l}, \e{l}, \thet{1})$, $\d{l} \apply \theta' = \e{l} \apply \theta'$, and $\thet{0} \moregen \theta'$, we have [by the \hyperlink{def-mgi}{definition of $\mgi$]}, $\thet{1} \moregen \theta'$.
   
   Because ${\mgi} (\thet{1}, \d{r}, \e{r}, \thet{2})$, $\d{r} \apply \theta' = \e{r} \apply \theta'$, and $\thet{1} \moregen \theta'$, we have [again by the definition of $\mgi$] $\thet{2} \moregen \theta'$, which is what we wanted to show.
   
  Because any nonatomic expression $e$ can be \hyperlink{prop:nonatom}{decomposed} into  $\lef(e)\cons \rig(e)$, we can phrase the above \hypertarget{prop:mgi-trans-LR}{\emph{transitivity property} of $\mgi$} as
    \[
    \begin{aligned}
    &\unot(\isatm(\e{1})) \uand \unot(\isatm(\e{2})) \uimplies\\
&\left[    \begin{aligned}
   &\left[\begin{aligned}
  &\mgi (\thet{0}, \lef(\e{1}), \lef(\e{2}), \thet{1})
  \, \uand \\
   &\mgi (\thet{1}, \rig(\e{1}), \rig(\e{2}), \thet{2}) 
   \end{aligned}\right] \uimplies \\
  &\mgi(\thet{0}, (\lef(\e{1}) {\cons} \rig(\e{1})), (\lef(\e{2}) {\cons} \rig(\e{2})), \thet{2})
   \end{aligned} \right].
   \end{aligned}
    \]
    \subsubsection{Reduction  Substitutions.}
To show the termination of the unification algorithms we were deriving, we found it necessary to strengthen the specification by requiring that the unifier we find be a \hypertarget{def:reduct}{\emph{reduction}}, in the sense that it maintains or reduces a certain set of variables in the environment or the unified expressions.  More precisely, for a given environment $\theta_0$ and expressions $e_1$ and $e_2$, we introduce the concept that a substitution is a \emph{reduction}, written $\reduce(\theta_0, vars(e_1, e_2)\!\apply\!\theta_0, \theta)$, where $\reduce$ is characterized by the 
\hypertarget{prop:red-def}{definition}
\begin{align*}
 \reduce(\thet{0}, v, \theta) \,\, \uiff \,\, 
\range(\theta) \subseteq\range(\theta_0) \cup  v,
 \end{align*}
 for any set of variable $v$.
 For example, for the environment $\{X \mapsto Y\}$ and variable set  $\{Y\}$, the substitution $\{X \mapsto Y\}$ itself is a reduction, because $\reduce(\{X \mapsto Y\},  \{Y\}, \{X \mapsto Y\} )$, that is, 
 \[\range(\{X \mapsto Y\})\ = \{Y\} \subseteq \{Y\} =\range(\{X \mapsto Y\}) \union \{Y\}. \]
 On the other hand, $\{Y \mapsto X\}$ is not a reduction, because 
   \[\range(\{Y \mapsto X\}) = \{X\} \not \subseteq \{Y\}.\]

 The reduction relation satisfies a kind of \hypertarget{prop:reflex-reduce}{\emph{reflexivity property}}, namely
 \[\reduce(\theta, v, \theta),\]
 since, for any substitution $\theta$ and set of variables $v$,
 \[\range(\theta) \subseteq \range(\theta) \union v.\]

 The reduction relation also satisfies a kind of \hypertarget{prop:trans-reduce}{\emph{transitivity property}}, namely
    \begin{alignat*}{1}
 &\left[\begin{aligned}
  &\!\reduce(\theta_0, v_1, \theta_1)
  \, \uand \\
   &\!\reduce(\theta_1, v_2, \theta_2) 
   \end{aligned}\right] \, \uimplies\\
   &\reduce(\theta_0, v_1 \union v_2, \theta_2)
  \end{alignat*} 
  for any substitutions $\theta_0$,  $\theta_1,$ and  $\theta_2$, and sets of variables $v_1$ and $v_2.$
To prove this, suppose $\reduce(\theta_0, v_1, \theta_1)$ and $\reduce(\theta_1, v_2, \theta_2)$.  We need to show 
$\reduce(\theta_0, v_1 \union v_2, \theta_2).$  We have
\begin{align*}
 \range(\theta_2) &\subseteq  \range(\theta_1) \union v_2  &&  
  \text{[by the \hyperlink{prop:red-def}{definition of $\reduce$}]} \\
    &\subseteq  \range(\theta _0) \union v_1 \union v_2  &&  
  \text{[by definition of $\reduce$, again]}.
 \end{align*} 
Hence [by the definition of $\reduce$, once again] $\reduce(\theta_0, v_1 \union v_2, \theta_2)$, as we wanted to show. 

We can also establish an \hypertarget{prop:instance-reduce}{\emph{instance property}} of the reduction relation, namely
\begin{align*}
    \var{}reduce(\theta_0, v,\, \theta) \uiff 
        \reduce(\theta_0, v \apply \theta_0, \theta).
\end{align*}
For
 \[\reduce(\theta_0, v \apply \theta_0, \theta) \]
 precisely when
 \[\range(\theta) \subseteq \range(\theta _0) \union v \apply \theta_0    
  \quad \text{[by the \hyperlink{prop:red-def}{definition of $\reduce$}]}\]
  precisely when 
 \[\range(\theta) \subseteq \range(\theta _0) \union \, v \, \union \,\range(\theta_0) \,
     \text{[by the \hyperlink{prop:vars-range-subset} {\emph{vars-range subset property}]}}\] 
    precisely when
 \[\range(\theta) \subseteq \range(\theta _0) \union v 
    \quad    \text{[by properties of sets]}\]
      precisely when
    \[reduce(\theta_0, v,\, \theta)\quad  \text{[by the definition of $\reduce$, again]},\]
as we wanted to show.

    Because $\range(\fail) = \uempty$, we have the 
    \hypertarget{prop-red-fail}{\emph{reduction fail property}}
 \[\reduce(\theta_0, v, \fail), \]
 for any substitution $\theta_0$ and set of variables $v$.
    \subsubsection{Most-General Idempotent Reducing Unifier.} 
Finally, we introduce the relation \hypertarget{def:mgiu} ${\mgiu} (\thet{0}, \e{1}, \e{2}, \theta)$ that will be used in the specification of the unification algorithm.  This relation holds when the substitution $\theta$ is a \hypertarget{prop:def-mgiu}{\emph{most-general idempotent reducing unifier}} of the expressions $\e{1}$ and $\e{2}$ with respect to the environment substitution $\thet{0}$; that is, 
\begin{align*} 
 {\mgiu}(&\thet{0}, \e{1}, \e{2}, \theta) \uiff \\
&\begin{aligned}
&\e{1} \apply \theta = \e{2} \apply \theta \, \uand 
&&\text{\qquad [$\theta$ is a unifier]} \\
 \,\, & \thet{0} \moregen \theta \, \uand
&&\text{\qquad [$\theta$ is an extension of $\thet{0}]$}\\
  \,\, & {\mgi}(\thet{0}, \e{1}, \e{2}, \theta)\, \uand 
 &&\text{\qquad [$\theta$ is most-general idempotent]}\\
 \,\, & {\reduce}(\theta_0, vars(e_1, e_2)\!\apply\! \theta_0, \theta)
 &&\text{\qquad [$\theta$ is a reduction].}
  \end{aligned}.\end{align*}

   \noindent In other words, $\theta$ is a unifier of the expressions $\e{1}$ and $\e{2}$, $\theta$ is an extension of the environment $\thet{0}$, $\theta$ is a most-general idempotent reduction for the argument expressions $\e{1}$ and $\e{2}$ with respect to $\thet{0}$, and $\theta$ is a reduction with respect to the environment and the variables of the arguments when instantiated by the environment substitution.

   For example, if $\thet{0}$ is the substitution  $\{\var{X} \mapsto \var{Y}\}$ and the two expressions $\e{1}$ and $\e{2}$ are $\var{Y}$ and $\var{Z}$, respectively, then the substitution $\theta = \{\var{X} \mapsto \var{Z}, \var{Y} \mapsto \var{Z}\}$ is a most-general idempotent reducing unifier, i.e., it satisfies 
   $\mgiu(\thet{0}, \e{1}, \e{2}, \theta)$.  First, $\theta$ is a unifier, because
   
   \begin{align*}
   \e{1} \apply \theta  &= \var{Y} \apply  \{\var{X} \mapsto \var{Z}, \var{Y} \mapsto \var{Z}\} \\
                        &= \var{Z} \\
                        &= \var{Z} \apply  \{\var{X} \mapsto \var{Z}, \var{Y} \mapsto \var{Z}\} \\
                        &= \e{2} \apply \theta.
    \end{align*}                     

Also $\theta$  is an extension of $\thet{0}$, because
  \begin{align*}
  \thet{0} \compose \theta &= \{\var{X} \mapsto \var{Y}\} \compose \{\var{X} \mapsto \var{Z}, \var{Y} \mapsto \var{Z}\} \\
                           &= \{\var{X} \mapsto \var{Z}, \var{Y} \mapsto \var{Z} \} \\
                           &= \theta.
   \end{align*}

Also, we can show that $\theta$ is most-general idempotent for this environment and expressions, that is,
$\mgi(\thet{0}, \e{1}, \e{2}, \theta)$.  For {\hypertarget{sup:mgi}{suppose}}, for some substitution $\theta'$, that
    $\e{1} \apply \theta' = \e{2} \apply \theta',$
that is,
\begin{align*}
    \var{Y} \apply \theta' = \var{Z} \apply \theta',  
\end{align*}
and that $\theta'$ is an extension of $\thet{0}$, that is,
\[\thet{0} \compose \theta' = \theta'.\]
Then, we want to show $\theta$ is \hyperlink{def:more-gen}{strongly more general than $\theta'$}, that is,
\begin{align*}
    \theta \compose \theta'  = \theta'.
\end{align*} 
To show this it suffices [by the \hyperlink{prop:subst-eq-vars}{\emph{variable equality property} for substitutions}] to show 
\begin{align*}
    \var{v} \apply {\theta \compose \theta' } = \var{v} \apply \theta,
\end{align*}
that is [by the \hyperlink{prop:subst-comp}{\emph{composition property}}],
\begin{align*}
    \var{v} \apply \theta \apply \theta'  = \var{v} \apply \theta,
\end{align*}
for any variable $\var{v}.$

When $\var{v}$ is $\var{X}$, we must show
\begin{align*}
    \var{X} \apply \theta \apply \theta'  = \var{X} \apply \theta' 
\end{align*}
or, equivalently [because $\var{X} \apply \theta = \var{Z}$], 
\begin{align*}
    \var{Z} \apply \theta'  = \var{X} \apply \theta' 
\end{align*}
or, equivalently [because $\thet{0} \compose \theta' =  \theta'$],
\begin{align*}
    \var{Z} \apply \thet{0} \apply \theta'  = \var{X} \apply \thet{0} \apply \theta' 
\end{align*}
or, equivalently [because $\var{Z} \apply \thet{0} =  \var{Z}$ and 
    $\var{X} \apply \thet{0} = \var{Y}$],
\begin{align*}
    \var{Z}  \apply \theta'  = \var{Y} \apply \theta' 
\end{align*}
which is true by {\hyperlink{sup:mgi}{our supposition}}.

When $\var{v}$ is $\var{Y}$, we must show
\begin{align*}
    \var{Y} \apply \theta \apply \theta'  = \var{Y} \apply \theta' 
\end{align*}
or, equivalently {[because $\var{Y} \apply \theta = \var{Z}$]}, 
\begin{align*}
    \var{Z} \apply \theta'  = \var{Y} \apply \theta'.
 \end{align*}   
 But this is true by {\hyperlink{sup:mgi}{the supposition}}.

 When $\var{v}$ is any variable other than $\var{X}$ or $\var{Y}$, we must show
 \begin{align*}
    \var{v} \apply \theta \apply \theta'  = \var{v} \apply \theta' 
\end{align*}
 or, equivalently [because $\theta$ misses $\var{v}$],
  \begin{align*}
    \var{v} \apply \theta'  = \var{v} \apply \theta'.
\end{align*}
But this is identically true.

Finally, we can show that $\theta = \{\var{X} \mapsto \var{Z}, \var{Y} \mapsto \var{Z}\}$ is a reduction,  that is, $\reduce(\{\var{X} \mapsto \var{Y}\}, \{Z\}, \{\var{X} \mapsto \var{Z}, \var{Y} \mapsto \var{Z}\})$ [by the \hyperlink{prop:red-def}{definition of $\reduce$}], because $\range(\theta) =  \{Z\} \subseteq \{Y\} \union  \{Z\}.$

It can similarly be shown that $\theta = \{\var{X} \mapsto \var{Y}, \var{Z} \mapsto \var{Y}\}$ is a most-general idempotent reducing unifier, that is,  $\mgiu(\thet{0}, \e{1}, \e{2}, \theta)$.  Thus, most-general idempotent reducing unifiers are not unique.

Although $\theta = \{\var{Y}  \mapsto \var{Z} \}$ is a unifier of $\var{Y}$ and $\var{Z}$, it is \emph{not} a most-general idempotent reducing unifier with respect to the environment $\thet{0} = \{\var{X} \mapsto \var{Y}\}$ because it is not an extension of $\thet{0}$.  That is because
\[\thet{0} \compose \theta = \{\var{X} \mapsto \var{Y}\} \compose \{\var{Y}  \mapsto \var{Z} \} = \{\var{X} \mapsto \var{Z}, \var{Y}  \mapsto \var{Z}\} \neq \{\var{Y}  \mapsto \var{Z} \} = \theta. \]

 \paragraph{Properties of most-general idempotent reducing unifiers.}  We can  show the \hypertarget{prop:idem-mgiu}{\emph{idempotence property}} of the relation $mgiu$, namely, if $\theta$ is a most-general idempotent reducing unifier of two expressions $\e{1}$ and $\e{2}$ with respect to an environment $\thet{0}$, then $\theta$ is indeed idempotent.  For suppose 
   \[\mgiu(\thet{0}, \e{1}, \e{2}, \theta).\]
Then [by the \hyperlink{def:mgiu}{definition of $\mgiu$}],
      \[\e{1} \apply \theta = \e{2} \apply \theta, \]
      \[\thet{0} \moregen \theta, \]  
            and
      \[{\mgi}(\thet{0}, \e{1}, \e{2}, \theta).\]
Hence [\hyperlink{def-mgi}{by the definition of $\mgi$}], for any substitution $\theta'$, 
\[
 \begin {aligned} &\e{1} \apply \theta' = \e{2} \apply \theta' \, \uand \,   
  \thet{0} \moregen \theta' \,\uimplies\\
                    &\theta \moregen \theta'.
   \end{aligned}
   \]
In particular, taking $\theta'$ to be $\theta$, we have
   \[\theta \moregen \theta;\]
that is, by the \hyperlink{prop:idem-more-gen}{\emph{idempotent-more-general property}}, $\theta$ is idempotent, as we wanted to show.

Again by the \hyperlink{prop:def-mgiu}{definition of $\mgiu$}, with have the \hypertarget{prop:mgiu-red}{\emph{reduction property}}, that every  most-general idempotent reducing unifier is a reduction;  that is,
\[\mgiu(\theta_0, e_1, e_2, \theta) \uimplies
\reduce(\theta_0, \vars(e_1, e_2)\!\apply\!\theta_0, \theta).\]

   Because the relation $\mgi$ has a \hyperlink{prop:sym-mgi}{\emph{symmetry property}} and the roles of $\e{1}$ and $\e{2}$ are otherwise interchangeable, we can establish the \hypertarget{prop:sym-mgiu}{\emph{symmetry property}} of the \emph{most-general idempotent reducing unifier relation} $\mgiu$, namely
    \begin{align*} 
 {\mgiu}(&\thet{0}, \e{1}, \e{2}, \theta) \uiff 
 {\mgiu}(\thet{0}, \e{2}, \e{1}, \theta). 
 \end{align*}

We earlier established an \hyperlink{prop:mgi-instance}{\emph{instance property}} for the most-general-idempotent relation ${\mgi}$; by similar reasoning, we can establish an \hypertarget{prop:mgiu-instance}{\emph{instance property}} of the most-general idempotent reducing unifier relation \emph{$mgiu$}, namely:
   \begin{align*}
 {\mgiu}(\thet{0}, \e{1}, \e{2}, \theta) \,\, \uiff \,\, 
  {\mgiu}(\thet{0}, \e{1}\!\apply\thet{0}, \e{2}\!\apply\thet{0}, \theta),
 \end{align*}
for any expressions $\e{1}$, $\e{2}$, and substitutions $\thet{0}$ and $\theta$, where \hypertarget{ass:$\thet{0}$-idem}{$\thet{0}$ is idempotent}. For
\[ {\mgiu}(\thet{0}, \e{1}\!\apply\thet{0}, \e{2}\!\apply\thet{0}, \theta)\]
\noindent is equivalent [by \hyperlink{def:mgiu}{definition}] to
\[\begin{aligned}
&\e{1}\!\apply\thet{0}\apply \theta = \e{2}\!\apply\thet{0} \apply \theta \, \uand 
 \\
 \,\, & \thet{0} \moregen \theta \, \uand
\\
  \,\, & {\mgi}(\thet{0}, \e{1}\!\apply\thet{0}, \e{2}\!\apply\thet{0}, \theta) \, \uand \\
  \,\, & {\reduce}(\theta_0, vars(e_1\!\apply\thet{0}, e_2\!\apply\thet{0})\apply\thet{0}, \, \theta),
\end{aligned}\]
\noindent which is equivalent [by the \hyperlink{prop:subst-comp}{\emph{composition property}}, 
\hyperlink{def:more-gen}{the definition of $\moregen$}, 
the \hyperlink{prop:mgi-instance}{\emph{instance property} of {$\mgi$}}, and the \hyperlink{prop-distr-vars}{\emph{distributivity property} of $vars$}] to
\[\begin{aligned}
&\e{1}\!\apply\thet{0}\compose \theta = \e{2} \!\apply\thet{0}\compose \theta \, \uand 
 \\
 \,\, & \thet{0} \compose \theta = \theta \, \uand
\\
  \,\, & {\mgi}(\thet{0}, \e{1}, \e{2}, \theta)\,  \uand  \\
  \,\, &{\reduce}(\theta_0, vars(e_1, e_2)\!\apply\thet{0}\!\apply\thet{0}, \, \theta),
\end{aligned}\]
\noindent which is equivalent [by properties of equality and the \hyperlink{ass:$\thet{0}$-idem}{idempotence of $\thet{0}$]} to
\[\begin{aligned}
&\e{1}\!\apply \theta = \e{2} \apply \theta \, \uand 
 \\
 \,\, & \thet{0} \compose \theta = \theta \, \uand
\\
  \,\, & {\mgi}(\thet{0}, \e{1}, \e{2}, \theta)\,  \uand  \\
  \,\, &{\reduce}(\theta_0, vars(e_1, e_2)\!\apply\thet{0}, \, \theta),
\end{aligned}\]
\noindent which is equivalent [by the \hyperlink{def:more-gen}{definitions of $\moregen$} and \hyperlink{def:mgiu}{$\mgiu$}, again]  to
\[{\mgiu}(\thet{0}, \e{1}, \e{2}, \theta).\]

   We have completed our introduction to the theory of expressions and substitutions;  we  introduce other properties of the theory as they are needed.  We now turn to our framework for program synthesis.

   \section*{A Brief Introduction to Deductive Program Synthesis}
   
   We begin by introducing the notion of the subject domain theory. We have used some of these ideas informally in presenting the theory of expressions.  We do not expect this section to serve as an introduction to logic, but we'll set down our terminology.

   \subsection*{Subject Domain Theory}
   
   Program synthesis requires a good deal of knowledge about the domain of application, which is provided here by the subject domain theory.  This is expressed by a set of logical formulas in mathematical logic.
   
   \subsubsection{Formulas and Terms.} We begin with a vocabulary comprising an infinite set of constants, an infinite set of variables, an infinite set of $n$-ary function symbols, and an infinite set of $n$-ary relation symbols, for each nonnegative integer $n$, as well as some logical connectives and a conditional operator.  
   
   The \emph{expressions} comprise \emph{terms} and \emph{formulas}; intuitively, the value of a formula is a truth value, either true or false, and the value of a term is an object, such as a number.  We define terms and formulas together inductively.  Terms comprise the constants, the variables, the result of applying an $n$-ary function symbol to $n$ terms, and the conditional term  $\cond {\mathcal{P}} {\expsub{t}{1}} {\expsub{t}{2}}$, where $\mathcal{P}$ is a formula and  $\expsub{t}{1}$ and $\expsub{t}{2}$ are terms.   We call $\expsub{t}{1}$ and $\expsub{t}{2}$ the \emph{then-branch} and the \emph{else-branch}, respectively, of the conditional term. (We sometimes write this as 
   $(\uif \mathcal{P} \uthen t_1 \uelse t_2)$.
   
   We include the equality symbol $=$ as a binary relation symbol in a single term or formula, but we also use it informally in English to relate two terms.
   
   Atomic formulas comprise $\emph{true}$, $\emph{false}$, and the result of applying an $n$-ary relation symbol to $n$ terms. Formulas comprise atomic formulas and the results of applying the logical connectives $\uand$,  $\uor$, $\unot$,  $\uimplies$, $\uimpliedby$, and  $\uiff$ to formulas. \emph{Ground} formulas are those that contain no variables. We freely switch between prefix and infix notation (e.g.,  ``$\uor(\mathcal{P},\, \mathcal{Q})$" and ``$\mathcal{P} \uor \mathcal{Q}$") as convenient.
   
   We also use the equivalence symbol $\uiff$ informally in English to relate two formulas. 
   Context will let us know if we are regarding $\uiff$ as part of a single formula or as an English abbreviation for “if and only if" to relate two formulas.
   We do not include explicit quantifiers, but, as in resolution theorem proving, we apply skolemization to remove quantifiers, so this will not cost us any expressive power.  A detailed treatment of skolemization appears in \citep{man:wal:book} and other introductions to automated reasoning.

  \subsubsection{Truth.}  A \emph{theory} consists of a finite set of formulas, called the \emph{axioms} of the theory. We do not assume that the axioms are consistent or logically independent. We start by introducing the notion of \emph{truth} of a formula in a theory.
   
   \subsubsection{Interpretation.} A (Herbrand) \emph{interpretation} is a possibly infinite list of ground atomic formulas, which are taken to be \emph{true}; those ground atomic formulas not on the list are taken to be \emph{false}.  With respect to an interpretation, we can determine the truth value of a nonatomic ground formula, which contains logical connectives, by applying the usual truth tables; for instance, 
   $\uor (\mathcal{P}, \mathcal{Q})$ is true if $\mathcal{P}$ is true or $\mathcal{Q}$ is true, and false otherwise;  $ \mathcal{P} \uimplies \mathcal{Q}$ is false if $\mathcal{P}$ is true and $\mathcal{Q}$ is false, and true otherwise  
   
   We then say that an arbitrary formula (not necessarily ground) is true under the interpretation if every instance of the formula is true, where an instance is obtained by replacing every occurrence of each variable in the formula by a corresponding term, that is, by applying a proper substitution. An arbitrary formula is false under an interpretation if some instance of the formula is false.   A formula is \emph{valid} if it is true under every interpretation. 
   
   \subsubsection{Equivalence and Equality.}
   Equivalence and equality are analogous; equivalence applies to formulas and equality to terms.
   
   \paragraph{Equivalence.}
   Two formulas $\expsub{\mathcal{P}}{1}$ and $\expsub{\mathcal{P}}{2}$ are \emph{equivalent} under an interpretation, written $\expsub{\mathcal{P}}{1} \uiff \expsub{\mathcal{P}}{2}$, if  $\expsub{\mathcal{P}}{1}$ and $\expsub{\mathcal{P}}{2}$ are either both true or both false under the interpretation.  
   For instance, the double negation sentence $\unot(\unot(\mathcal{P}))$ is equivalent to $\mathcal{P}.$
   The implication sentence $\expsub{\mathcal{P}}{1} \uimplies  \expsub{\mathcal{P}}{2}$  is equivalent to the sentence $ \uor(\unot(\expsub{\mathcal{P}}{1}), \expsub{\mathcal{P}}{2}).$ 
   The equivalence sentence $\expsub{\mathcal{P}}{1} \uiff \expsub{\mathcal{P}}{2}$ is equivalent to the sentence
   \begin{align*}
   \uand (\expsub{\mathcal{P}}{1} \uimplies  \expsub{\mathcal{P}}{2},\expsub{\mathcal{P}}{2} \uimplies  \expsub{\mathcal{P}}{1}),
   \end{align*}
   to the sentence
   \begin{align*}
   \uand (\uor(\unot(\expsub{\mathcal{P}}{1}),  \expsub{\mathcal{P}}{2}),\uor(\unot(\expsub{\mathcal{P}}{2}), \expsub{\mathcal{P}}{1})),
   \end{align*}
   and to the sentence
   \begin{align*}
   \uor (\uand(\expsub{\mathcal{P}}{1},  \expsub{\mathcal{P}}{2}), 
   \uand(\unot (\expsub{\mathcal{P}}{1}), \unot (\expsub{\mathcal{P}}{2}))).
  \end{align*} 
    Equivalence has a \hypertarget{prop:equiv-subst}{\emph{substitutivity property}}; namely, for any formula $\mathcal{S}\langle\expsub{\mathcal{P}}{1}\rangle$, if $\expsub{\mathcal{P}}{1}$ and $\expsub{\mathcal{P}}{2}$ are equivalent, so are $\mathcal{S}\langle\expsub{\mathcal{P}}{1}\rangle$ and $\mathcal{S}\langle \expsub{\mathcal{P}}{2}\rangle,$ where 
    $\mathcal{S}\langle \expsub{\mathcal{P}}{2}\rangle$ is obtained from $\mathcal{S}\langle\expsub{\mathcal{P}}{1}\rangle$ by replacing zero, one, or more occurrences of $\expsub{\mathcal{P}}{1}$ with $\expsub{\mathcal{P}}{2}$.
  \paragraph{Equality.} 
   Two terms $\expsub{t}{1}$ and $\expsub{t}{2}$ are \emph{equal} under an interpretation, written $\expsub{t}{1} = \expsub{t}{2}$, if they satisfy the \hypertarget{prop:eql-subst}{\emph{substitutivity property}}, that is, if, for any formula $\mathcal{F}\langle\var{x}\rangle$,  $\mathcal{F}\langle\expsub{t}{1}\rangle$ is true precisely when  $\mathcal{F}\langle\expsub{t}{2}\rangle$ is true under the interpretation.  It follows that equality has the usual properties, such as the \hypertarget{prop:eq-ref}{\emph{reflexivity property}}, \[t = t,\] for all terms $t$, the \hypertarget{prop:eq-sym}{\emph{ symmetry property}}, \[(\expsub{t}{1} = \expsub{t}{2}) \uimplies (\expsub{t}{2} = \expsub{t}{1}),\] for all terms $\expsub{t}{1}$ and $\expsub{t}{2}$, and the \hypertarget{prop:eq-trans}{\emph{transitivity property}}, \[((\expsub{t}{1} = \expsub{t}{2}) \uand (\expsub{t}{2} = \expsub{t}{3})) \uimplies (\expsub{t}{1} = \expsub{t}{3}),\] for all terms  $\expsub{t}{1}$, $\expsub{t}{2}$, and $\expsub{t}{3}$.
   
   \subsubsection{Conditional Terms.} Under a given interpretation, the conditional term \[\cond {\mathcal{P}} {\expsub{t}{1}} {\expsub{t}{2}}\] is equal to $\expsub{t}{1}$ if $\mathcal{P}$ is true under the interpretation, and equal to $\expsub{t}{2}$ if $\mathcal{P}$ is false. Regardless of the truth value of $\mathcal{P}$, if $\expsub{t}{1}$ and $\expsub{t}{2}$ are equal, the value of the conditional is the value of $\expsub{t}{1}$. Thus the conditional term $\cond {\mathcal{P}} t t$ is equal to $t$. (We call this the \hypertarget{prop-cond-same}{\emph{same-branch property}}.) 
   
   Our Snark implementation systematically applies equivalences and equalities to simplify formulas, and it paraphrases implication and equivalence sentences in terms of the other connectives. In our presentation, we often leave implications and equivalences in place for clarity.

    \subsubsection{Expressions versus S-Expressions.} In our section on the theory of expressions and substitutions, we dealt only with S-expressions, which contain occurrences of constants but no function or relation symbols, other than $\cons$ (\emph{cons}). In the balance of this paper we deal with conventional expressions.  We apply substitutions to these expressions and we assume that all operators $\emph{op}$, including both relation symbols and logical operators, and all function symbols $\uf$ and proper substitutions $\theta$ satisfy the \hypertarget{prop-subst-dist}{\emph{distributivity  property}},  
    \[\emph{op}(\expsub{t}{1}, \expsub{t}{2}, \ldots, \expsub{t}{n})\apply \theta \uiff \, \emph{op}(\expsub{t}{1}\apply \theta, \expsub{t}{2}\apply \theta, \ldots, \expsub{t}{n}\apply \theta),\]
    \noindent and 
     \[\uf(\expsub{t}{1}, \expsub{t}{2}, \ldots, \expsub{t}{n})\apply \theta = \, \uf(\expsub{t}{1}\apply \theta, \expsub{t}{2}\apply \theta, \ldots, \expsub{t}{n}\apply \theta),\] for any expressions $\expsub{t}{i}$ and proper substitutions $\theta$. 
But, as we remarked earlier,  conventional expressions can be represented as S-expressions, as in Lisp, so this is not a substantive extension.

   \subsubsection{Models and Validity for a Theory} A \emph{model} of a theory is an interpretation under which every axiom of the theory is true. A formula is \emph{valid in a theory} if it is true under every model of the theory. A theory is \emph{consistent} if it has a model.

 \subsection*{Mathematical Induction}\begin{quote}\emph{A hundred bottles of beer on the wall, \\
 A hundred bottles of beer. \\
 If one of those bottles should happen to fall, \\
 Ninety-nine bottle of beer on the wall! \\
 \ldots .
 }
 \\---Traditional American and Canadian Song
 \end{quote}
   
 As we have mentioned previously, recursive calls are introduced to the program being derived by application of the mathematical induction principle.  
 
 \subsubsection{Reification.} Although we are working in a first-order logic, it will be useful to quantify over relations and to have functions that apply to relations and produce new ones.  To allow this, we employ \emph{reification};
 in other words,  we allow variables and complex terms other than relation symbols to stand for relations.
Because first-order logic does not allow anything other than a relation symbol at the head of an atomic formula, we can chose to represent a formula ${r}(\x{1}, \x{2}, \ldots, \x{n})$, where ${r}$ is a (perhaps complex) term that stands for a relation, as 
\emph{wf}\!$(r, \x{1}, \x{2}, \ldots, \x{n})$, where   \emph{wf} is a relation symbol. In discussing binary relations, and, especially, well-founded relations and orderings, we will use the conventional notation $x \expsub{\prec}{r}\, y$.

\subsubsection{Sequences.}  A \emph{sequence}  $\langle\expsub{a}{0}, \expsub{a}{1}, \expsub{a}{2}, \ldots\rangle $ is a unary (one-argument) function mapping the natural numbers into terms. We say that a sequence $a' = \langle\expsub{a}{0}',\, \expsub{a}{1}',\, \expsub{a}{2}',\, \ldots\rangle  $ is a subsequence of a sequence $a =\langle\expsub{a}{0}, \expsub{a}{1}, \expsub{a}{2}, \ldots \rangle $ if all the elements of $a'$ occur distinctly in $a$ in the same order, perhaps with intervening elements; we say that $a'$ is an \emph{initial} subsequence of $a$ if $\expsub{a}{0}' = \expsub{a}{0}.$  A sequence  $a = \langle\expsub{a}{0}, \expsub{a}{1}, \expsub{a}{2}, \ldots\rangle$ will be said to \emph{converge} if eventually all its elements are equal; that is, if, for some natural number $j$, $\expsub{a}{j} = \expsub{a}{j+1} = \expsub{a}{j+2} =\ldots .$  If $j$  is the least such natural number, the sequence may be said to \emph{converge at $j$}.

 \subsubsection{Well-Founded Relations.} 
 \begin{quote}
 \emph{
A well-known scientist \textup{(}some say it was Bertrand Russell\textup{)} once gave a public lecture on astronomy \ldots.  At the end of the lecture, a little old lady at the back of the room got up and said: “What you have told us is rubbish. The world is really a flat plate supported on the back of a giant tortoise." The scientist gave a superior smile before replying, ''What is the tortoise standing on?" “You're very clever, young man, very clever," said the old lady. “But it's turtles all the way down!"
}
---[Stephen Hawking, \emph{A Brief History of Time}, 1988]
 \end{quote}

 The form of induction we use is well-founded induction, that is, induction over a well-founded relation.  A relation  $\expsub{\prec}{w}$ is well-founded if it admits no infinite (strictly) decreasing sequences, i.e., no sequences $a =\langle\expsub{a}{0}, \expsub{a}{1}, \expsub{a}{2}, \ldots\rangle $ such that $\expsub{a}{0} \, \expsub{\succ}{w} \, \expsub{a}{1} \, \expsub{\succ}{w} \, \expsub{a}{2} \, \expsub{\succ}{w} \,\ldots.$ (Here, $\expsub{\succ}{w}$ is the \emph{inverse} of $\expsub{\prec}{w}$, defined by $\var{x} \expsub{\succ}{w} \var{y} \uiff \var{y} \expsub{\prec}{w} \,\var{x}.$) For example, the less-than relation $<$ is well-founded over the natural numbers (because any infinite decreasing sequence would eventually hit 0), but $<$ is not well-founded over all the integers: if we include the negative integers, we can go down forever.  Another example of a well-founded relation is the proper subset relation over the finite sets.

  A well-founded relation $\prec$ is \hypertarget{prop:wf-irref}{\emph{irreflexive}}; that is,  $\var{x} \expsub{\nprec}{\,w} \var{x}$, because if $x \, \expsub{\prec}{w} \,x$  for some $x$,   $\langle x, x, x, \ldots \rangle$ would constitute an infinite decreasing sequence. It is also \hypertarget{prop-wf-antisym}{\emph{antisymmetric}}; that is,  $\unot(\var{x} \,\expsub{\prec}{w} \,\var{y} \uand \var{y} \,\expsub{\prec}{w} \,\var{x})$ because if $x \,\expsub{\prec}{w} \,y$ and $y \, \expsub{\prec}{w} \,x$ for some $x$ and $y$,  $\langle x, y, x,y, x, y \ldots \rangle$ would constitute an infinite decreasing sequence.  We do not assume that well-founded relations are transitive;  for instance, the predecessor relation $\texttt{prec}$, defined on the natural numbers by $\texttt{prec}(\var{x}, \var{y})$ if $\var{x} + 1 = \var{y}$, is well-founded but not transitive.

 According to the \hypertarget{prop:wf-ind}{\emph{induction principle} over a well-founded relation} $\expsub{\prec}{w}$, if we are trying to prove a proposition $\mathcal{P}[a]$, we can assume inductively that the proposition $\mathcal{P}[\var{x}]$ holds for any entity $\var{x}$ such that $\var{x} \expsub{\prec}{w} \,a$, that is, such that \var{x} is less than $a$ with respect to $\expsub{\prec}{w}$.  In other words, we prove $\mathcal{P}[a]$ under the induction hypothesis $\var{x} \expsub{\prec}{w} \,a  \, \uimplies \mathcal{P}[\var{x}]$.
 
 Well-founded induction is also called \emph{Noetherian induction}.  Over the natural numbers, well-founded induction over $<$ is usually called \emph{complete induction}. Our reason for using well-founded induction is that it is domain independent and extremely general; the same induction principle applies to nonnegative integers, expressions, substitutions, and other sorts of entities.
 
 We can give an informal justification for the well-founded induction principle as follows.  Suppose that, contrary to the principle, we have a proposition $\mathcal{P}[\expsub{a}{0}]$ that is false under a given model but for which the induction hypothesis  $\var{x} \expsub{\prec}{w} \,\expsub{a}{0} \, \uimplies \mathcal{P}[\var{x}]$ is true.  If $\mathcal{P}[\var{x}]$ were true for every entity $\var{x}$ such that $\var{x} \expsub{\prec}{w} \,\expsub{a}{0}$, the induction hypothesis would imply that $\mathcal{P}[\expsub{a}{0}]$ were true too, but it isn't.  Hence there must be some entity $\expsub{a}{1}$ such that $\expsub{a}{1} \expsub{\prec}{w} \,\expsub{a}{0}$, that is, $\expsub{a}{0} \expsub{\succ}{w} \, \expsub{a}{1}$  but $\mathcal{P}[\expsub{a}{1}]$ is false.  Repeating the same reasoning on $\expsub{a}{1}$, we establish the existence of an entity  $\expsub{a}{2}$ such that  $\expsub{a}{1} \expsub{\succ}{w} \, \expsub{a}{2}$  but $\mathcal{P}[\expsub{a}{2}]$ is false. In this way, we can construct an infinite decreasing sequence $\langle\expsub{a}{0}, \expsub{a}{1}, \expsub{a}{2}, \ldots \rangle $, that is, one such that  $\expsub{a}{0} \, \expsub{\succ}{w} \, \expsub{a}{1} \, \expsub{\succ}{w} \, \expsub{a}{2} \, \expsub{\succ}{w} \,\ldots ,$ contradicting the well-foundedness of $\expsub{\prec}{w}$.
 
 In a well-understood theory, such as the theory of expressions, there will be an arsenal of well-founded relations. 

 \paragraph{Proper-Subexpression Relation.} The \hypertarget{prop:wf-subex}{\emph{proper-subexpression}} relation $\occursin$ is well-founded in the theory of expressions, since we only consider finite expressions.  To see this, recall that a proper subexpression of a given expression must be strictly smaller that the expression; in other words, 
  \[\e{1} \occursin \e{2} \uimplies \size(\e{1})  < \size(\e{2}).\]
 \noindent Thus, if there were an infinite decreasing sequence of expressions with respect to the proper-subexpression relation, their sizes would form an infinite decreasing sequence of nonnegative integers, which is impossible.

 Constructing nontrivial programs, including the unification algorithm, requires combining known well-founded relations to construct new ones peculiarly appropriate to the program being constructed. We now discuss ways of doing this.
     \subsubsection{Induced Relation.} For any unary function symbol $g$, we define the well-founded relation $\expsub{\prec}{g(w)}$ \hypertarget{def:wf-induced}{\emph{induced by $g$ and $\expsub{\prec}{w}$}} by the property
 \[\var{x} \expsub{\prec}{g(w)} \var{y} \uiff g(\var{x}) \, \expsub{\prec}{w} \, g(\var{y}).\]
 \noindent If $\expsub{\prec}{w}$ is well-founded, so is  $\expsub{\prec}{g(w)}.$  For, if $\expsub{\prec}{w}$ is well-founded but $\expsub{\prec}{g(w)}$ is not.   there is an infinite sequence $\langle\expsub{a}{0}, \expsub{a}{1}, \expsub{a}{2}, \ldots \rangle$ such that  \[\expsub{a}{0} \, \expsub{\succ}{g(w)} \, \expsub{a}{1} \, \expsub{\succ}{g(w)} \, \expsub{a}{2} \, \expsub{\succ}{g(w)} \,\ldots .\]  But then, by the definition of the \hyperlink{def:wf-induced}{induced relation},  \[g(\expsub{a}{0}) \, \expsub{\succ}{w} \, g(\expsub{a}{1}) \, \expsub{\succ}{w} \, g(\expsub{a}{2}) \, \expsub{\succ}{w} \,\ldots;\] that is, there is an infinite decreasing sequence with respect to $\expsub{\prec}{w}$, contradicting the well-foundedness of $\expsub{\prec}{w}$.

 This is a case in which we use a complex term to stand for a relation.  In the implementation, we use reification to squeeze this into first-order logic. 

 \paragraph{Vars and Size Relations.} The \hypertarget{def:wf-vars}{{\emph{vars} relation $\expsub{\prec}{\vars(\subset)}$}}  is defined as the relation induced by the function $\vars$ and the subset ordering $\subset.$  For two expressions $\e{1}$ and $\e{2}$, we define the relation $\expsub{\prec}{\vars(\subset)}$ by 
\[   \expsub{e}{1}    \expsub{\prec}{\vars(\subset)} \expsub{e}{2} \uiff \vars(\expsub{e}{1}) \, \subset \, \vars(\expsub{e}{2}).\]
\noindent Because the subset ordering is well-founded on finite sets, the relation $\expsub{\prec}{\vars(\subset)}$ is well-founded on expressions.  
We abbreviate this ordering as $\expsub{\prec}{\vars}$.

Similarly, the \hypertarget{def:wf-size}{\emph{size}} ordering $\expsub{\prec}{\size(<)}$  is defined for expressions $\e{1}$ and $\e{2}$ by
\[   \expsub{e}{1}    \expsub{\prec}{\size(<)} \expsub{e}{2} \uiff \size(\expsub{e}{1}) \, < \, \size(\expsub{e}{2}).\]
Because the less-than ordering is well-founded on the nonnegative integers, the \emph{size} ordering is well-founded on expressions. We abbreviate this ordering  as $\expsub{\prec}{\size}$. 

If $e$ is a nonatomic expression, we can establish that
 $\lef(e) \expsub{\prec}{\size}\, e$ and \allowbreak $\rig(e) \expsub{\prec}{\size}\, e$. This follows because $\lef(e)$ and $\rig(e)$ are proper subexpressions of $e$.

\subsubsection{Weakly Decreasing Sequences.}

  For any relation $\expsub{\prec}{w}$, we define its \hypertarget{def:rel-ref-clo}{\emph{reflexive closure}} $\preccurlyeq$ by the property
 \[\var{x} \expsub {\preccurlyeq}{w} \var{y} \uiff 
 [\var{x} \expsub{\prec}{ w} \, \var{y} \uor  \var{x} = \var{y}].\] 

 \noindent For instance, the reflexive closure $\expsub{\preccurlyeq}{\vars}$ of the vars relation $\expsub{\prec}{\vars}$ is defined by
 \[ \expsub{e}{1} \expsub{\preccurlyeq}{\vars}\, \expsub{e}{2} \uiff  
[\expsub{e}{1} \expsub{\prec}{ \vars} \, \expsub{e}{2} \uor  \expsub{e}{1} = \expsub{e}{2}] \uiff
[\vars(\expsub{e}{1}) \subset \vars(\expsub{e}{2}) \uor  \expsub{e}{1} = \expsub{e}{2}]. \]

A sequence $a = \langle\expsub{a}{0}, \expsub{a}{1}, \expsub{a}{2}, \ldots\rangle$ is \hypertarget{def:seq-weak-dec}{\emph{weakly decreasing}}
 with respect to a relation $\expsub{\prec}{w}$ if 
 $\expsub{a}{0} \, \expsub{\succcurlyeq}{w} \, \expsub{a}{1} \, \expsub {\succcurlyeq}{w} \, \expsub{a}{2} \, \expsub {\succcurlyeq}{w} \, \expsub{a}{3} \ldots. $  

 We can prove that a sequence $a = \langle\expsub{a}{0}, \expsub{a}{1}, \expsub{a}{2}, \ldots\rangle$ that is weakly decreasing with respect to a relation  $\expsub{\prec}{w}$ either converges or contains a strictly decreasing  subsequence. For suppose that $a$ does not converge. We construct a strictly decreasing subsequence $b$ as follows.

 Take $b_0 = a_0$. Since $a$ is does not converge, there must be a subsequent element $a_j$ such that $a_0 = a_1 = \ldots = a_{j-1} \neq a_j$.  
 Because $a$ is weakly decreasing, we  have $a_0 {\succcurlyeq}_w a_j$, that is, because $a_0 \neq a_j$, we have $a_0 {\succ}_{w} \, a_j$.  Take $b_1$ to be $a_j$. Thus we have $b_0 \succ_w b_1$.
 
Similarly, if we have found elements of $a$ such that 
$b_0 \succ b_1 \succ \ldots \succ b_k$, 
we can find a subsequent element $b_{k+1}$ such that $b_k \succ b_{k+1}$.
In this way, we can we can construct an infinite strictly decreasing subsequence of $a$.
 
 In the case in which $\prec_w$ is well-founded, infinite strictly decreasing sequences do not exist.  Therefore, we have the
\hypertarget{prop:wf-seq-weak-dec}{\emph{weakly decreasing sequence property}} of well-founded relations, that if $\prec_w$ is well-founded, all weakly decreasing sequences over $\preccurlyeq$ must converge.

\subsubsection{Well-Founded Relations on Tuples.} For the synthesis of recursive programs with more than one argument, we form them into a tuple that we treat as a single entity.  For this situation, it is necessary to discuss well-founded relations on tuples, which can be constructed from well-founded relations on single arguments.  We begin discussing well-founded relations on pairs (2-tuples), with the understanding that this can be extended to triples or general $n$-tuples, i.e.,  tuples of $n$ elements.


 If we have a well-founded relation, we can extend it to a well-founded relation on pairs simply by ignoring one of the components.  For instance, if $\expsub{\prec}{w}$ is a relation, then the relation $\expsub{\prec}{\ufirst(w)}$ on pairs induced by the function $\ufirst$ satisfies the condition
 \[\langle \varsub{x}{1}, \,  \varsub{x}{2}\rangle \, \expsub{\prec}{\ufirst(w)} 
    \langle \varsub{y}{1}, \,  \varsub{y}{2}\rangle, \]
    that is,
    \[\ufirst(\langle \varsub{x}{1}, \,  \varsub{x}{2}\rangle) \, \expsub{\prec}{w} 
    \ufirst(\langle \varsub{y}{1}, \,  \varsub{y}{2}\rangle) \]
    that is,
    \[\varsub{x}{1}\expsub{\prec}{w} 
   \varsub{y}{1}.\]

We have shown that if a relation is well-founded, the relation induced on it by a function is also well-founded.  Thus, if $\expsub{\prec}{w}$ is well-founded, so is $\expsub{\prec}{\ufirst(w)}$. Intuitively speaking, any infinite sequence of pairs that is decreasing if we ignore the second elements will correspond to an infinite decreasing sequence of individual elements; thus if the sequence of individual elements cannot exist, neither can the sequence of pairs.

Similarly, if we have a well-founded relation on pairs, we can induce a corresponding well-founded relation on triples, and so on.

In the theory of expressions, the \emph{vars} relation, satisfying the property
 \[ \langle  \expsub{d}{1}, \expsub{d}{2} \rangle   
  \expsub{\prec}{\vars} 
      \langle  \expsub{e}{1}, \expsub{e}{2} \rangle
      \uiff \vars(\langle  \expsub{d}{1}, \expsub{d}{2} \rangle ) \, \subset \, \vars(\langle  \expsub{e}{1}, \expsub{e}{2} \rangle),\] 
      and the \emph{size} relation, satisfying the property
      \[ \langle  \expsub{d}{1}, \expsub{d}{2} \rangle   
  \expsub{\prec}{\size} 
      \langle  \expsub{e}{1}, \expsub{e}{2} \rangle
      \uiff \size(\langle  \expsub{d}{1}, \expsub{d}{2} \rangle ) < \size(\langle  \expsub{e}{1}, \expsub{e}{2} \rangle),\] 
      are well-founded on tuples of expressions. These are the relations induced on pairs of expressions by the functions $\vars$ and $\size$ and the well-founded relations $\subset$ and $<$, respectively.

      Because we know that, if the expression $e$ is nonatomic, we have $\lef(e) \occursin e$ and $\rig(e) \occursin e,$ we can establish that
      \[\langle  \lef(\e{1}), \lef(\e{2}) \rangle \expsub{\prec}{\vars} 
      \langle  \e{1}, \e{2} \rangle
      \] 
      and     
      \[\langle  \rig(\e{1}), \rig(\e{2}) \rangle \expsub{\prec}{\vars} 
      \langle  \e{1}, \e{2} \rangle. \]

\begin{samepage}
\subsubsection{Lexicographic Combination of Well-Founded Relations.}
\nopagebreak
\begin{quote}\emph{
Little man can beat a big man every time if he's in the right and keeps a-coming.}
\\---Texas Rangers' creed.
\end{quote}
\end{samepage}

 For any two binary relations $\expsub{\prec}{\expsub{w}{1}}$ and  $\expsub{\prec}{\expsub{w}{2}}$, we define their \hypertarget{def:pairs-lex}{\emph{lexicographic combination}} $\expsub{\prec}{\lex(\expsub{w}{1}, \expsub{w}{2})}$ on pairs by the property
 \begin{alignat*}{2}
  \begin{split}
  &\langle \varsub{x}{1}, \,  \varsub{x}{2}\rangle \, \expsub{\prec}{\lex(\expsub{w}{1}, \expsub{w}{2})} \langle \varsub{y}{1}, \,  \varsub{y}{2}\rangle
  \uiff \\
  &\left(\begin{aligned}
  &\varsub{x}{1} \expsub{\prec}{\expsub{w}{1}} \, \varsub{y}{1} \, \uor \, \\
  &\varsub{x}{1} = \varsub{y}{1} \uand
  \varsub{x}{2} \expsub{\prec}{\expsub{w}{2}} \, \varsub{y}{2}  
  \end{aligned}\right).
  \end{split}
\end{alignat*}

  Sometimes we find it more convenient to use the logically equivalent 
  \hypertarget{prop:lex-refl}{\emph{reflexive lexicographic property}}
\begin{alignat*}{2}
   \begin{split}
  &\langle \varsub{x}{1}, \,  \varsub{x}{2}\rangle \, \expsub{\prec}{\lex(\expsub{w}{1}, \expsub{w}{2})} 
   \langle \varsub{y}{1}, \,  \varsub{y}{2}\rangle
  \uiff \\
 &\left(\begin{aligned}
  &\varsub{x}{1} \expsub{\prec}{\expsub{w}{1}} \, \varsub{y}{1} \, \uor \, \\
  &\varsub{x}{1} \expsub{\preccurlyeq}{\expsub{w}{1}} \, \expsub{\var{y}}{1}\uand
  \varsub{x}{2} \expsub{\prec}{\expsub{w}{2}} \, \varsub{y}{2}
  \end{aligned}\right).
\end{split}
\end{alignat*}
  
  \noindent It is easy to see that the definition implies the reflexive lexicographic property, because $\varsub{x}{1} = \varsub{y}{1}$ implies $\varsub{x}{1} \expsub{\preccurlyeq}{\expsub{w}{1}} \, \expsub{\var{y}}{1}$. 
  In the other direction, if the right side of the reflexive lexicographic property holds and the first disjunct $\varsub{x}{1} \expsub{\prec}{\expsub{w}{1}} \, \varsub{y}{1}$ is true, then  the right side of the definition holds.
  But suppose the right side of the reflexive lexicographic property holds but the first disjunct $\varsub{x}{1} \expsub{\prec}{\expsub{w}{1}} \, \expsub{\var{y}}{1}$ is false;  then the second disjunct \[[\varsub{x}{1} \expsub{\preccurlyeq}{\expsub{w}{1}} \, \expsub{\var{y}}{1} \uand
  \varsub{x}{2} \expsub{\prec}{\expsub{w}{2}} \, \expsub{\var{y}}{2}]\bigr]\] is true and hence [because $\varsub{x}{1} \expsub{\prec}{\expsub{w}{1}} \, \expsub{\var{y}}{1}$ is false]
  \[ [\varsub{x}{1} = \expsub{\var{y}}{1} \uand \varsub{x}{2} \expsub{\prec}{\expsub{w}{2}} \, \expsub{\var{y}}{2}]. \]
Hence the right side of the definition holds.

 We can extend this definition to apply to three or more binary relations. For instance, for three relations the definition is
  \begin{alignat*}{2}
  \begin{split}
  &\langle \varsub{x}{1}, \,  \varsub{x}{2}, \,  \varsub{x}{3}\rangle \, \expsub{\prec}{\lex(\expsub{w}{1}, \expsub{w}{2}, \expsub{w}{3})} \langle \varsub{y}{1}, \,  \varsub{y}{2}, \,  \varsub{y}{3}\rangle
  \uiff \\
  &\left(\begin{aligned}
  &\varsub{x}{1} \expsub{\prec}{\expsub{w}{1}} \, \varsub{y}{1} \, \uor \, \\
  &\varsub{x}{1} = \varsub{y}{1} \uand
  \varsub{x}{2} \expsub{\prec}{\expsub{w}{2}} \, \varsub{y}{2}\, \uor \, \\
  &\varsub{x}{1} = \varsub{y}{1} \uand \varsub{x}{2} = \varsub{y}{2} \uand
  \varsub{x}{3} \expsub{\prec}{\expsub{w}{3}} \, \varsub{y}{3}
  \end{aligned}\right).
  \end{split}
\end{alignat*}

 If $\expsub{\prec}{\expsub{w}{1}}$ and $\expsub{\prec}{\expsub{w}{2}}$ are both well-founded relations, so is their lexicographic combination $\expsub{\prec}{\lex(\expsub{w}{1}, \expsub{w}{2})}$.  For suppose, to the contrary, there is an infinite decreasing sequence of pairs $\{\langle\expsub{a}{0}, \expsub{b}{0}\rangle, \langle\expsub{a}{1}, \expsub{b}{1}\rangle, \langle\expsub{a}{2}, \expsub{b}{2}\rangle, \ldots \rangle$, that is, one such that
  \[\langle\expsub{a}{0}, \expsub{b}{0}\rangle \, \expsub{\succ}{\lex(\expsub{w}{1}, \expsub{w}{2})} \, \langle\expsub{a}{1}, \expsub{b}{1}\rangle \, \allowbreak \expsub{\succ}{\lex(\expsub{w}{1}, \expsub{w}{2})} \,\allowbreak \langle\expsub{a}{2}, \expsub{b}{2}\rangle \, \expsub{\succ}{\lex(\expsub{w}{1}, \expsub{w}{2})} \,\ldots .\]   By the definition of the \hyperlink{def:pairs-lex}{lexicographic combination on pairs}, we know that, for each natural number $i$, we have either
   \begin{align*}
       & \expsub{a}{i} \expsub{\succ}{\expsub{w}{1}} \, \expsub{a}{i+1} \uor  && [\text{the $\expsub{\prec}{\expsub{w}{1}}$ case for $i$}]\\ 
       & \expsub{a}{i} = \expsub{a}{i+1}  \uand
 \expsub{b}{i} \expsub{\succ}{\expsub{w}{2}} \,\expsub{b}{i+1})  && [\text{the $\expsub{\prec}{\expsub{w}{2}}$ case for $i$}].
   \end{align*}
   \noindent In either case, we have $\expsub{a}{i} \expsub{\succcurlyeq}{\expsub{w}{1}} \, \expsub{a}{i+1}$ for each natural number $i$.  But since $\expsub{\prec}{\expsub{w}{1}}$ is well-founded, we know (from the \hyperlink{prop:wf-seq-weak-dec}{\emph{weakly decreasing sequence property}}) that the sequence cannot weakly decrease indefinitely and hence must converge at some natural number $j$; thus for each $k$ such that $k \geq j,$ we have $\expsub{a}{k}  =  \expsub{a}{k+1}.$   But then (by the \hyperlink{prop:wf-irref}{irreflexivity of well-founded relations}), for each $k$ such that $k \geq j,$ the  $\expsub{\prec}{\expsub{w}{1}}$ case for $k$ cannot hold, and therefore the $\expsub{\prec}{\expsub{w}{2}}$ case for $k$ must hold.
   Hence, for each $k$ such that $k \geq j,$ we have $\expsub{b}{k} \expsub{\succ}{\expsub{w}{2}} \,\expsub{b}{k+1},$ but this contradicts the well-foundedness of $\expsub{\prec}{\expsub{w}{2}}$.  Therefore no decreasing sequence of pairs can exist under the lexicographic combination, so the lexicographic combination must be well-founded.
   
   \subsubsection{Lexicographic  Combination of Induced Relations.}  In the derivation of the unification algorithm, we find it convenient to use a 
   \hypertarget{def:rel-induce-lex}{\emph{lexicographic combination of induced relations.}} While the construction is quite general, we describe it in a special case that resembles its use in the unification derivation.  We define the lexicographic combination \hypertarget{def:rel-lex-vars-size}{$\expsub{\prec}{\lex(\vars,\size)}$} of the \emph{vars} relation $\expsub{\prec}{\vars}$ and the \emph{size} relation  $\expsub{\prec}{\size}$ by the property
   \begin{alignat*}{2}
  \begin{split}
  &\e{1} \, \expsub{\prec}{\lex(\vars,\size)}\, \e{2}
  \uiff \\
  &\left(\begin{aligned}
  &\vars(\e{1}) \subset \, \vars(\e{2})\, \uor \, \\
  &\vars(\e{1}) = \, \vars(\e{2}) \uand
 \size(\e{1}) < \, size(\e{2})\,
  \end{aligned}\right).
  \end{split}
  \end{alignat*}
  In other words, with respect to the lexicographic combination, 
$\e{1}$ is less than $\e{2}$ if the variable set of $\e{1}$ is a proper subset of variable set of $\e{2}$, or if the variable set of $\e{1}$ equals the variable set of $\e{2}$ and the size of $\e{1}$ is  less than the size of $\e{2}$.

This is well-founded, because it is the relation induced by a function on simpler well-founded relation.  The function is the function \hypertarget{def:fn-var-size}{$\varssize$},  defined by 
\[vars\!\!-\!\!size(e) = \langle vars(e), size(e) \rangle,\]
which maps any expression into the pair of its variable set and its size. 
The simpler well-founded relation is the lexicographic relation \hypertarget{def:rel-lex-subset-size}{ $\expsub{\prec}{\lex(\subset, <)}$,} defined by 
\begin{alignat*}{2}
  \begin{split}
  &\langle s_{1}, n_{1}\rangle \,  \expsub{\prec}{\lex(\subset, <)}\, \langle s_{2},n_{2} \rangle
  \uiff \\
  &\left(\begin{aligned}
  &s_{1} \subset \, s_{2}\, \uor \, \\
  &s_{1} = \, s_{2} \uand
 n_{1} < \, n_{2}\,
  \end{aligned}\right),
  \end{split}
  \end{alignat*} 
 which is the lexicographic combination of the proper-subset relation $\subset$ on the finite sets and the less-than relation $<$ on the nonnegative integers. 

 The relation $\expsub{\prec}{\lex(\vars,\size)}$ is equivalent to the relation  $\expsub{\prec}{\varssize(\lex(\subset, <))}$ induced by function 
$\varssize$ on the above relation $\expsub{\prec}{\lex(\subset, <)}$, because
\[\e{1}  \expsub{\prec}{\varssize(\lex(\subset, <))} \e{2} \]
is equivalent [by the \hyperlink{def:wf-induced}{definition of induced relations}] to
\[\varssize(\e{1}) \expsub{\prec}{\lex(\subset, <)} \varssize(\e{2}),\]
which is equivalent [by the \hyperlink{def:fn-var-size}{definition of $\varssize$}] to
\[\langle\vars(\e{1}), \size(\e{1})\rangle \, \expsub{\prec}{\lex(\subset, <)} \langle \vars(\e{2}), \size(\e{2})\rangle,\]
which is equivalent [by the \hyperlink{def:rel-lex-subset-size}{definition of $\lex(\subset, <)$}] to
\begin{alignat*}{2}
  \begin{split}
  &\left(\begin{aligned}
  &\vars(\e{1}) \subset \, \vars(\e{2})\, \uor \, \\
  &\vars(\e{1}) = \, \vars(\e{2}) \uand
 \size(\e{1}) < \, \size(\e{2})\,
  \end{aligned}\right),
  \end{split}
  \end{alignat*}
 which is equivalent [by the \hyperlink{def:rel-lex-vars-size}{definition of $\expsub{\prec}{\lex(\vars,\size)}$}] to
\[\e{1} \, \expsub{\prec}{\lex(\vars,\size)}\, \e{2}.\]

Because the component relations $<$ and $\subset$ are well founded on finite sets and nonnegative integers, their lexicographic combination is well-founded on the pairs, and so is the relation induced by the function $\varssize$ onto the lexicographic relation.

In our derivation, we actually use the equivalent \hyperlink{prop:lex-refl}{\emph{reflexive lexicographic }}
version of the lexicographic combination of relations,

\begin{alignat*}{2}
  \begin{split}
  &\e{1} \, \expsub{\prec}{\lex(\vars,\size)}\, \e{2}
  \uiff \\
  &\left(\begin{aligned}
  &\vars(\e{1}) \subset \, \vars(\e{2})\, \uor \, \\
  &\vars(\e{1}) \subseteq \, \vars(\e{2}) \uand
 \size(\e{1}) < \, size(\e{2})\,
  \end{aligned}\right).
  \end{split}
  \end{alignat*}

   \subsection*{Specifications and Programs} 
   For this paper, we deal with a particular form of specification and program.

   \subsubsection{Specifications.}  We restrict our attention to the synthesis of applicative (side-effect-free) programs from declarative specifications.  We begin with a specification of the form 
  \begin{equation*} \uf(a) \Lleftarrow \operatorname{find} \var{Z} \operatorname{such} \operatorname{that} \mathscr{Q}[a, \, \var{Z}],
  \end{equation*}
  where $a$ is the input parameter and $\var{Z}$ is the output.  In other words, for a given input entity $a$, we want to find output entity $\var{Z}$ that satisfies the input-output condition $\mathscr{Q}[a, \, \var{Z}]$; that entity will be the output produced by the program $\uf(a)$ we are synthesizing.   The input-output condition $\mathscr{Q}[a, \, \var{Z}]$ is a formula that contains no variable other than $\var{Z}$. (Because we allow tuples of entities, all this extends to allow multiple inputs and outputs, too, or none at all.)
 \subsubsection{Primitive Symbols.}  We assume that logical connectives, the conditional operator, and certain of the constant, function, and relation symbols of the theory have been declared to be \emph{primitive}; intuitively, this means that we assume we know how to compute them, and we allow them to appear in the program we are constructing.  A term or formula is said to be \emph{primitive} if every constant, function, or relation symbol it contains is primitive.   A primitive term can turn out to be equal to a nonprimitive term, even though we know how to compute one but not the other.

 Most skolem functions are not primitive; they yield an entity we know exists but may not know how to find.   There may also  be other symbols that we include in the specification or in an axiom that we do not expect to appear in a final program.  For instance, if we are constructing a program to test if a natural number is prime, we might include the symbol $\emph{prime}$ in the specification but expect it to be deconstructed and not to appear explicitly in the final program. Thus, a computable term may be nonprimitive.

  \subsubsection{Programs.} Our programs are of the form 
  \begin{equation*}
      \uf(a) =  t,
  \end{equation*}
  
  \noindent where $t$ is a primitive term, called the \emph{body} of the program. Such a program is said to satisfy the above specification if, for every model of the theory, the corresponding instance $\mathscr{Q}[a, \, t]$ of the input-output condition is true.

We require that the output entries be primitive to ensure that the program is something we know how to compute. 
We do not require the body $t$ of the program to be ground (variable-free); if the body contains a variable, that variable can be replaced by any primitive ground term. By not instantiating the variable, we leave the program open to a wider variety of subsequent refinements and optimizations.

      \subsection*{Deductive Tableaus}

            Resolution theorem provers such as Snark are refutation procedures, in which our conjectured theorem, and all its subgoals, are negated and put into a clausal form, in which $\uand$, $\uor$, and $\unot$ are the only logical connectives.  While this transformation allows efficient implementation, it also makes proofs difficult for people to follow. In our presentation, we introduce a notation, the \emph{deductive tableau}, which retains a distinction between assertions and goals, and allows the implication ($\uimplies$), reverse implication ($\uimpliedby$), and equivalence ($\uiff$) in formulas. The tableau also provides a representation for the incremental construction of a program as the proof proceeds.
      
        \subsubsection{The Structure.}   
            A derivation from a specification   \[ \uf(a) \Lleftarrow \operatorname{find} \var{Z} \operatorname{such} \operatorname{that} \mathscr{Q}[a, \, \var{Z}] \]
   is carried out with a tableau structure, a set of \emph{assertions} and \emph{goals}, which are formulas, each with an optional associated \emph{output entry}, a term:   
   \begin{center}
 \begin{tabular}{T}
 \hline
 \begin{center} Assertions \end{center} &  \begin{center} Goals \end{center} & \begin{center} $\,\uf(a)\,$\end{center}  \\ \hhline{|=|=|=|}
 \begin{center}$\expsub{\mathcal{A}}{i}$ \end{center} &  & \begin{center}$\expsub{t}{i}$\end{center} \\  \hline
         & \begin{center}$\expsub{\mathcal{G}}{j}$\end{center} & \begin{center}$\expsub{t}{j}$\end{center} \\
 
 \hline
\end{tabular}
\end{center}
 We introduce some apparatus to provide the meaning of a tableau with respect to the given specification.

\subsubsection{Allowable Outputs.} For any interpretation, a  term is an \hypertarget{def:tab-allow-out}{\emph{allowable output}} of the tableau if it is primitive and is an instance  $\expsub{t}{i}\apply \theta$ of an output entry  $\expsub{t}{i}$ corresponding to either 

\begin{itemize}
\item[$\bullet$] a goal $\expsub{\mathcal{G}}{i}$ of the tableau such that the corresponding instance $\expsub{\mathcal{G}}{i}\apply \theta$ is true under the model, or
\item[$\bullet$] an assertion $\expsub{\mathcal{A}}{i}$ of the tableau such that the corresponding instance $\expsub{\mathcal{A}}{i}\apply \theta$ is false under the model.

\end{itemize}
If a goal row has no output entry, any term is allowable under an interpretation for which the goal is true.  If an assertion has no output entry, any term is allowable under an interpretation for which the assertion is false.

A transformation on tableaus preserves allowable output if, for a given interpretation, the allowable outputs before the transformation equal the allowable outputs afterwards.

\subsubsection{Correctness Condition of a Tableau.} In a given theory, with respect to a given specification, a tableau is said to satisfy the \hypertarget{def:tab-correct}{\emph{correctness condition}} for a given model $\mathcal{M}$ for the theory if any term that is allowed under $\mathcal{M}$ satisfies the input-output condition. A tableau is \hypertarget{def:tab-corr-th}{\emph{correct}} for a theory and specification if it satisfies the correctness condition for any model of the theory.

A transformation on tableaus \hypertarget{def:tr-pres-coff}{\emph{preserves correctness with respect to a theory and specification}} if, for a given model, a term has the correctness condition before the transformation precisely when it has the correctness condition afterwards.  A transformation that preserves output entries for a model will also preserve the correctness condition. Consequently, if one tableau is correct in the theory with respect to the specification, so is the other.


\subsubsection{Duality.}  There is a \hypertarget{prop:tab-dual}{\emph{duality property}} between assertions and goals; in fact, any row
 \begin{center}
\begin{tabular}{T}
\hline
 \begin{center}$\mathcal{A}$\end{center} & \begin{center}$\qquad$\end{center} & \begin{center}$t$\end{center}\\
 
 \hline
\end{tabular}
\end{center}
 
\noindent can be replaced by the row

   \begin{center}
\begin{tabular} {T}
\hline
    \begin{center}$\qquad$\end{center} & \begin{center}$ \unot(\mathcal{A})$\end{center} & \begin{center}$t$\end{center} \\
 
 \hline
\end{tabular}
\end{center}

\noindent while preserving the allowable outputs.  For, if there is a term that is an instance $t\apply \theta$ of the output entry, the corresponding instance of the goal, $ (\unot \,(\mathcal{A}))\apply \theta$, that is,  $ \unot (\mathcal{A}\apply \theta)$, is true precisely when the corresponding instance of the assertion, $\mathcal{A}\apply \theta$, is false.

Because of duality, we could have phrased a derivation using only assertions, or only goals.  Indeed, the {Snark}  implementation of the deductive tableau uses only assertions; goals are negated, translated into clausal form, and put into the assertion column. But the distinction between assertions and goals can make a derivation easier to follow. Typically, forward reasoning is conducted in the assertion column and backward reasoning in the goal column.

\subsubsection{Splitting Properties.} We can split assertions that are conjunctions and goals that are disjunctions or implications into their components without changing the outputs allowed.  For instance, according to the \hypertarget{rul:tab-impl-spl}{\emph{implication splitting property}}, a goal row

 \begin{center}
\begin{tabular}{T}
\hline
 \begin{center}$\qquad$\end{center}  &  \begin{center}$\mathcal{A}\uimplies\mathcal{G}$\end{center} & \begin{center}$t$\end{center} \\
  \hline
\end{tabular}
\end{center}

\noindent can be replaced by the two rows

 \begin{center}
\begin{tabular}{T}
\hline
 \begin{center}$\mathcal{A}$\end{center} &   \qquad & \begin{center} $t$ \end{center} \\
  \hline
     &  \begin{center}$ \mathcal{G}$ \end{center} & \begin{center}$t$\end{center} \\
  \hline
\end{tabular}
\end{center}

\noindent without altering allowable outputs. For if, under a given model, $t \apply \theta$ is an instance of the output entry for which the corresponding instance of the original goal, $(\mathcal{A}\uimplies\mathcal{G})\apply \theta $, that is,  $\mathcal{A}\apply \theta \uimplies\mathcal{G}\apply \theta, $ is true, then (by the truth table for $\uimplies$) either the corresponding instance of the assertion, $\mathcal{A}\apply \theta $, is false or the corresponding instance of the goal, $\mathcal{G}\apply \theta $, is true, and vice versa.  

There is also an analogous
\hypertarget{rul:tab-and-spl}{\emph{and-splitting property}} that decomposes an assertion that contains a conjunction into two (or more) assertions, and an \hypertarget{rul:tab-or-spl}{\emph{or-splitting rule}} that decomposes a goal that contains a disjunction into two (or more) goals, without altering the allowable output entries.  In our {Snark} implementation, and in resolution theorem provers in general, splitting is performed automatically in the transformation to clausal form. 

While in implementations, implication ($\uimplies$) and equivalence ($\uiff$) are paraphrased in terms of conjunction ($\uand$), disjunction ($\uor$), and negation ($\unot$) for efficiency,  our presentation  often leaves them intact to make the proof easier to understand, as we have remarked.

\subsubsection{Instance Property.} According to the \hypertarget{prop:tab-inst}{\emph{instance property}}, if a tableau contains the assertion row

 \begin{center}
\begin{tabular}{T}
\hline
 \begin{center}$\mathcal{A}$\end{center} & $\qquad$ & \begin{center}$t$\end{center} \\
  \hline
\end{tabular}
\end{center}

\noindent we can add any instance of that row,

 \begin{center}
\begin{tabular}{T}
\hline
 \begin{center}$\mathcal{A}\apply \theta$\end{center} & $\qquad$ & \begin{center}$t\apply\theta$\end{center} \\
  \hline
\end{tabular}
\end{center}
\noindent without changing the allowable outputs. This is not itself a rule of our system---it would allow the introduction of many irrelevant rows---but it is used to justify rules.

To prove this, suppose (under a given interpretation) $(t\apply\theta)\apply \theta'$ is an instance of the new row for which the corresponding instance of the assertion $(\mathcal{A}\apply \theta) \apply \theta'$ is false.
In other word, by the composition property, $t\apply(\theta \compose \theta')$  is an instance of the new row for which the corresponding instance of the assertion $\mathcal{A}\apply (\theta \compose \theta')$ is false. Then, $t\apply(\theta \compose \theta')$ is an instance of the output entry $t$ for which the corresponding instance of the given assertion $\mathcal{A}$ is false.


Variables in goals behave as if they were existentially quantified, while in assertions they appear to be universally quantified; we use upper-case symbols for variables, to distinguish them. During a proof, in assertions or goals, our rules will be able to replace variables with other terms.

When the substitution is a permutation $\pi$, the row and its instance are interchangeable; if either is present, we can add the other without affecting the allowable outputs.  We have remarked that a permutations substitution has an inverse substitution $\pi^{-1}$.  If the tableau contains the instance row

 \begin{center}
\begin{tabular}{T}
\hline
  \begin{center}$\mathcal{A}\apply \pi$\end{center} & \qquad &  \begin{center}$t\apply\pi$\end{center} \\
  \hline
\end{tabular},
\end{center}
we can also add an instance of that instance, i.e., 
 \begin{center}
\begin{tabular}{T}
\hline
  \begin{center}$(\mathcal{A}\apply \pi) \apply \pi^{-1} $\end{center} &   $\qquad$ &  \begin{center}$(t\apply\pi) \apply \pi^{-1}$\end{center} \\
  \hline
\end{tabular},
\end{center}
which (by the composition property) is
 \begin{center}
\begin{tabular}{T}
\hline
  \begin{center}$\mathcal{A}\apply (\pi \compose \pi^{-1}) $\end{center} &   $\qquad$ &  \begin{center}$t\apply (\pi \compose \pi^{-1})$\end{center} \\
  \hline
\end{tabular},
\end{center}
which (because $\pi^{-1}$ is an inverse) is just the original row
 \begin{center}
\begin{tabular}{T}
\hline
  \begin{center}$\mathcal{A} $\end{center} &  \qquad &  \begin{center}$t$\end{center} \\
  \hline
\end{tabular}.
\end{center}

By \hyperlink{prop:tab-dual}{duality}, the instance properties hold for goals as well as assertions.  Thus, if a tableau contains a goal row, an instance of that row can be added without changing the allowable outputs, and the row may be replaced by a row to which a permutation substitution has been applied to its goal and its output entry.

In presenting a derivation, we freely rename the variables in a row to improve clarity.

\subsubsection{Orphaned Output Entries and Valid Assertions.} If the output entry of a goal row 
   \begin{center}
\begin{tabular} {T}
\hline
    $\qquad$ &  \begin{center}$ \mathcal{G}$\end{center} &  \begin{center}$\var{V}$\end{center} \\
 \hline
\end{tabular}
\end{center}
consists of a  new variable $\var{V}$, one that does not occur elsewhere in that row, then, according to the \hypertarget{prop:tab-orph}{\emph{orphaned output-entry property}}, we can drop the output entry and replace the row with
   \begin{center}
\begin{tabular} {T}
\hline
    $\qquad$ &  \begin{center}$ \mathcal{G}$\end{center} & $\qquad$\\
 \hline
\end{tabular}.
\end{center}
For, if $\mathcal{G}\apply \theta$ is true under an interpretation, the new row will allow any primitive term $t.$ However, \hyperlink{def:subst-add}{we know} $\mathcal{G}\apply 
 (\{\var{V} \mapsto t\} \addto \theta) $ is also true, since $\var{V}$ does not occur in $\mathcal{G}.$ Since $\var{V}\apply (\{\var{V} \mapsto t\} \addto \theta)$ is {t}, the replaced row also allows any primitive term.

 If a tableau row has multiple output entries, we can drop any of them if it is a variable that does not occur elsewhere in the row (including another output column.)

A formula that is valid in the theory can never be false in any model for the theory; hence that formula can be added to a tableau as an assertion without changing the correctness of the tableau with respect to the theory and given specification. It never allows any new output entries in any model. 
Therefore, for a given theory, by the \hypertarget{prop:tab-valid-into}{\emph{valid-formula-introduction property}}, we may include in our tableau any formula that is a definition or axiom or has been previously proved in that theory. For instance, we have established in the theory of substitutions the \hyperlink{prop:idem-more-gen}{\emph{idempotent-more-general property}}, that a substitution is idempotent precisely when it is more-general-idempotent than itself.  Thus, we can include in a tableau for the theory the assertion

\noindent  
 \begin{center}
\begin{tabular}{T}
 \hline
\begin{center}
{\begin{align*}
&\idem(\Theta) \uiff   \\   
&\Theta \moregen \Theta
\end{align*}}
\end{center}
      &  & \\
\hline
\end{tabular}.
\end{center}
\noindent  We have made the substitution $\Theta$ a variable so the assertion will behave as if the substitution is universally quantified; $\Theta$ is the upper-case $\theta$.

For the derivation of the unification algorithm, we have proved the validity of all the added assertions that were not axiomatic, either by hand or by a separate deductive-tableau proof.  Adding a nonvalid assertion compromises the correctness of the final program.

\subsection*{The Derivation Process}
\subsubsection{Initial Tableau.} For a given theory and a specification 
  \begin{equation*} \uf(a) \Lleftarrow \operatorname{find}\var{Z} \operatorname{such} \operatorname{that} \mathscr{Q}[a, \,\var{Z}],
  \end{equation*}
our \hypertarget{def:tab-init-row}{\emph{initial tableau}} is

      \begin{center}
\begin{tabular}{T}
 \hline
  \begin{center}Assertions\end{center} &  \begin{center}Goals\end{center} &  \begin{center}$\uf(a)$\end{center} \\ \hhline{|=|=|=|}
 &   \begin{center}$\mathscr{Q}[a, \, \var{Z}] $\end{center} &  \begin{center}$\var{Z}$\end{center} \\ 
 
 \hline
\end{tabular}.
\end{center}
\noindent This is a tableau with a single row, a goal, with a single variable, the output entry. The tableau is correct with respect to the specification, for if $\theta$ is a substitution for which the instance of the goal $\mathscr{Q}[a, \, \var{Z}] \apply \theta$  is true under a model of the theory, the corresponding instance of the output entry is $\var{Z} \apply \theta$ itself. But this expression is [by \hyperlink{prop-subst-dist}{\emph{distributivity}}, because $\var{Z}$ is the only variable in the input-output condition] $\mathscr{Q}[a, \, \var{Z} \apply \theta]$.  Thus the initial tableau is correct with respect to the specification. The fact that we have used an upper-case symbol $\var{Z}$ in the goal indicates that it is a variable, and has tacit existential quantification; our rules can instantiate it freely in the course of the search for a proof.

  We actually allow multiple inputs and outputs. If the specification has multiple outputs $Z_1, \ldots, Z_n$, the tableau will have $n$ output columns $f_1, \ldots, f_n$.  When the input-output condition $\mathscr{Q}$ has no outputs, the tableau can be used as a simple theorem prover, to show the validity of $\mathcal{Q[a]}$ in the theory. But for the unification algorithm, the tableau has a single output entry.

\subsubsection{Final Tableau} A deductive-tableau program derivation framework applies \emph{rules} to the tableau, which add or remove rows while maintaining correctness.  The process continues until we develop a \hypertarget{def:tab-fin-row}{{\emph{final row}}}, which consists (in the \emph{assertion case}) of an assertion $\false$,

 \begin{center}
\begin{tabular}{T}
\hline
 \begin{center}$\false$\end{center} & $\qquad$ & \begin{center}$t$\end{center} \\
 
 \hline
\end{tabular},
\end{center}

\noindent or (in the \emph{goal case}) a goal $\true$,

   \begin{center}
\begin{tabular} {T}
\hline
    $\qquad$ & \begin{center}$\true$\end{center} & \begin{center}$t$\end{center} \\
 
 \hline
\end{tabular}.
\end{center}

We maintain correctness throughout;  hence (in the goal case), if any instance of the goal is true under a model, the corresponding instance of the output entry will satisfy the input-output condition. For any substitution $\theta$, the instance of the goal,  $\true \apply \theta$, that is, $\true$, is true under any model; hence the corresponding instance of the input-output condition   $\mathscr{Q}[a, \, t \apply \theta ]$, is satisfied.  (The assertion case is similar.) 
From this derivation, we can extract the \hypertarget{def:tab-fin-prog}{\emph{final program}}
 \begin{equation*} \uf(a) =  t.
 \end{equation*}
 The program may still contain variables, but the correctness condition guarantees that any instance of the program will satisfy the specification. 
 
There may be many derivations from the same initial tableau, leading to different final tableaus and final programs, each satisfying the specification. 

 \subsection*{Case Analysis and Conditional Program Formation} In a deductive approach, a conditional program structure is introduced as a result of a case analysis in the proof. 
 We describe the resolution rule and the equivalence and equality replacement rules, all of which perform case analysis and can introduce conditionals.
 
  \subsubsection{Resolution Rule.} 
\begin{quote}\emph{Once you eliminate the impossible, whatever remains \ldots  must be the truth.}
 \\---Arthur Conan Doyle, \emph{The Case-Book of Sherlock Holmes}, 1927
 \end{quote}

 Paradoxically, our derivation of the unification algorithm requires the resolution and equivalence and equality replacement rules, but these rules depend on the unification algorithm.
 
 We will first describe the resolution rule as it applies to two goals, but, by \hyperlink{prop:tab-dual}{duality}, it will also be applied to two assertions and to an assertion and a goal. We shall assume that the rows have no variables in common;  they are standardized apart if necessary.  By the instance property, this maintains the correctness of the tableau.

 Suppose we have developed two goal rows, of the form

   \begin{center}
\begin{tabular} {T}
\hline
    $\qquad$ & \begin{center}$\uand\!(\boxed{\expsub{\mathcal{P}}{1}},\expsub{\mathcal{R}}{1})$\end{center} & \begin{center}$\expsub{t}{1}$\end{center} \\
  \hline
\end{tabular}
\end{center}
\noindent and
  \begin{center}
\begin{tabular} {T}
\hline
    $\qquad$ &  \begin{center}$ \uand\!(\!\unot(\boxed{\expsub{\mathcal{P}}{2}}),\expsub{\mathcal{R}}{2})$\end{center} &  \begin{center}$\expsub{t}{2}$\end{center} \\
  \hline
\end{tabular},
 \end{center}

\noindent where the two rows are standardized apart, and $\expsub{\mathcal{P}}{1}$ and $\expsub{\mathcal{P}}{2}$ are unifiable, with the most-general unifier $\theta$; let $ \mathcal{P} = \expsub{\mathcal{P}}{1} \apply \theta = \expsub{\mathcal{P}}{2} \apply \theta$.  (The boxes are to highlight the unified subexpressions, but have no logical significance.)

Then, according to the \hypertarget{rule:tab-res-GG}{\emph{goal-goal resolution rule,}} we can infer the new goal row

  \begin{center}
\begin{tabular} {T}
\hline
    $\qquad$ & $\uand(\expsub{\mathcal{R}}{1} \apply \theta, \ \expsub{\mathcal{R}}{2} \apply \theta)$ &  $\cond {\mathcal{P}} {\expsub{t}{1} \apply \theta} {\expsub{t}{2} \apply \theta}$ \\
  \hline
\end{tabular}.
\end{center}

\noindent The rule has introduced a conditional term into the output entry;  the test is the unified formulas, and the then-branch and else-branch are instances under $\theta$ of the corresponding output entries  $\expsub{t}{1}$, and $\expsub{t}{2}$, respectively, of the given goal rows.

If the first goal is simply $\expsub{\mathcal{P}}{1}$, we treat is as if it were $\uand(\expsub{\mathcal{P}}{1},\true)$. We then automatically perform propositional-logic simplifications, such as $\uand(\mathcal{P}, \true) \rightarrow \mathcal{P}$, to the inferred row whenever possible.  If the first goal is actually 
$\uand(\expsub{\mathcal{P}}{1},\expsub{\mathcal{R}}{1},\allowbreak \expsub{\mathcal{R}}{2}, \ldots)$, with more than two conjuncts, we treat it as if it were $\uand(\expsub{\mathcal{P}}{1}, \uand(\expsub{\mathcal{R}}{1},\allowbreak \expsub{\mathcal{R}}{2}, \ldots))$.

We show that, for a given interpretation, any term allowed by the new row is also allowed by the given tableau. For, suppose some instance  
\begin{align*}
&\uand(
\expsub{\mathcal{R}}{1}\apply \theta,
\expsub{\mathcal{R}}{2}\apply \theta) \apply \theta' \  \, 
\end{align*}
that is, by \hyperlink{prop-subst-dist}{\emph{distributivity}}, 
\begin{align*}
&\uand(
(\expsub{\mathcal{R}}{1}\apply \theta)\apply \theta',
(\expsub{\mathcal{R}}{2}\apply \theta)\apply \theta') 
\end{align*}
of the new goal  is true; hence (by the truth table for $\uand$) both $(\expsub{\mathcal{R}}{1}\apply \theta)\apply \theta'$ and $(\expsub{\mathcal{R}}{2}\apply \theta)\apply \theta'$ are true. We show that the corresponding instance  
of the new output entry,
 \begin{align*}
   &\cond 
   {\mathcal{P}} 
   {\expsub{t}{1} \apply \theta} 
   {\expsub{t}{2} \apply \theta} \apply \theta'\
      &&=  \text{[by \hyperlink{prop-subst-dist}{\emph{distributivity}}]} \\
     &\cond 
     {\mathcal{P}\apply \theta'} 
     {(\expsub{t}{1} \apply \theta) \apply \theta'} 
     {(\expsub{t}{2} \apply \theta) \apply \theta'} \ 
       &&=  \text{[by the \hyperlink{prop:subst-comp}{\emph{composition property}}]} \\
     &\cond 
     {\mathcal{P}\apply \theta'} 
     {\expsub{t}{1} \apply (\theta \compose \theta')} 
     {\expsub{t}{2} \apply (\theta \compose \theta')}, \ 
 \end{align*}
 is already allowed by one of the two given rows.

The proof distinguishes between two cases.  In the first case, in which the instance of the test, 
$\mathcal{P}\apply \theta'$, that is
[because $\mathcal{P} = 
\expsub{\mathcal{P}}{1} \apply \theta)$],
$({\expsub{\mathcal{P}}{1} \apply \theta})\apply \theta'$, is true, we have
\begin{align*}
&\uand(({\expsub{\mathcal{P}}{1} \apply \theta})\apply \theta',
(\expsub{\mathcal{R}}{1}\apply \theta)\apply \theta'
)  \, &&\uiff 
  \text{[by \hyperlink{prop-subst-dist}{\emph{distributivity}}]} \\
&\uand(({\expsub{\mathcal{P}}{1} \apply \theta}),
(\expsub{\mathcal{R}}{1}\apply \theta))\apply \theta' \, &&\uiff 
   \text{[by \hyperlink{prop-subst-dist}{\emph{distributivity}}, again]} \\
  &\uand(({\expsub{\mathcal{P}}{1}},
\expsub{\mathcal{R}}{1})\apply \theta) \apply \theta' \, &&\uiff  
 \text{[by the \hyperlink{prop:subst-comp}{\emph{composition property}}]} \\
&\uand({\expsub{\mathcal{P}}{1}},
\expsub{\mathcal{R}}{1})\apply (\theta \compose \theta')
   \end{align*}
is true.  But this is an instance of the first row.  We have assumed the row allows the corresponding instance of its output entry, $\expsub{t}{1} \apply (\theta \compose \theta').$ But in this case, because the test
$\mathcal{P}\apply \theta'$ is true, the conditional answer \[\cond 
     {\mathcal{P}\apply \theta'} 
     {\expsub{t}{1} \apply (\theta \compose \theta')} 
     {\expsub{t}{2} \apply (\theta \compose \theta')}\] is equal to its then-branch $\expsub{t}{1} \apply (\theta \compose \theta')$;  that is, the conditional is allowed by the first row.

In the second case, in which the instance of the test, 
$\mathcal{P}\apply \theta'$, that is
[because $\mathcal{P} = 
\expsub{\mathcal{P}}{2} \apply \theta)$],
$({\expsub{\mathcal{P}}{2} \apply \theta})\apply \theta'$, is false, we have [by the truth table for $\unot$] $\unot(({\expsub{\mathcal{P}}{2} \apply \theta})\apply \theta')$ is true, and hence
\begin{align*}
&\uand(\unot(({\expsub{\mathcal{P}}{2} \apply \theta})\apply \theta'),
(\expsub{\mathcal{R}}{2}\apply \theta)\apply \theta'
)  \,  &&\uiff 
  \text{[by \hyperlink{prop-subst-dist}{\emph{distributivity}}]} \\
&\uand(\unot({\expsub{\mathcal{P}}{2} \apply \theta}),
(\expsub{\mathcal{R}}{2}\apply \theta))\apply \theta' \,  &&\uiff 
  \text{[by \hyperlink{prop-subst-dist}{\emph{distributivity}}, again]} \\
  &\uand((\unot{\expsub{\mathcal{P}}{2}},
\expsub{\mathcal{R}}{2})\apply \theta) \apply \theta' \, &&\uiff  
 \text{[by the \hyperlink{prop:subst-comp}{\emph{composition property}}]} \\
&\uand(\unot{\expsub{\mathcal{P}}{2}},
\expsub{\mathcal{R}}{2})\apply (\theta \compose \theta')
   \end{align*}
is true.  But this is an instance of the second row.  We have assumed the row allows the corresponding instance of its output entry, $\expsub{t}{2} \apply (\theta \compose \theta').$ However,  in this case, because the test
$\mathcal{P}\apply \theta'$ is false, the conditional answer $\cond 
     {\mathcal{P}\apply \theta'} 
     {\expsub{t}{1} \apply (\theta \compose \theta')} 
     {\expsub{t}{2} \apply (\theta \compose \theta')}$ is equal to its else-branch $\expsub{t}{2} \apply (\theta \compose \theta')$;  that is, the conditional is allowed by the second row.

This concludes the proof of correctness of the resolution rule and its introduction of conditional terms.  

\subsubsection{Missing Output Entries and the Resolution Rule.}  When one of the two goal rows has no output entry, we do not form a conditional.  For instance, suppose we have two goal rows,

   \begin{center}
\begin{tabular} {T}
\hline
    $\qquad$ & \begin{center}$\uand(\boxed{\expsub{\mathcal{P}}{1}},\expsub{\mathcal{R}}{1})$ \end{center}& \begin{center}$\expsub{t}{1}$\end{center} \\
  \hline
\end{tabular}
\end{center}
\noindent and
  \begin{center}
\begin{tabular} {T}
\hline
    $\qquad$ & \begin{center}$ \uand(\unot (\boxed{\expsub{\mathcal{P}}{2}}),\expsub{\mathcal{R}}{2})$ \end{center}&  $\qquad$ \\
  \hline
\end{tabular},
\end{center}
\noindent where the same conditions apply but the second row has no output entry.  But we have mentioned that a row with no output entry can be replaced by a row with a new output variable $\var{V}$:

  \begin{center}
\begin{tabular} {T}
\hline
    $\qquad$ & \begin{center}$ \uand(\unot (\boxed{\expsub{\mathcal{P}}{2}}),\expsub{\mathcal{R}}{2})$\end{center} & \begin{center} $\var{V} $\end{center} \\
  \hline
\end{tabular}.
\end{center}

\noindent Applying the resolution rule, we obtain
 \begin{center}
\begin{tabular} {T}
\hline
    $\qquad$ & \begin{center}$\uand(\expsub{\mathcal{R}}{1} \apply \theta, \ \expsub{\mathcal{R}}{2} \apply \theta)$\end{center} &  \begin{center}$\cond {\mathcal{P}} {\expsub{t}{1} \apply \theta} {\var{V} \apply \theta}$\end{center}\\
  \hline
\end{tabular}.
\end{center}

\noindent By the instance property, we can add a row in which new variable $\var{V}$ is replaced by $\expsub{t}{1}$ (or any term), obtaining

 \begin{center}
\begin{tabular} {T}
\hline
    $\qquad$ & \begin{center}$\uand(\expsub{\mathcal{R}}{1} \apply \theta, \ \expsub{\mathcal{R}}{2} \apply \theta)$ \end{center} & \begin{center} $\cond {\mathcal{P}} {\expsub{t}{1} \apply \theta} {\expsub{t}{1} \apply \theta}$\end{center} \\
  \hline
\end{tabular}
\end{center}


\noindent or, equivalently (by the \hyperlink{prop-cond-same}{\emph{same-branch property}} of the conditional), 
   \begin{center}
\begin{tabular} {T}
\hline
    $\qquad$ & \begin{center}$\uand(\expsub{\mathcal{R}}{1} \apply \theta, \ \expsub{\mathcal{R}}{2} \apply \theta) $ \end{center} & \begin{center}$\expsub{t}{1}\apply \theta$\end{center} \\
  \hline
\end{tabular}.
\end{center}
\noindent Avoiding the introduction of a conditional when the test $\mathcal{P}$ is nonprimitive may enable the rule to yield a primitive output entry; otherwise, application of the rule would have been blocked.

If neither row contains an output entry, the resulting row also has no output entry. To  prove this, replace each row with one containing new variables $\var{U}$ and $\var{V}$ as output entries; the resolution rule will allow us to derive a new row with output entry $\cond {\mathcal{P}} {\var{U} \apply \theta} {\var{V} \apply \theta}$, that is [because $\var{U}$ and $\var{V}$ are new variables, not in the domain of $\theta$], $\cond {\mathcal{P}} {\var{U} } {\var{V}}$. Then, take an instance of the row with $\var{V}$ in the output entry replaced by $\var{U}$, yielding  $\cond {\mathcal{P}} {\var{U} } {\var{U}}.$ By the \hyperlink{prop-cond-same}{\emph{same-branch property}}, this output entry is equal to the variable $\var{U}$. But a row with a new variable as its output entry can be replaced by a row with no output entry at all.

\subsubsection{Dual Forms of the Resolution Rule.}  We have called the above the \hyperlink{rule:tab-res-GG}{\emph{goal-goal version}} of the rule.  By \hyperlink{prop:tab-dual}{duality}, we can apply the resolution rule to two assertion rows (the \hypertarget{rule:tab-res-AA}{\emph{assertion-assertion version}}), or to an assertion and a goal (the \hypertarget{rule:tab-res-AG}{\emph{assertion-goal version}}).  For two assertions, the \emph{assertion-assertion version} is as follows. Suppose we have developed two assertion rows, of the form

   \begin{center}
\begin{tabular}{T}
\hline
   \begin{center} $\uor(\!\unot(\boxed{\expsub{\mathcal{P}}{1}}),\expsub{\mathcal{R}}{1})$\end{center} & 
   $\qquad$ 
   & \begin{center}$\expsub{t}{1}$\end{center}\\
  \hline
\end{tabular}
\end{center}

\noindent and
  \begin{center}
\begin{tabular} {T}
\hline
  \begin{center}  $ \uor(\boxed{\expsub{\mathcal{P}}{2}},\expsub{\mathcal{R}}{2})$ \end{center} 
    &  $\qquad$ 
    &\begin{center} $\expsub{t}{2}$ \end{center}\\
  \hline
\end{tabular},
\end{center}
 
\noindent where, as in the  \hyperlink{rule:tab-res-GG}{\emph{goal-goal version}}, the two rows are standardized apart, and $\expsub{\mathcal{P}}{1}$ and $\expsub{\mathcal{P}}{2}$ are unifiable, with the most-general unifier $\theta$, and $ \mathcal{P} = \expsub{\mathcal{P}}{1} \apply \theta = \expsub{\mathcal{P}}{2} \apply \theta$.  Then, we can infer the new assertion row

\begin{center}
\begin{tabular}{T}
\hline
  \begin{center} $\uor(\expsub{\mathcal{R}}{1} \apply \theta, \ \expsub{\mathcal{R}}{2} \apply \theta) $ \end{center}& &
   \begin{center} $\cond {\mathcal{P}} {\expsub{t}{1} \apply \theta} {t_2 \apply \theta}$ \end{center}
  \\
  \hline
\end{tabular}.
\end{center}

\noindent To see this, we can (by \hyperlink{prop:tab-dual}{duality}) pass both assertions to the goal column by negating them, apply the  \hyperlink{rule:tab-res-GG}{\emph{goal-goal version}}  of the rule, then (by duality, again) push the resulting goal to the assertion column by negating it, and then simplify (by propositional-logic equivalences) the resulting assertion.

As in the \emph{goal-goal version} of the rule, if the first row has no output entry, no conditional is formed; the output entry for the new row is simply ${\expsub{t}{2} \apply \theta}$.  If the second row has no output entry, the new row has output entry  ${\expsub{t}{1} \apply \theta}.$  If neither given row has an output entry, neither does the new row.

Similarly, in the \hyperlink{rule:tab-res-AG}{\emph{assertion-goal version}} of the rule, we assume we are given the assertion row
   
   \begin{center}
\begin{tabular} {T}
\hline
    \begin{center}$\uor(\boxed{\expsub{\mathcal{P}}{1}},\expsub{\mathcal{R}}{1})$ \end{center}& $\qquad$ & \begin{center}$\expsub{t}{1}$\end{center} \\
  \hline
\end{tabular}
\end{center}
\noindent and the goal row
  \begin{center}
\begin{tabular} {T}
\hline
   $\qquad$ & \begin{center}$ \uand (\boxed{\expsub{\mathcal{P}}{2}},\expsub{\mathcal{R}}{2})$ \end{center}   & \begin{center}$\expsub{t}{2}$\end{center} \\
  \hline
\end{tabular},
\end{center}
 
\noindent where, as in the other versions, the two rows are standardized apart, and $\expsub{\mathcal{P}}{1}$ and $\expsub{\mathcal{P}}{2}$ are unifiable, with the most-general unifier $\theta$, and $ \mathcal{P} = \expsub{\mathcal{P}}{1} \apply \theta = \expsub{\mathcal{P}}{2} \apply \theta$.  Then, we can infer the new goal row
  \begin{center}
\begin{tabular} {T}
\hline
   $\qquad$ & 
   \begin{center}
   $\uand(\!
 \begin{aligned}[t]
 &\!\!\unot(\expsub{\mathcal{R}}{1}) \apply \theta,  \\
 &\expsub{\mathcal{R}}{2} \apply \theta)
 \end{aligned}$ 
 \end{center}
 &   \begin{center}$\cond {\mathcal{P}} {\expsub{t}{2} \apply \theta} {\expsub{t}{1} \apply \theta}$\end{center}\\
  \hline
\end{tabular}.
\end{center}
\noindent The justification is similar to those for the other versions.


\subsubsection{Equivalence and Equality Replacement Rules.}  Simple equivalences can be treated as rewrite rules:  replace any instance of the left side with the corresponding instance of the right side.  More generally, we state the \emph{equivalence replacement} rule here.  As with the resolution rule, there are \emph{assertion-assertion}, \emph{goal-goal} and \emph{assertion-goal versions}, but we show only the  \hypertarget{rule:tab-equiv-AG}{\emph{assertion-goal}}  form.

We assume we are given the assertion row
   \begin{center}
\begin{tabular} {T}
\hline
    \begin{center}$\uor(\boxed{\expsub{\mathcal{P}}{1}}\uiff \mathcal{Q},{\expsub{\mathcal{R}}{1}})$ \end{center}& $\qquad$ & \begin{center}$\expsub{t}{1}$\end{center} \\
  \hline
\end{tabular}
\end{center}
\noindent and the goal row
  \begin{center}
\begin{tabular} {T}
\hline
   $\qquad$ &  \begin{center}$ \uand (\mathcal{S}\langle\boxed{\expsub{\mathcal{P}}{2}}\rangle,\expsub{\mathcal{R}}{2})$\end{center}   & \begin{center} $\expsub{t}{2}$\end{center} \\
  \hline
\end{tabular},
\end{center}

\noindent where, as in the other rules, the two rows are standardized apart and $\expsub{\mathcal{P}}{1}$ and $\expsub{\mathcal{P}}{2}$ are unifiable, with the most-general unifier $\theta$, and $\mathcal{P} = \expsub{\mathcal{P}}{1} \apply \theta = \expsub{\mathcal{P}}{2} \apply \theta$.  Here, $\mathcal{S}\langle\expsub{\mathcal{P}}{2}\rangle$ is a formula that contains at least one occurrence of $\expsub{\mathcal{P}}{2}.$
Then, the \emph{equivalence replacement} rule allows us to infer the new goal row

\begin{center} 
\begin{tabular}{T}
\hline
$\qquad$  & 
  \begin{center}

$\begin{aligned}
\uand(&\!\unot({\expsub{\mathcal{R}}{1} \apply \theta}),\\
&\mathcal{S}\langle\mathcal{Q}\rangle\apply \theta, \\ 
&\expsub{\mathcal{R}}{2} \apply \theta)
\end{aligned}$
\end{center}
& 
\begin{center} $\begin{aligned}\uif\,(&{\mathcal{P} \uiff \mathcal{Q}\apply \theta}, \\
&{\expsub{t}{2} \apply \theta}, \\
&{\expsub{t}{1} \apply \theta})\end{aligned}$\end{center}
\\
\hline
\end{tabular}.
\end{center}

\noindent Here, $\mathcal{S}\langle\mathcal{Q}\rangle$ is the result of replacing one or more occurrences of $\expsub{\mathcal{P}}{2}$ in $\mathcal{S}\langle\expsub{\mathcal{P}}{2}\rangle$ with $\mathcal{Q}$. We omit the justification, which depends on the \hyperlink{prop:equiv-subst}{\emph{substitutivity property}} of equivalence.

We have shown the \emph{left-to-right version} of the rule, in which an instance of the left side of the equivalence is replaced by the corresponding instance of the right side;  there is also an analogous \emph{right-to-left version}.

In our {Snark} implementation, equivalences are replaced by other connectives by  translating to clause form, so the equivalence rules do not appear in proofs.  However, in exposition the rules make some proofs easier to understand. 

The  \hypertarget{rule:tab-eql-AG}{\emph{equality replacement}} rule is analogous to the \emph{equivalence replacement} rule. It corresponds to the paramodulation rule in resolution theorem proving.

We assume we are given the assertion row
   \begin{center}
\begin{tabular} {T}
\hline
\begin{center}
    $\uor(\boxed{\expsub{s}{1}}= t,\expsub{\mathcal{R}}{1})$ 
  \end{center}  
    & $\qquad$ &\begin{center} $\expsub{t}{1}$\end{center}  
    \\
  \hline
\end{tabular}
\end{center}
\noindent and the goal row
  \begin{center}
\begin{tabular} {T}
\hline
   $\qquad$ &  
   \begin{center}
 $\uand (\mathcal{P}\langle\boxed{\expsub{s}{2}}\rangle,\expsub{\mathcal{R}}{2})$
\end{center}
     &\begin{center} $\expsub{t}{2}$\end{center} \\
  \hline
\end{tabular},
\end{center}

\noindent where, as in the other rules, the two rows are standardized apart, and $\expsub{s}{1}$ and $\expsub{s}{2}$ are unifiable, with the most-general unifier $\theta$, and $s = \expsub{s}{1} \apply \theta = \expsub{s}{2} \apply \theta$.  Here, $\mathcal{P}\langle\expsub{s}{2}\rangle$ is a formula that contains at least one occurrence of $\expsub{s}{2}.$

Then, the \emph{equality replacement} rule allows us to infer the new goal row
  \begin{center}
\begin{tabular} {T}
\hline
   $\qquad$ & 
   \begin{center}
     $\begin{aligned}
\uand(&\!\unot({\expsub{\mathcal{R}}{1} \apply \theta}),\\
&\mathcal{P}\langle t\rangle\apply \theta, \\  
&\expsub{\mathcal{R}}{2} \apply \theta)
\end{aligned}$
\end{center} & 
\begin{center}
$
\begin{aligned}
\uif(&{s = t\apply \theta}, \\
&{\expsub{t}{2} \apply \theta}, \\
&{\expsub{t}{1} \apply \theta})
\end{aligned}
$
\end{center}\\
  \hline
\end{tabular}.
\end{center}


\noindent Here, $\mathcal{P}\langle t\rangle$ is the result of replacing one or more occurrences of $\expsub{s}{2}$ in $\mathcal{P}\langle\expsub{s}{2}\rangle$ with $t$. We omit the justification, which depends on the \hyperlink{prop:eql-subst}{\emph{substitutivity property}} of equality. 

We have shown the \emph{left-to-right version} of the rule, in which an instance of the left side of the equality is replaced by the corresponding instance of the right side;  there is also an analogous \emph{right-to-left version}. In the {Snark} implementation, the version to be applied depends on a symbol ordering strategy.

This rule corresponds to the \emph{paramodulation} rule in ordinary resolution theorem proving.

\subsection*{Mathematical Induction and Recursive Program Formation}   We now see how the well-founded induction principle is integrated into the deductive tableau. In the deductive tableau framework, we begin with the initial tableau

   \begin{center}
\begin{tabular}{T}
 \hline
 \begin{center} Assertions\end{center} &
 \begin{center} Goals\end{center}
 &\begin{center} $\uf(a)$ \end{center}\\ 
 \hhline{|=|=|=|} 
 &\begin{center} $ \mathscr{Q}[a,  \var{Z}] $ \end{center}& \begin{center}$\var{Z}$ \end{center}\\ 
 
 \hline
\end{tabular}.
\end{center}
\noindent In other words, for a given input entity $a$, we are constructing a program $\uf$ that will yield an output $\var{Z}$ that satisfies the input-output condition $ \mathscr{Q}[a, \var{Z}] $. For a well-founded relation $\expsub{\prec}{w}$, we can conduct the proof under the induction hypothesis that recursive calls to the program $\uf(\var{x})$ we are constructing will satisfy the input-output condition $ \mathscr{Q}[\var{x}, \, \uf(\var{x})]$ for any input $\var{x}$ such that $\var{x}\,\expsub{\prec}{w} \, a.$  We add this \hypertarget{ass:ind-hyp}{\emph{induction hypothesis}} to the initial tableau as an assertion:

      \begin{center}
\begin{tabular}{T}
 \hline
 \begin{center} Assertions \end{center} 
 & \begin{center} Goals \end{center}
 &\begin{center} $\uf(a)$ \end{center}\\ \hhline{|=|=|=|}
 & \begin{center} $ \mathscr{Q}[a,  \var{Z}] $ \end{center}
 & \begin{center} $\var{Z}$ \end{center} \\ 
 \hline 
 \begin{center} $ \var{X} \expsub{\prec}{w}\,a   \, \uimplies \mathscr{Q}[\var{X},\, \uf(\var{X})]$ \end{center} &  &  \\ 
 \hline
\end{tabular}.
\end{center}
In other words, instead of proving the desired conclusion itself, we prove the inductive step, which says that the induction hypothesis implies the desired conclusion. 

Using the induction hypothesis in a proof can introduce a recursive call into the program being derived.  However,  including the induction hypothesis in the tableau does not force us to derive a recursive program. If we don't use the induction hypothesis in the derivation proof, the program we derive won't use recursion.

Use of the well-founded induction principle provides a termination argument for the derived program.  If the program $\uf(a)$ invokes a recursive call $\uf(\var{X})$, we require that $ \var{X} \expsub{\prec}{ w} \, a $, that is,  $ a \expsub{\succ}{w} \, \var{X}$. If there were a nonterminating computation in which $\uf(\expsub{a}{0})$ invokes $\uf(\expsub{a}{1})$, which invokes $\uf(\expsub{a}{2})$, and so forth, the sequence of arguments $\expsub{a}{0}, \expsub{a}{1}, \expsub{a}{2}, \ldots  $ would constitute an infinite decreasing sequence, because $\expsub{a}{0} \expsub{\succ}{w}\, \expsub{a}{1} \expsub{\succ}{w} \, \expsub{a}{2}, \ldots ,$ contradicting the well-foundedness of $\expsub{\prec}{w}$.  The program may still fail to terminate for cases in which the specification does not restrict the output, e.g, if the programs's input is illegal.

The choice of the well-founded relation is not straightforward.  Generally, we cannot anticipate what relation to use until the proof is underway. \citet{man:wal} developed a complete proof in which the well-founded relation was unspecified; only then, when one could see what properties the relation had to satisfy, was it possible to construct the appropriate relation.  

For an automatic system, we can expect the formulator of a subject domain theory to provide properties of the most basic well-founded relations in that theory; e.g., for the theory of expressions and substitutions, we would be given  properties of the $\vars$  relation $\expsub{\prec}{\vars(\subset)}$  and the $\size$ relation $\expsub{\prec}{\size(<)}$.  But any particular algorithm may require some combination of the basic relations for its derivation and termination proof. We can reasonably hope  an automatic system would discover this combination, but we have not yet introduced the necessary apparatus; we have unpublished work in this direction.  

Finding these well-founded relations can be challenging for people.  For instance, the associative-commutative unification algorithm,  which allows us to take into account the associativity and commutativity of function symbols in the unified expressions \citep{sti:acu,  liv-siek:acu} was published (and used) years before the well-founded relation was discovered that allowed its termination to be established \citep{fag:acu}.

For the derivation discussed in this paper, we provide a successful well-founded relation as an unautomated “eureka" step. 

We only apply induction to the initial tableau;  if induction needs to be applied to subsequent rows, this must be done in a separate subsidiary tableau, and the theorem established by that derivation may be added to the main tableau as a new assertion.  We have not automated this step; 
the proliferation of subsidiary tableaus could put a drain on the search.  This may be regarded as a topic for future research.

\section*{The Derivation of the Unification Algorithm}

We illustrate the use of the deductive tableau framework with examples selected from the derivation of a unification algorithm.


 
\subsection*{The Specification and First Steps}\hypertarget{sec:spec-first}{} 
For our specification for the unification algorithm, we will assume that the environment substitution is idempotent; this allows a simpler algorithm and termination proof. Thus,  the \hypertarget{unif:cond-IO}{\emph{input-output condition}} for the unification algorithm will be
   \begin{align*}
       &\idem(\thet{0}) \uimplies \, \\
       &\,{\mgiu}(\thet{0}, \e{1}, \e{2}, \Theta)
   \end{align*}
   
   \noindent where the inputs are $\thet{0}$, $\e{1}$, and $\e{2}$, and the output is $\Theta$.

   Our primitive instructions include the failure indicator $\fail$, the empty substitution $\emptysubst$, the expression decomposers $\lef$ and $\rig$, the substitution application $\apply$, the composition operator $\compose$, the  replacement operator $\mapsto$ and the relation symbols $=$, $\issubst$, $\isatm$, $\iscnst$, $\isvar$, $\occursin$, and $\misses$.  The logical connectives, the conditional operator $(\uif\_\uthen \_\uelse\_)$ and recursive calls are primitive as usual.

We are given as inputs an idempotent environment substitution $\thet{0}$ and two expressions $\e{1}$ and $\e{2},$ and want to find an output $\Theta$ that is an extension of $\thet{0}$ and is a most-general idempotent reducing unifier of $\e{1}$ and $\e{2}.$  Our \hypertarget{unif:spec}{\emph{specification}} is thus

\begin{alignat*}{2}
  \begin{split}
   \unify(&\thet{0}, \e{1}, \e{2}) \Lleftarrow  \\ 
&\operatorname{find} \, \Theta \,
     \operatorname{such}\operatorname{that} \\
&\begin{aligned}
  &&&&& \idem(\thet{0}) \uimplies  \\
  &&&&& \mgiu(\thet{0}, \e{1}, \e{2}, \Theta). 
   \end{aligned}
  \end{split}
\end{alignat*}

\vspace*{.2cm}
\noindent Our \hypertarget{unif:tab-init}{\emph{initial tableau}} is then

\begin{center}  
\begin{tabular}{T}
 \hline
\begin{center}  Assertions \end{center}&
\begin{center}  Goals  \end{center}  &
\begin{center}  $\unify(\thet{0}, \e{1}, \e{2})$  \end{center}\\
 \hline\hline
  & \begin{center}$\begin{aligned}[t] &\idem(\thet{0}) \uimplies \\  &\,{\mgiu}(\thet{0}, \e{1}, \e{2} , \Theta) \\ \end{aligned}$ \end{center} & \begin{center}$\Theta$ \end{center} \\

\hline
\end{tabular}.
\end{center}

\noindent In other words, for given input entities $\thet{0}$, $\e{1}$, and $\e{2},$ we are constructing a program  $\,\unify(\thet{0}, \e{1}, \e{2})$ that will yield an output $\Theta$ that satisfies the input-output condition  $\idem(\thet{0}) \uimplies {\mgiu}(\thet{0}, \e{1}, \e{2} , \Theta)$.  Here $\Theta$ is a variable; it behaves as if it were existentially quantified, and it can be instantiated with other terms during the proof process.

 By the \hyperlink{rul:tab-impl-spl}{\emph{implication splitting}} property, we can replace this goal with the separate assertion and goal
\begin{center}
\begin{tabularx}{1.0\textwidth} { 
  | >{\raggedright\arraybackslash}X 
  | >{\raggedright\arraybackslash}X 
  || >{\raggedright\arraybackslash}X | }
 \hline 
 \vspace{-.3cm}\begin{center}$\idem(\thet{0})$\end{center} &  & \vspace{-.3cm}\begin{center}$\Theta$ \end{center} \\
\hline
  & \vspace{-.3cm}\begin{center}$ {\mgiu}(\thet{0}, \e{1}, \e{2} , \Theta)$  \end{center}  & \vspace{-.3cm}\begin{center}$\Theta$\end{center} \\
\hline
\end{tabularx}.
\end{center}
Because the output entry $\Theta$ is a variable that does not occur in the above assertion, it may be dropped (by the \hyperlink{prop:tab-orph}{\emph{orphaned output-entry property}}), leaving

\begin{center}
\begin{tabularx}{1.0\textwidth} { 
  | >{\raggedright\arraybackslash}X 
  | >{\raggedright\arraybackslash}X 
  || >{\raggedright\arraybackslash}X | }
 \hline 
 \vspace{-.3cm}\begin{center}$\boxed{\idem(\thet{0})}$\end{center} &  & \\
\hline
\end{tabularx}.
\end{center}
We shall subsequently refer to the above assertion (without the output entry) and goal as the \hypertarget{ass:init}{\emph{initial assertion}} and the \hypertarget{goal:init}{\emph{initial goal}}, respectively. 

We have remarked, as the \hyperlink{prop:idem-more-gen}{\emph{idempotent-more-general property}}, that a substitution is idempotent precisely when it is an extension of itself; thus our tableau contains the valid assertion

\begin{center}
\begin{tabularx}{1.0\textwidth} { 
  | >{\raggedright\arraybackslash}X 
  | >{\raggedright\arraybackslash}X 
  || >{\raggedright\arraybackslash}X | }
 \hline 
 \begin{center}
    \vspace{-.25cm}
$\begin{aligned}\boxed{\idem(\Theta)} \uiff  \Theta \moregen \Theta \end{aligned}$
\end{center}
  & 
  &  \\ \hline
\end{tabularx}.
\end{center}
 The symbol $\Theta$ in this assertion is a variable, which makes it behave as if it were universally, rather than existentially, quantified, because it is in an assertion rather than a goal. The two assertions are already standardized apart, since they have no variables in common. We can apply the \emph{assertion-assertion version} of the  \hyperlink{rule:tab-equiv-AG}{\emph{equivalence replacement}} rule to this assertion and the previous assertion,  taking $\expsub{\mathcal{P}}{1}$ and  $\expsub{\mathcal{P}}{2}$ to be the boxed subexpressions  $\idem(\Theta)$ and $\idem(\thet{0})$, respectively, and taking the most-general unifier to be $\{\Theta \mapsto \thet{0}\}$. We obtain the new assertion

\begin{center}
\begin{tabularx}{1.0\textwidth}{V}
 \hline 
 \begin{center}
 \vspace{-.25cm}
$\begin{aligned} \thet{0} \moregen \thet{0} \end{aligned}$
\end{center}
  & 
  &  \\ \hline
\end{tabularx}.
\end{center}
\noindent Since neither of the given assertions has an output entry, the new assertion has none either.  We shall later call this phrasing of our \hyperlink{ass:init}{\emph{initial assertion}} the \hypertarget{ass:idem}{\emph{paraphrased idempotence assumption}}, that the environment is more-general than itself. 
 
  We defined the \hyperlink{prop:def-mgiu}{most-general idempotent reducing unifier relation} by an equivalence, which we can, by the  \hyperlink{prop:tab-valid-into}{\emph{valid-formula-introduction property}}, incorporate as a valid assertion in our tableau.
    \begin{center}
  \begin{tabular} {|m{0.45\textwidth}|m{0.20\textwidth}||m{0.15\textwidth}|}
\hline
 \begin{center} 
${\begin{aligned} 
&\boxed{\mgiu(\Thet{0}, \E{1}, \E{2}, \Theta)} \uiff \\
&\left[\begin{aligned}
&\E{1} \apply \Theta = \E{2} \apply \Theta \, \uand 
 \\
 \,\, & \Thet{0} \moregen \Theta \, \uand
\\
  \,\, & {\mgi}(\Thet{0}, \E{1}, \E{2}, \Theta)\uand
\\
  \,\, &{\reduce}(\Thet{0}, vars(\E{1}, \E{2})\!\apply\!\Thet{0}, \Theta)
\end{aligned}\right]\end{aligned}}$\hspace{1cm} 
\end{center}& &  \\  \hline
\end{tabular}.
\end{center}

\noindent   We use this definition to expand the \hyperlink{goal:init}{\emph{initial goal}}: 
\begin{center}
\begin{tabular}{T}
\hline
  & \begin{center}$ \boxed{{\mgiu}(\thet{0}, \e{1}, \e{2} , \Theta)}$  \end{center}  & \begin{center}$\Theta$\end{center} \\
\hline
\end{tabular}.
\end{center}
We apply the \emph{assertion-goal version} of the \hyperlink{rule:tab-equiv-AG}{\emph{equivalence replacement}} rule to this assertion and the \hyperlink{goal:init}{\emph{initial goal}} (standardizing the two rows apart by renaming the variable $\Theta$ in the goal to $\Thet{1}$, taking  $\expsub{\mathcal{P}}{1}$ and  $\expsub{\mathcal{P}}{2}$ to be the boxed subexpressions ${\mgiu}(\Thet{0}, \E{1}, \E{2}, \Theta)$ and $ {\mgiu}(\thet{0}, \e{1}, \e{2} , \Thet{1})$, respectively, and taking the most-general unifier to be $\{\Thet{0} \mapsto \thet{0}, \E{1} \mapsto \e{1}, \E{2} \mapsto \e{2}, \Thet{1} \mapsto \Theta\}$. We expand our goal to obtain

    \begin{center}
  \begin{tabular}{S}
\hline
 & \begin{center}
${\begin{aligned} 
&\begin{aligned}
&\boxed{\e{1} \apply \Theta = \e{2} \apply \Theta} \, \uand 
 \\
 \,\, & \thet{0} \moregen \Theta \, \uand
\\
  \,\, & {\mgi}(\thet{0}, \e{1}, \e{2}, \Theta)\, \uand
\\
  \,\, & {\reduce}(\thet{0},\vars(\e{1}, \e{2})\!\apply\thet{0}, \Theta)
\end{aligned}\end{aligned}}$\hspace{1cm} 
\end{center} & \begin{center}$\Theta$\end{center} \\  \hline
\end{tabular}.
\end{center}
Since only one of the given rows has an output entry, no conditional term is formed in the new output entry.  We shall refer to this as the \hypertarget{goal-exp-init}{\emph{expanded initial goal}}. Here, $\Theta$ is still a variable; the other variables in the assertion have been instantiated with constants, the program's input parameters.

\subsection*{The Equal-Expression Case}\hypertarget{sec:eq-exps}{}

We have (as a property of equality) the valid assertion
\begin{center}
  \begin{tabular}
  {|m{0.40\textwidth}|m{0.20\textwidth}||m{0.20\textwidth}|}
\hline
\begin{center}
\hspace{-.5cm}
{\begin{align*}
&  \E{1} = \E{2}  \, \uimplies \\
& \boxed{\E{1}\apply\Theta = \E{2}\apply\Theta}
\end{align*}}
\end{center}
 & &
\\
\hline
\end{tabular}.
\end{center}
We standardize this row apart from the above \hyperlink{goal-exp-init}{\emph{expanded initial goal}}  by renaming the variable $\Theta$ in the goal to $\Thet{1}$. We apply the \hyperlink{rule:tab-res-AG}{\emph{assertion-goal version}} of the resolution rule to the above assertion and the \emph{expanded initial goal}. For the unified subexpressions, we take  $\expsub{\mathcal{P}}{1}$ to be the boxed subexpression $\E{1}\apply\Theta = \E{2}\apply\Theta$ in the assertion and  $\expsub{\mathcal{P}}{2}$ to be the boxed subexpression (after renaming) $\e{1} \apply \Thet{1} = \e{2} \apply \Thet{1}$ in the goal. The most-general unifier is $\{\varsub{E}{1} \mapsto \expsub{e}{1}, \varsub{E}{2} \mapsto \expsub{e}{2}, \Thet{1} \mapsto \Theta\}$.  The resulting goal is
\begin{center}
\begin{tabular}{S}
 \hline 
  & 
  \begin{center}
${\begin{aligned} 
&\begin{aligned}
&\e{1} = \e{2}  \, \uand 
 \\
 \,\, & \thet{0} \moregen \Theta \, \uand
\\
  \,\, & \boxed{{\mgi}(\thet{0}, \e{1},  \e{2}, \Theta)}\, \uand
\\
  \,\, & reduce(\thet{0},\vars(\e{1}, \e{2})\!\apply\thet{0}, \Theta)
\end{aligned}.\end{aligned}}$\hspace{1cm} 
\end{center} & \begin{center}$\Theta$ \end{center} \\
\hline
\end{tabular}.
\end{center}

We established that the most-general-idempotent relation $\mgi$ has a \hyperlink{prop:reflex-mgi}{\emph{reflexivity property}}, expressed in the valid assertion
\begin{center}
  \begin{tabular}{|m{0.40\textwidth}|m{0.20\textwidth}||m{0.20\textwidth}|}
\hline
\begin{center}
\[\hspace{-.5cm}
\begin{aligned}
& \Thet{1} = \Thet{2} \,\, \uimplies \\
& \boxed{\mgi(\Thet{1}, \E{1}, \E{2}, \Thet{2})}  
\end{aligned}
\]
\end{center}
 & &
\\
\hline
\end{tabular}.
\end{center}
In other words, the environment $\Thet{1}$ itself is most-general idempotent for any two expressions (though it may not be a unifier).  This assertion and the most recent goal are already standardized apart---they have no variables in common.  By \hyperlink{rule:tab-res-AG}{\emph{assertion-goal resolution}}, unifying the boxed subexpressions with the most-general unifier 
$\{\varsub{E}{1} \mapsto \expsub{e}{1}, \varsub{E}{2} \mapsto \expsub{e}{2}, \Thet{1} \mapsto \thet{0}, \Thet{2} \mapsto \Theta\}$,
we obtain the goal
\begin{center}
\begin{tabular}{S}
 \hline 
  & 
  \begin{center}
${\begin{aligned} 
&\begin{aligned}
&\boxed{\theta_0 = \Theta} \, \uand \\
&\e{1} = \e{2}  \, \uand 
 \\
 \,\, & {\thet{0} \moregen \Theta} \, \uand
\\
  \,\, &
{\reduce}(\thet{0},\vars(\e{1}, \e{2})\!\apply\thet{0}, \Theta)
\end{aligned}.\end{aligned}}$\hspace{1cm} 
\end{center} & \begin{center}$\Theta$ \end{center} \\
\hline
\end{tabular}.
\end{center}
By \hyperlink{rule:tab-res-AG}{\emph{assertion-goal resolution}} between the \hyperlink{prop:eq-ref}{\emph{reflexivity property}} of equality, 
\begin{center}
\begin{tabular}{S}
 \hline 
  \begin{center}
${\begin{aligned} 
&\begin{aligned}
 \boxed{X = X}
\end{aligned}\end{aligned}}$\hspace{2cm} 
\end{center} 
  & 
 &  \\
\hline
\end{tabular},
\end{center}
and the above goal, with the most-general unifier $\{X \mapsto \thet{0}, \Theta \mapsto \thet{0}\}$, we obtain the goal 
\begin{center}
\begin{tabular}{S}
 \hline 
  & 
  \begin{center}
${\begin{aligned} 
&\begin{aligned}
&\e{1} = \e{2}  \, \uand 
 \\
 \,\, & \boxed{\thet{0} \moregen \thet{0}} \, \uand
\\
  \,\, &
{\reduce}(\thet{0},\vars(\e{1}, \e{2})\!\apply\thet{0}, \thet{0})
\end{aligned}.\end{aligned}}$\hspace{1cm} 
\end{center} & \begin{center}$\thet{0}$ \end{center} \\
\hline
\end{tabular}.
\end{center}
Note that the output variable  $\Theta$ has now been replaced by the environment $\thet{0}$ in the output column.

By \hyperlink{rule:tab-res-AG}{\emph{assertion-goal resolution}} between the  
\hyperlink{ass:idem}{\emph{paraphrased idempotence assumption}},
\begin{center}
\begin{tabular}{S}
 \hline 
  \begin{center}
${\begin{aligned} 
&\begin{aligned}
 \boxed{\thet{0} \moregen \thet{0}}
\end{aligned}\end{aligned}}$\hspace{2cm} 
\end{center} 
  & 
 & \begin{center} \end{center} \\
\hline
\end{tabular}.
\end{center}
and the above goal, with most-general unifier the empty substitution $\emptysubst$, we reduce the goal to
\begin{center}
\begin{tabular}{S}
 \hline 
  & 
  \begin{center}
${\begin{aligned} 
&\begin{aligned}
&\e{1} = \e{2} \, \uand
\\
  \,\, &\boxed{\reduce(\thet{0}, vars(\e{1}, \e{2})\!\apply\thet{0}, \thet{0})}
\end{aligned}\end{aligned}}$\hspace{1cm} 
\end{center} & \begin{center}$\thet{0}$ \end{center} \\
\hline
\end{tabular}.
\end{center}
By \hyperlink{rule:tab-res-AG}{\emph{assertion-goal resolution}} between the \hyperlink{prop:reflex-reduce}{\emph{reflexivity property}} of $\reduce$,\textbf{}
\begin{center}
\begin{tabular}{S}
 \hline 
  \begin{center}
${\begin{aligned} 
&\begin{aligned}
 \boxed{\reduce(\Theta, V, \Theta)}
\end{aligned}\end{aligned}}$\hspace{2cm} 
\end{center} 
  & 
 & \begin{center}$\thet{0}$ \end{center} \\
\hline
\end{tabular},
\end{center}
we have the goal 
\begin{center}
\begin{tabular}{S}
 \hline 
  & 
    \begin{center}
${\begin{aligned} 
\e{1} = \e{2}\end{aligned}}$
\end{center} 
 & \begin{center}$\thet{0}$ \end{center} \\
\hline
\end{tabular}.
\end{center}
According to this row, when the two expression arguments are equal, the environment substitution $\thet{0}$ itself will satisfy the input-output condition for the desired program.  By \hyperlink{prop:tab-dual}{duality}, we could regard its negation as an assertion,
\begin{center}
\begin{tabular}{S}
 \hline 
  \begin{center}
${\begin{aligned} 
&\begin{aligned}
\unot (\e{1} = \e{2})
\end{aligned}\end{aligned}}$\hspace{2cm} 
\end{center} 
  & 
 & \begin{center}$\thet{0}$ \end{center} \\
\hline
\end{tabular}.
\end{center}
That is, henceforth we can use the \hypertarget{ass:neq}{\emph{non-eq}} case assumption, that the two expression arguments are \emph{not} equal. Otherwise, the environment substitution $\thet{0}$ would be a satisfactory output.

\subsection*{Failure Environment}\hypertarget{sec:fail-env}{}
We earlier introduced a special constant $\blk$, the \emph{black hole}, and we assumed that the failure substitution $\fail$ maps any expression into the black hole; that is, $\var{E} \apply \fail = \blk$, for any expression $\var{E}$.  Hence our tableau includes the \hyperlink{prop:unif-blk}{valid assertion}
 \begin{center}
  \begin{tabular}{S}
\hline
 \begin{center} 
$\begin{aligned}\!
\boxed{\varsub{E}{1} \apply \fail = \varsub{E}{2} \apply \fail}\,\,
 \end{aligned}$
\end{center}& &  \\  \hline
\end{tabular}.
\end{center}
We apply the \hyperlink{rule:tab-res-AG}{\emph{assertion-goal resolution}} to this assertion and the \hyperlink{goal-exp-init}{\emph{expanded initial goal}},
\begin{center}
\begin{tabular}{S}
 \hline 
  & 
  \begin{center}
${\begin{aligned} 
&\begin{aligned}
&\boxed{\e{1} \apply \Theta = \e{2} \apply \Theta} \, \uand 
 \\
 \,\, & \thet{0} \moregen \Theta \, \uand
\\
  \,\, & {\mgi}(\thet{0}, \e{1}, \e{2}, \Theta)\, \uand
\\
  \,\, & {\reduce}(\thet{0},\vars(\e{1}, \e{2})\!\apply\thet{0}, \Theta).
\end{aligned}.\end{aligned}}$\hspace{1cm} 
\end{center} & \begin{center}$\Theta$ \end{center} \\
\hline
\end{tabular}.
\end{center}
The rows are already standardized apart---they have no variables in common. For the unified subexpressions, we take  $\expsub{\mathcal{P}}{1}$ to be the entire (boxed) assertion $\varsub{E}{1} \apply \fail = \varsub{E}{2} \apply \fail$, and $\expsub{\mathcal{P}}{2}$ to be the boxed subexpression of the goal, $\e{1} \apply \Theta = \e{2} \apply \Theta.$ The most-general unifier is $\{\varsub{E}{1} \mapsto \expsub{e}{1}, \varsub{E}{2} \mapsto \expsub{e}{2}, \Theta \mapsto \fail\}$.  The resulting goal is
\begin{center}
\begin{tabular}{S}
 \hline 
  & 
  \begin{center}
${\begin{aligned} 
&\begin{aligned}
 \,\, & {\thet{0} \moregen \fail} \, \uand
\\
  \,\, & {\mgi}(\thet{0}, \e{1}, \e{2}, \fail)\, \uand
\\
  \,\, & \boxed{{\reduce}(\thet{0},\vars(\e{1}, \e{2})\!\apply\thet{0}, \fail)}
\end{aligned}\end{aligned}}$\hspace{1cm} 
\end{center} & \vspace{-.3cm}\begin{center}$\fail$ \end{center} \\
\hline
\end{tabular}.
\end{center}
Here we have introduced the failure indicator $\fail$ into the output entry.  Henceforth, we will be a bit more brisk in describing how each rule is applied.

By \emph{assertion-goal resolution} with the \hyperlink{prop-red-fail}{\emph{reduction fail property}},
\begin{center}
\begin{tabular}{S}
 \hline 
 \begin{center}\hspace{-.31cm} 
$\boxed{{\reduce}(\Theta_0, V, \fail)}$
 \end{center}
  & 
  & \begin{center}$\fail$ \end{center} \\
\hline
\end{tabular},
\end{center}
taking $\Theta_0$ to be $\theta_0$ and $V$ to be $\vars(\e{1}, \e{2})\!\apply\thet{0}$, we obtain

\begin{center}
\begin{tabular}{S}
 \hline 
  & 
  \begin{center}
${\begin{aligned} 
&\begin{aligned}
 \,\, & \boxed{\thet{0} \moregen \fail} \, \uand
\\
  \,\, & {\mgi}(\thet{0}, \e{1}, \e{2}, \fail)
\end{aligned}.\end{aligned}}$\hspace{1cm} 
\end{center} & \vspace{-.3cm}\begin{center}$\fail$ \end{center} \\
\hline
\end{tabular}.
\end{center}
We established that any substitution $\Theta$ is more-general idempotent than the failure substitution $\fail$, because $\Theta \compose \fail = \fail$; therefore our tableau contains the assertion

\begin{center}
\begin{tabular}{S}
 \hline 
 \begin{center}
$\boxed{\Theta \moregen \fail} $ 
\end{center} & & \\
\hline
\end{tabular}.
\end{center}
\noindent By the \hyperlink{rule:tab-res-AG}{\emph{assertion-goal version}} of the resolution rule (taking $\Theta$ to be $\thet{0}$), we obtain the new goal

\begin{center}
\begin{tabular}{S}
 \hline 
  & 
  \begin{center}
${\begin{aligned} 
&\begin{aligned}
  \,\, & \boxed{{\mgi}(\thet{0}, \e{1}, \e{2}, \fail)}
\end{aligned}\end{aligned}}$\hspace{1cm} 
\end{center} & \begin{center}$\fail$ \end{center} \\
\hline
\end{tabular}.
\end{center}
Because we will have need to refer back to this goal later, we name it the \hypertarget{goal-mgi-failure}{\emph{most-general idempotent failure goal}}.

We have also established (by the \hyperlink{prop:reflex-mgi}{\emph{reflexivity property}} of the $\mgi$ relation) that the environment substitution is most-general idempotent with respect to itself, for any expressions; therefore, our tableau contains the assertion

\begin{center}
  \begin{tabular}{|m{0.32\textwidth}|m{0.32\textwidth}||m{0.16\textwidth}|}
\hline
\begin{center}
\[\hspace{-.5cm}
\begin{aligned}
& \Thet{1} = \Thet{2} \,\, \uimplies \\
& \boxed{\mgi(\Thet{1}, \E{1}, \E{2}, \Thet{2})}  
\end{aligned}
\]
\end{center}
 & &
\\
\hline
\end{tabular}.
\end{center}
\noindent Again, by applying the \hyperlink{rule:tab-res-AG}{\emph{assertion-goal version}} of the resolution goal, omitting the details but taking $\Thet{1}$ to be $\thet{0}$, $E_{1}$ and $E_{2}$ to be $e_{1}$ and $e_{2}$, respectively, and $\Theta_2$ to be $\fail$, we obtain the new goal

\begin{center}
\begin{tabular}{|m{0.32\textwidth}|m{0.32\textwidth}||m{0.16\textwidth}|}
 \hline 
  & 
  \begin{center}
${\begin{aligned} 
&\begin{aligned}
 \boxed{\thet{0} = \fail} \end{aligned}\end{aligned}}$\hspace{2cm} 
\end{center} & \begin{center}$\fail$ \end{center} \\
\hline
\end{tabular}.
\end{center}

We have introduced the relation $\isprop$ to characterize the proper substitutions; consequently, we include the assertion 
\begin{center}
 \begin{tabular}{|m{0.32\textwidth}|m{0.32\textwidth}||m{0.16\textwidth}|}
 \hline 
 \begin{center}
 {\begin{align*}
 \begin{aligned}
 &\boxed{\Theta = \bot} \uiff \\ &\unot(\isprop(\Theta))
 \end{aligned}
\end{align*}}
\end{center} 
& & \\
\hline
\end{tabular}.
\end{center}

\noindent By applying the \emph{assertion-goal version} of the \hyperlink{rule:tab-equiv-AG}{\emph{equivalence replacement}} rule to the most recent assertion and goal, taking $\Theta$ to be $\thet{0}$,
we obtain the new goal

\begin{center}
\begin{tabular}{|m{0.32\textwidth}|m{0.32\textwidth}||m{0.16\textwidth}|}
 \hline 
  & 
  \begin{center}
$\begin{aligned} \hspace{-5pt}
&\begin{aligned}
\unot(\isprop(\thet{0}))
  \end{aligned}\end{aligned}$ 
\end{center} & \begin{center}$\fail$ \end{center} \\
\hline
\end{tabular}.
\end{center}

According to this row, when the environment  $\thet{0}$ is not a proper substitution, the failure substitution $\fail$ itself will satisfy the input-output condition for the desired program.  By \hyperlink{prop:tab-dual}{duality}, we could regard its negation as an assertion

\begin{center}
\begin{tabular}{|m{0.32\textwidth}|m{0.32\textwidth}||m{0.16\textwidth}|}
\hline
  \begin{center}
${\begin{aligned} 
&\begin{aligned}
\isprop(\thet{0})
  \end{aligned}\end{aligned}}$
\end{center} 
  & 
 & \begin{center}$\fail$ \end{center} \\
\hline
\end{tabular}.
\end{center}
That is, henceforth we can use the \hypertarget{ass:prop}{\emph{is-prop}} case assumption, that the environment is a proper substitution. Otherwise, the failure substitution $\fail$ would be a satisfactory output.

\subsection*{The Constant Cases.} Because applying the resolution rule does not remove the given assertion or goal, our tableau still contains the \emph{most-general idempotent failure goal},
\begin{center}
\begin{tabular}{|m{0.32\textwidth}|m{0.32\textwidth}||m{0.16\textwidth}|}
 \hline 
  & 
  \begin{center}
${\begin{aligned} 
&\begin{aligned}
  \,\, & \boxed{{\mgi}(\thet{0}, \e{1}, \e{2}, \fail)}
\end{aligned}\end{aligned}}$\hspace{1cm} 
\end{center} & \begin{center}$\fail$ \end{center} \\
\hline
\end{tabular}.
\end{center}
\noindent We have seen that any substitution is most-general idempotent for two expressions that are ununifiable;  in particular, if the two expressions are distinct constants, any substitution will satisfy the condition.   Thus,  our tableau contains the assertion
\begin{center}
\begin{tabular}{|m{0.32\textwidth}|m{0.32\textwidth}||m{0.16\textwidth}|}
 \hline 
 \begin{center}
 \hspace{-3cm}
 {\begin{align*}  \hspace{-.7cm}
{\begin{aligned} 
 &\,\boxed{{\mgi}(\Thet{0}, \E{1}, \E{2}, \Theta)} \,\, \uimpliedby \,\, \\
 &\iscnst(\E{1}) \uand \\
 &\iscnst(\E{2}) \uand \\
 &\unot (\E{1} = \E{2})
 \end{aligned}}
\end{align*}}
\end{center} 
& & \\
\hline
\end{tabular}.
\end{center}
\noindent By the \hyperlink{rule:tab-res-AG}{\emph{assertion-goal version}} of the resolution rule, 
taking $\Thet{0}$ and $\Theta$ to be $\thet{0}$ and $\fail$, respectively, and $\E{1}$ and $\E{2}$ to be $e_1$ and $e_2$, respectively,
we can derive the new goal 
  \begin{center}
\begin{tabular}
{|m{0.32\textwidth}|m{0.32\textwidth}||m{0.16\textwidth}|}
  \hline 
  
 & \begin{center}
{\begin{align*} \hspace{-.7cm}
 \begin{aligned}
&\iscnst(\e{1}) \uand \\ &\iscnst(\e{2}) \uand \\ &\unot (\e{1} = \e{2})
 \end{aligned}
\end{align*}}
\end{center}
 &  \begin{center}$\fail$ \end{center} \\
\hline
\end{tabular}.
\end{center}

\noindent In other words, when $\e{1}$ and $\e{2}$ are distinct constants, the failure indicator satisfies the input-output relation for the unification algorithm.

\subsection*{Introduction of the Occurs Check}
\hypertarget{sec:occurs-check}{}
A unification algorithm must fail if one of its argument expressions is a proper subexpression of the other; unless infinite expressions are allowed, there is no proper substitution that unifies them.   (This is not so for the unification algorithms employed in the  logic programming language Prolog, in which it is possible to include infinite terms in the unifier.) We now see how the occurs check is introduced into our algorithm.

We have established that, in any environment, if one expression is a proper subexpression of another, no substitution will be a unifier but any substitution will be most-general idempotent.  Thus, we can use the assertion 

\begin{center}
\begin{tabular}
{|m{0.32\textwidth}|m{0.32\textwidth}||m{0.16\textwidth}|}
 \hline 
 \begin{center}
 ${\begin{aligned}
 &\boxed{{\mgi}(\Thet{0}, \E{1}, \E{2}, \Theta)} \,\, \uimpliedby \, \\
&\E{1} \occursin \E{2}.
   \end{aligned}}$
   \end{center}
  & 
  &  \\
\hline
\end{tabular}.
\end{center}
We also developed the \hyperlink{goal-mgi-failure}{\emph{most-general idempotent failure goal}},
\begin{center}
\begin{tabular}
{|m{0.32\textwidth}|m{0.32\textwidth}||m{0.16\textwidth}|}
 \hline 
  & 
  \begin{center}
${\begin{aligned} 
&\begin{aligned}
  \,\, & \boxed{{\mgi}(\thet{0}, \e{1}, \e{2}, \fail)}
\end{aligned}\end{aligned}}$\hspace{1cm} 
\end{center} & \begin{center}$\fail$ \end{center} \\
\hline
\end{tabular}.
\end{center}
By the \hyperlink{rule:tab-res-AG}{\emph{assertion-goal version}} of the resolution rule, taking the unifier to be
$\{\Thet{0} \mapsto \thet{0}, \E{1} \mapsto \e{1}, \E{2} \mapsto \e{2}, \Theta \mapsto \fail\}$, we obtain the new goal
\begin{center}
\begin{tabular}{T}
 \hline 
  & 
  \begin{center}
${\begin{aligned} 
&\begin{aligned}
  \,\, & {\e{1} \occursin \e{2}}
\end{aligned}\end{aligned}}$\hspace{1cm} 
\end{center} & \begin{center}$\fail$ \end{center} \\
\hline
\end{tabular}.
\end{center}
(We have skipped over the paraphrase of the implication  \[\begin{aligned}
 {\mgi}(\Thet{0}, \E{1}, \E{2}, \Theta)  \uimpliedby \,
(\E{1} \occursin \E{2})
   \end{aligned}\]
   as the disjunction \[\begin{aligned}
 {\mgi}(\Thet{0}, \E{1}, \E{2}, \Theta)  \uor 
\unot(\E{1} \occursin \E{2})
   \end{aligned}\]
prior to application of the resolution rule.)

In other words, when the first expression argument is a proper subexpression of the second, the failure indicator $\fail$ satisfies the input-output relation for the unification algorithm. Use of this row in subsequent proof steps accounts for the introduction of the occurs check into the derived algorithm.

By duality, we can view the above goal as an assertion
\begin{center}
\begin{tabular}{T}
 \hline 
  
  \begin{center}
${\begin{aligned} 
\begin{aligned}
   \unot ({\e{1} \occursin \e{2}}) 
\end{aligned} & \end{aligned}}$\hspace{1cm} 
\end{center} & \,\, & \begin{center}$\fail$ \end{center} \\
\hline
\end{tabular}.
\end{center}
In other words, elsewhere in the derivation we may use the 
\hypertarget{ass:not-occ}{\emph{non-sub} case assumption}, that the first expression argument is not a proper subexpression of the second.

\subsection*{Introduction of a Conditional}

We have seen some early stages of the derivation.  Now, to give a simple example of the formation of a conditional construct, we skip ahead and show the final stage.  

We have described the formation of the portion of the program that handles the case in which the environment is not a proper substitution. In the final stage, we will have also developed a goal that corresponds to the case in which the environment is indeed a proper substitution:

\begin{center}
\begin{tabular}{|m{0.22\textwidth}|m{0.36\textwidth}||m{0.22\textwidth}|}
 \hline 
  & 
  \begin{center}
  \begin{align*}
\boxed{\isprop(\thet{0})}
  \end{align*}
\end{center} & \begin{center}
\begin{align*}\mathcal{T} 
\end{align*}
\end{center} \\
\hline
\end{tabular}.
\end{center}

\noindent
Here, $\mathcal{T}$ stands for the portion of the program that handles the case in which $\theta$ is proper.  So as not to spoil the surprise, we do not yet show this program segment.

In the earlier stages, we obtained the goal row 

\begin{center}
\begin{tabular}{|m{0.22\textwidth}|m{0.36\textwidth}||m{0.22\textwidth}|}
 \hline 
  &  \begin{center}
\begin{align*}
\unot(\boxed{\isprop(\thet{0})})
\end{align*}
\end{center} 
    & \begin{center}
    \begin{align*}
    \fail 
    \end{align*}\end{center} \\
\hline
\end{tabular}.
\end{center}
Applying the  \hyperlink{rule:tab-res-GG}{\emph{goal-goal version}}  of the resolution rule, taking the most-general unifier to be the empty substitution $\{\}$ (since the boxed subformulas are already identical), we develop the final program

\noindent
\begin{center}
\begin{tabular}{S} 
\hline
     & 
     \begin{center}
\begin{align*}
\emph{true}
\end{align*}
\end{center} 
& \begin{center}
 $\uif
(\begin{aligned}[t]&{\isprop(\thet{0})},\\
               &\mathcal{T}, \\
            &{\fail})
 \end{aligned}$ 
 \end{center}
\\
  \hline
\end{tabular}.
\end{center}

\noindent 
When $\isprop(\thet{0})$ is true, $\mathcal{T}$ is a suitable output. 
When $\isprop(\thet{0})$ is false, $\fail$ is a suitable output. So, in either case, the conditional $\cond{\isprop(\thet{0})}{\mathcal{T}}{\fail}$ is a suitable output.
\subsection*{Introduction of the Replacement}\hypertarget{sec:replacement}
This section shows how the environment substitution is composed with a replacement substitution to yield a new unifier.  This occurs under a special circumstance, in which the first expression argument is a variable that is missed by the environment and that does not occur in the second expression argument, and the second expression argument is also missed by the environment.  

We do this in several stages:  we manipulate the case assumptions, we develop the unifier, we show that the unifier is an extension of the environment, and we establish that the unifier is most-general idempotent and a reduction.

\subsubsection{Managing the Case Assumptions.}
In what follows, we treat the case assumptions as isolated assertions; in the context of the larger derivation, they might be subformulas of larger expressions, which would be dealt with by other proof steps.
We have assumed, as the \hyperlink{ass:prop}{\emph{is-prop}}  case assumption for the remainder of the derivation, that 
the environment $\thet{0}$ is a proper substitution, i.e.,
\begin{center}
  \begin{tabular}{|m{0.40\textwidth}|m{0.20\textwidth}||m{0.20\textwidth}|}
\hline
\begin{center}
$\boxed{\isprop(\thet{0})}$
\end{center}& & \begin{center} $\fail$ \end{center} \\  \hline
\end{tabular}.
\end{center}
Otherwise, the failure indicator $\fail$ meets the conditions of the specification.

In the subsection \hyperlink{sec:occurs-check}{\emph{Introduction of the Occurs Check}}, we introduced the \hyperlink{ass:not-occ}{\emph{non-sub} case assumption} that the first argument expression
 $\e{1}$ is not a proper subexpression of the second, $\e{2}$; that is,
\begin{center}
  \begin{tabular}{|m{0.40\textwidth}|m{0.20\textwidth}||m{0.20\textwidth}|}
\hline
\begin{center}
$\boxed{\unot(\e{1} \occursin \e{2})}$
\end{center}& & \begin{center} $\fail$ \end{center} \\  \hline
\end{tabular}.
\end{center}
Otherwise, the argument expressions are not unifiable and, again, the failure indicator $\fail$ meets the conditions of the specification.

Recall the \hyperlink{prop:occurs-ref}{\emph{reflexive-closure property}} of the $\occursin$ relation,
\[E_1 \occurseq E_2 \uiff (E_1 \occursin E_2 \uor \E{1} = \E{2}). \]
This implies the left-to-right implication 
\begin{center}
  \begin{tabular}{|m{0.40\textwidth}|m{0.20\textwidth}||m{0.20\textwidth}|}
\hline
\begin{center}
$
\begin{aligned}
&E_1 \occurseq E_2 \uimplies  \\
&\boxed{E_1 \occursin E_2} \uor \E{1} = \E{2} 
\end{aligned}
$
\end{center}& &  \\  \hline
\end{tabular},
\end{center} 
which can be included in our tableau as a valid assertion. (In our Snark implementation, the assertion would be translated into clausal form, as  
\[
\begin{aligned}[b]
&\unot(E_1 \occurseq E_2) \uor  \\
&\boxed{E_1 \occursin E_2} \uor \E{1} = \E{2} 
\end{aligned}).
\]
By the resolution rule applied to this assertion and the  above \hyperlink{ass:not-occ}{\emph{non-sub}} case assumption, taking $E_1$ and $E_2$ to be $e_1$ and $e_2$, respectively, we obtain
\begin{center}
  \begin{tabular}{|m{0.40\textwidth}|m{0.20\textwidth}||m{0.20\textwidth}|}
\hline
\begin{center}
\[
\begin{aligned}
&e_1 \occurseq e_2 \uimplies 
\boxed{e_{1} = e_2}
\end{aligned}
\]
\end{center}& & 
\begin{center} $\fail$ \end{center}
\\  \hline
\end{tabular}.
\end{center} 
Here no conditional has been introduced into the output entry, because one of the two resolved assertions has no output entry.

In the subsection \hyperlink{sec:eq-exps}{\emph{The Equal-Expression Case}}, we introduced the \hyperlink{ass:neq}{\emph{non-eq}} case assumption, that $\e{1}$ and $\e{2}$ are not the same, i.e.,
\begin{center}
  \begin{tabular}{|m{0.40\textwidth}|m{0.20\textwidth}||m{0.20\textwidth}|}
\hline
\begin{center}
$\boxed{\unot (\e{1} = \e{2})}$
\end{center}& & \begin{center} $\thet{0}$ \end{center} \\  \hline
\end{tabular}.
\end{center}
(Otherwise, the argument expressions are already equal, and the environment substitution $\thet{0}$ meets the conditions of the specification.)
By the resolution rule applied to the new assertion and the \hyperlink{ass:not-eq}{\emph{not-eq}} case assumption (the unifying substitution is empty, because the boxed subformulas are identical), we obtain 
\begin{center}
\begin{tabular}{|m{0.40\textwidth}|m{0.20\textwidth}||m{0.20\textwidth}|}
 \hline 
\begin{center}
$\boxed{\unot(\e{1} \occurseq \e{2})}$
\end{center}
  & 
& 
\begin{center}
\begin{center}
 $\uif
(\begin{aligned}[t]&{e_1 = e_2},\\
               &{\thet{0}}, \\
            &{\fail})
 \end{aligned}$ 
 \end{center}
 \end{center}
\\
\hline
\end{tabular}.
\end{center}
Here we have introduced a conditional term into the output entry, because both assertions have output entries.  

Thus, we can assume henceforth the \hypertarget{ass:non-sub}{\emph{non-sub}} case assumption that $e_1$ is not a subexpression of $e_2$;  otherwise, the conditional term satisfies the conditions of the specification.

We also assume the \hypertarget{ass:is-var}{\emph{is-var}} case assumption, that $\e{1}$ is a variable}, i.e.
\begin{center}
  \begin{tabular}{|m{0.40\textwidth}|m{0.20\textwidth}||m{0.20\textwidth}|}
\hline
\begin{center}
$\boxed{\isvar(\e{1})}$
\end{center}& & \begin{center} $\expsub{\mathcal{T}}{1}$ \end{center} \\  \hline
\end{tabular},
\end{center}
and the \hypertarget{ass:miss}{\emph{missed input}} case assumptions, that the environment substitution $\thet{0}$ misses $\e{1}$ and $\e{2}$ , i.e.,
\begin{center}
  \begin{tabular}{|m{0.40\textwidth}|m{0.20\textwidth}||m{0.20\textwidth}|}
\hline
\begin{center}
 $\boxed{\misses(\thet{0},\e{1})}$
\end{center} 
& & \begin{center} $\expsub{\mathcal{T}}{2}$ \end{center} \\  \hline
\end{tabular}
\end{center}
and 
\begin{center}
  \begin{tabular}{|m{0.40\textwidth}|m{0.20\textwidth}||m{0.20\textwidth}|}
\hline
\begin{center}
 $\boxed{\misses(\thet{0},\e{2})}$
\end{center} 
& & \begin{center} $\expsub{\mathcal{T}}{3}$ \end{center} \\  \hline
\end{tabular}.
\end{center}
These rows have output entries ($\expsub{\mathcal{T}}{1}$, $\expsub{\mathcal{T}}{2}$ and $\expsub{\mathcal{T}}{3}$) that correspond to program segments that satisfy the conditions of the specification when the corresponding assertions are false.  These entries will be derived in other sections of the derivation.  Snark will introduce conditional expressions when each of these assumptions is used, but we ignore them for the moment. 

We also can assert the definition of the \hyperlink{def:miss}{$\misses$ relation}, 

\begin{center}
  \begin{tabular}{|m{0.40\textwidth}|m{0.20\textwidth}||m{0.20\textwidth}|}
\hline
\begin{center}
{$\begin{aligned}
\boxed{\misses(\Theta, \var{E})} \uiff  \var{E} \apply \Theta = \var{E}
\end{aligned}$} \hspace{1cm} 
\end{center}& &  \\  \hline
\end{tabular}.
\end{center}

Applying the \emph{equivalence replacement rule} to this and two \hyperlink{ass:misses}{\emph{misses} case assumptions}, that the environment misses the two expression arguments (taking $\Theta$ to be  $\thet{0}$ and $\var{E}$ to be $\e{1}$ and $\e{2}$ respectively), we have the two \hypertarget{ass:miss-para}{\emph{paraphrased misses}} case assumptions, 
\begin{center}
  \begin{tabular}{|m{0.40\textwidth}|m{0.20\textwidth}||m{0.20\textwidth}|}
\hline
\begin{center}
{$\e{1} \apply \thet{0} = \e{1} $}\hspace{1cm} 
 \hspace{1cm} 
\end{center}& &  \\  \hline
\end{tabular}
\end{center}
and
\begin{center}
  \begin{tabular}{|m{0.40\textwidth}|m{0.20\textwidth}||m{0.20\textwidth}|}
\hline
\begin{center}
{$\e{2} \apply \thet{0} = \e{2} $}\hspace{1cm} 
 \hspace{1cm} 
\end{center}& &  \\  \hline
\end{tabular}.
\end{center}
For simplicity we have omitted the output entries ($\expsub{\mathcal{T}}{2}$ and $\expsub{\mathcal{T}}{3}$) that accompany these assertions.

This concludes our handling of the case assumptions; we now look at how the unifier is found.
\subsubsection{Finding the Unifier.} Earlier we developed the \hyperlink{goal-exp-init}{\emph{expanded initial goal}}
\begin{center}
\begin{tabular}{|m{0.20\textwidth}|m{0.40\textwidth}||m{0.20\textwidth}|}
 \hline 
  & 
  \begin{center}
${\begin{aligned} 
&\begin{aligned}
&\boxed{\e{1} \apply \Theta} = \boxed{\e{2} \apply \Theta} \, \uand 
 \\
 \,\, & \thet{0} \moregen \Theta \, \uand
\\
  \,\, & {\mgi}(\thet{0}, \e{1}, \e{2}, \Theta)\, \uand
\\
  \,\, & {\reduce}(\thet{0},\vars(\e{1}, \e{2})\!\apply\thet{0}, \Theta)
\end{aligned}.\end{aligned}}$\hspace{1cm} 
\end{center} & \begin{center}$\Theta$ \end{center} \\
\hline
\end{tabular}.
\end{center}
We first apply the \emph{equality replacement rule} twice in succession to the \hyperlink{prop:subst-comp}{\emph{composition property}} of substitutions,
\begin{center}
  \begin{tabular}{|m{0.40\textwidth}|m{0.20\textwidth}||m{0.20\textwidth}|}
\hline
\begin{center}
{$\begin{aligned}
\boxed{\var{E} \apply (\Thet{1}\! \compose \Thet{2})} = (\var{E} \apply \Thet{1}) \apply \Thet{2}
\end{aligned}$} \hspace{1cm} 
\end{center}& &  \\  \hline
\end{tabular},
\end{center}
and this goal, taking  $\var{E}$ to be $\e{1}$ and $e_2$, respectively, and $\Theta$ to be $\Thet{1}\! \compose \Thet{2}$, to obtain
\begin{center}
\begin{tabular}{|m{0.16\textwidth}|m{0.44\textwidth}||m{0.20\textwidth}|}
 \hline 
  &\begin{center} 
$
\begin{aligned}
&\begin{aligned}&\boxed{(\e{1} \apply \Thet{1})} \apply \Thet{2} = 
                \boxed{(\e{2} \apply \Thet{1})} \apply \Thet{2}  \end{aligned}   \\
  \,\, & \uand    \\          
 \,\, & \thet{0} \moregen \Thet{1}\! \compose \Thet{2} \, \uand
\\
  \,\, & {\mgi}(\thet{0}, \e{1}, \e{2}, \Thet{1}\! \compose \Thet{2}) \uand
\\
  \,\, & {\reduce}(\thet{0},\vars(\e{1}, \e{2})\!\apply\thet{0}, \Thet{1}\! \compose \Thet{2})
\end{aligned}$
\end{center}
& 
\begin{center}$\Thet{1}\! \compose \Thet{2}$ \end{center}\\
\hline
\end{tabular}.
\end{center}                                
The composition of two (yet to be determined) substitutions has been introduced into the output entry. By two applications of the \emph{equality replacement rule}  to this goal and our \hyperlink{ass:miss-para}{\emph{paraphrased misses} case assumptions}, that $\thet{0}$ misses $\e{1}$ and $\e{2}$, that is
\begin{center}
  \begin{tabular}{|m{0.40\textwidth}|m{0.20\textwidth}||m{0.20\textwidth}|}
\hline
\begin{center}
{$\boxed{\e{1} \apply \thet{0}}= \e{1} $}\hspace{1cm} 
 \hspace{1cm} 
\end{center}& &  \\  \hline
\end{tabular}
\end{center}
and
\begin{center}
  \begin{tabular}{|m{0.40\textwidth}|m{0.20\textwidth}||m{0.20\textwidth}|}
\hline
\begin{center}
{$\boxed{\e{2} \apply \thet{0}}= \e{2} $}\hspace{1cm} 
 \hspace{1cm} 
\end{center}& &  \\  \hline
\end{tabular},
\end{center}
respectively, taking $\Thet{1}$ to be $\thet{0}$, we obtain

\begin{center}
\begin{tabular}{|m{0.16\textwidth}|m{0.44\textwidth}||m{0.20\textwidth}|}
 \hline 
  &\begin{center} 
$
\begin{aligned}
&\begin{aligned}&\boxed{\e{1} \apply \Thet{2}} = \e{2} \apply \Thet{2} \uand \end{aligned}   \\
 \,\, & \thet{0} \moregen (\thet{0}\! \compose \Thet{2}) \, \uand
\\
  \,\, & {\mgi}(\thet{0}, \e{1}, \e{2}, (\thet{0}\! \compose \Thet{2})) \uand
\\
  \,\, & {\reduce}(\thet{0},\vars(\e{1}, \e{2})\!\apply\thet{0}, \thet{0}\! \compose \Thet{2})
\end{aligned}.$
\end{center}
& 
\begin{center}$\thet{0}\! \compose \Thet{2}$ \end{center}\\
\hline
\end{tabular}.
\end{center}
 Here, the first of the two substitutions in the composition, $\Thet{1}$, has been found to be $\thet{0}$.

We  defined the replacement substitution $\{\E{1} \mapsto \E{2}\}$ to satisfy the    \hyperlink{prop:repl-ident}{\emph{replacement identity property}}  
\begin{center}
  \begin{tabular}{|m{0.40\textwidth}|m{0.22\textwidth}||m{0.18\textwidth}|}
\hline
\begin{center}
$
\begin{aligned}
\hspace{-.5cm}
{\begin{aligned}
&\boxed{\E{1} \apply \{\E{1} \mapsto \E{2} \}} = \E{2} \quad \uimpliedby \\ &\boxed{\isvar(\E{1}) }
\end{aligned}}
\end{aligned}
$
\end{center}& &  \\  \hline
\end{tabular}.
\end{center}
By the \emph{equality replacement rule}, taking $\E{1}$ to be $\e{1}$ and $\Theta_2$ to be $\{e_1 \mapsto E_2\}$, followed by resolution with the \hyperlink{ass:is-var}{\emph{is-var}} case assumption
\begin{center}
  \begin{tabular}{|m{0.40\textwidth}|m{0.22\textwidth}||m{0.18\textwidth}|}
\hline
\begin{center}
$
\boxed{\isvar(\e{1})} 
$
\end{center}& &  \\  \hline
\end{tabular},
\end{center}
we obtain
\begin{center}
\begin{tabular}{|m{0.10\textwidth}|m{0.52\textwidth}||m{0.18\textwidth}|}
 \hline 
  &\begin{center} 
$
\begin{aligned}
&\begin{aligned}&\boxed{\E{2} = \e{2} \apply \{\e{1} \mapsto  \E{2}\}} \uand \end{aligned}   \\
 \,\, & \thet{0} \moregen (\thet{0} \compose \{\e{1} \mapsto  \E{2}\} \, \uand
\\
  \,\, & {\mgi}(\thet{0}, \e{1}, \e{2}, (\thet{0} \compose \{\e{1} \mapsto  \E{2}\}))\uand
\\
  \,\, & {\reduce}(\thet{0},\vars(\e{1}, \e{2})\!\apply\thet{0}, \thet{0}\! \compose \{\e{1} \mapsto  \E{2}\})
\end{aligned}$
\end{center}
& 
\begin{center}$\thet{0} \compose \{\e{1} \mapsto  \E{2}\}$ \end{center}\\
\hline
\end{tabular}.
\end{center}
Here, a replacement has been introduced into the output entry.  
The replaced variable is the first input expression $\e{1}$; the introduced expression $\E{2}$ has yet to be determined.

Earlier, we established a \hyperlink{prop:repl-nonocc}{\emph{replacement nonoccur property}}
\begin{center}
  \begin{tabular}{|m{0.40\textwidth}|m{0.22\textwidth}||m{0.18\textwidth}|}
\hline
\begin{center}
$
\begin{aligned}
&\boxed{\var{D} \apply \{\E{1} \mapsto \E{2} \} = \var{D}} \quad \uimpliedby \\
&\left[\begin{aligned}  
&\!\isvar(\E{1}) \uand \\
&\!\unot(\E{1} \occurseq \var{D})
\end{aligned}\right]
\end{aligned}
$
\end{center}& &  \\  \hline
\end{tabular},
\end{center}
which means that a replacement has no effect on an expression if the replaced variable does not occur in the expression.  We apply the \emph{resolution rule}---Snark can reverse the arguments of the equality symbol---taking $\var{D}$ to be $\e{2}$, $\E{1}$ to be $\e{1}$, and $E_2$ to be $e_2$, and again resolving with the above \hyperlink{ass:is-var}{\emph{is-var}} case assumption $\isvar(\e{1})$ and the \hyperlink{ass:non-sub}{\emph{non-sub}} case assumption $\unot(\e{1} \occurseq \e{2})$,  to obtain
\begin{center}
\begin{tabular}{|m{0.10\textwidth}|m{0.52\textwidth}||m{0.18\textwidth}|}
 \hline 
  &\begin{center} 
$
\begin{aligned}
&     \boxed{\thet{0} \moregen (\thet{0} \compose \{\e{1} \mapsto  \e{2}\})} \, \uand
\\
  \,\, & {\mgi}(\thet{0}, \e{1}, \e{2}, (\thet{0} \compose \{\e{1} \mapsto  \e{2}\}))\uand
\\
  \,\, & {\reduce}(\thet{0},\vars(\e{1}, \e{2})\!\apply\thet{0}, \thet{0}\! \compose \{\e{1} \mapsto  \e{2}\}) \\
  \end{aligned}$
\end{center}
& 
\begin{center}$\thet{0} \compose \{\e{1} \mapsto  \e{2}\}$ \end{center}\\
\hline
\end{tabular}.
\end{center}
Here, the introduced expression $E_2$ in the output entry has been found to be the second input expression, 
$e_2$.

\subsubsection{Establishing the More-Generality Property.}  The \hyperlink{prop:more-gen-comp}{\emph{composition property}} of the relation $\moregen$ told us that
\begin{center}
  \begin{tabular}{|m{0.40\textwidth}|m{0.22\textwidth}||m{0.18\textwidth}|}
\hline
\begin{center}
{$
 {\begin{aligned}  
 &\Thet{0} \moregen \Thet{1} \, \uimplies \\
 &\boxed{\Thet{0} \moregen \Thet{1}\! \compose \Thet{2}}
 \end{aligned}}
$}
\end{center}& &  \\  \hline
\end{tabular}.
\end{center}
By applying the resolution rule, taking $\Thet{0}$ and $\Thet{1}$ to be $\thet{0}$, and $\Thet{2}$ to be $\{\e{1} \mapsto  \e{2}\},$ we obtain
\begin{center}
\begin{tabular}{|m{0.10\textwidth}|m{0.52\textwidth}||m{0.18\textwidth}|}
 \hline 
  &\begin{center} 
$\begin{aligned}
  \,\, & {\mgi}(\thet{0}, \e{1}, \e{2}, (\thet{0} \compose \{\e{1} \mapsto  \e{2}\})) \uand \\
  &\boxed{\thet{0} \moregen \thet{0}}\uand
\\
  \,\, & {\reduce}(\thet{0},\vars(\e{1}, \e{2})\!\apply\thet{0}, \thet{0}\! \compose \{\e{1} \mapsto  \e{2}\})
  \end{aligned}$
\end{center}
& 
\begin{center}$\thet{0} \compose \{\e{1} \mapsto  \e{2}\}$ \end{center}\\
\hline
\end{tabular}.
\end{center}
The boxed subsentence says that $\thet{0}$ is idempotent---exactly our   \hyperlink{ass:idem}{\emph{paraphrasedidempotence assumption}};  it is whisked away by the resolution rule, leaving us with the goal
\begin{center}
\begin{tabular}{|m{0.10\textwidth}|m{0.52\textwidth}||m{0.18\textwidth}|}
 \hline 
  &\begin{center} 
$\begin{aligned}
  \,\, & \boxed{{\mgi}(\thet{0}, \e{1}, \e{2}, (\thet{0} \compose \{\e{1} \mapsto  \e{2}\}))} \uand
\\
  \,\, & {\reduce}(\thet{0},\vars(\e{1}, \e{2})\!\apply\thet{0}, \thet{0}\! \compose \{\e{1} \mapsto  \e{2}\})
  \end{aligned}$
\end{center}
& 
\begin{center}$\thet{0} \compose \{\e{1} \mapsto  \e{2}\}$ \end{center}\\
\hline
\end{tabular}.
\end{center}

\subsubsection{Establishing Most-General Idempotence and Reduction.} We have shown the \emph{mgi replacement property}, that the replacement substitution is  most-general idempotent; that is,
\begin{center}
  \begin{tabular}{|m{0.46\textwidth}|m{0.16\textwidth}||m{0.18\textwidth}|}
\hline
\begin{center}$
\begin{aligned}
    &\boxed{\mgi(\Thet{0}, \E{1}, \E{2}, (\Thet{0} \compose \{\E{1} \mapsto \E{2}\})} \, \uimpliedby \\
    &\isvar(\E{1})
\end{aligned}
$\end{center}
& &  \\  \hline
\end{tabular}.
\end{center}
Applying the resolution rule to this assertion and the most recent goal, 
taking $\Thet{0}$ to be $\thet{0}$ and $\E{1}$ and $\E{2}$ to be $\e{1}$ and $\e{2}$, respectively, followed by resolution with the \hyperlink{ass:is-var}{\emph{is-var}} case assumption $\isvar(\e{1})$, we obtain the goal
\begin{center}
\begin{tabular}{|m{0.10\textwidth}|m{0.52\textwidth}||m{0.18\textwidth}|}
 \hline 
  &\begin{center} 
$\begin{aligned}
  {\reduce}(\thet{0},\boxed{{\vars(\e{1}, \e{2})}\!\apply\thet{0}}, \thet{0}\! \compose \{\e{1} \mapsto  \e{2}\})
  \end{aligned}$
\end{center}
& 
\begin{center}$\thet{0} \compose \{\e{1} \mapsto  \e{2}\}$ \end{center}\\
\hline
\end{tabular}.
\end{center}
Because, by the \hyperlink{ass:miss}{\emph{misses}} case assumptions, $\theta_0$ misses $e_1$ and $e_2$, and  by the \hyperlink{prop-vars-union}{definition of the multi-argument $vars$ notation}, we may reduce the above goal to
\begin{center}
\begin{tabular}{|m{0.10\textwidth}|m{0.52\textwidth}||m{0.18\textwidth}|}
 \hline 
  &\begin{center}$
  {\reduce}\begin{aligned}[t]
(&\thet{0},\vars(e_1) \cup\vars(e_2),\\ 
   &\thet{0}\! \compose \{\e{1} \mapsto  \e{2}\}))
\end{aligned}
  $
   \end{center}
& 
\begin{center}$\thet{0} \compose \{\e{1} \mapsto  \e{2}\}$ \end{center}\\
\hline
\end{tabular}.
\end{center}
Then, because (by the \hyperlink{ass:is-var}{\emph{is-var}} case assumption) $e_1$ is a variable (i.e., $\isvar(e_1)$ and hence $\vars(e_1) = \{e_1\}$, we have
\begin{center}
\begin{tabular}{|m{0.10\textwidth}|m{0.52\textwidth}||m{0.18\textwidth}|}
 \hline 
  &\begin{center} 
$\begin{aligned}
  \boxed{{\reduce}(\thet{0},\{e_1\} \cup\vars(e_2),\, \thet{0}\! \compose \{\e{1} \mapsto  \e{2}\})}
  \end{aligned}$
\end{center}
& 
\begin{center}$\thet{0} \compose \{\e{1} \mapsto  \e{2}\}$ \end{center}\\
\hline
\end{tabular}.
\end{center}
Expanding the goal by the \emph{equivalence replacement rule} and the \hyperlink{prop:red-def}{\emph{definition of reduction}},
\begin{center}
  \begin{tabular}{|m{0.46\textwidth}|m{0.16\textwidth}||m{0.18\textwidth}|}
\hline
\begin{center}$\begin{aligned}
 &\boxed{{\reduce}(\Thet{0}, V, \Theta)} \,\, \uiff  \\
 &\range(\Theta) \subseteq \range(\Theta_0) \cup  V
 \end{aligned}$\end{center}
& &  \\  \hline
\end{tabular},
\end{center}
taking $\Thet{0}$ to be $\thet{0}$, $V$ to be $\{e_1\} \cup\vars(e_2)$, and $\Theta$ to be $\thet{0}\! \compose \{\e{1} \mapsto  \e{2}\}$, we obtain
\begin{center}
\begin{tabular}{|m{0.10\textwidth}|m{0.52\textwidth}||m{0.18\textwidth}|}
 \hline 
  &\begin{center}$\begin{aligned}
  &\boxed{\range(\thet{0}\! \compose \{\e{1} \mapsto  \e{2}\}) }\subseteq \\&\range(\thet{0}) \cup  \{e_1\} \cup\vars(e_2)
   \end{aligned}$\end{center}
& 
\begin{center}$\thet{0} \compose \{\e{1} \mapsto  \e{2}\}$ \end{center}\\
\hline
\end{tabular}.
\end{center}
By the transitivity of set inclusion and the \hyperlink{prop-range-comp}{\emph{range property} of composition},
\begin{center}
  \begin{tabular}{|m{0.46\textwidth}|m{0.16\textwidth}||m{0.18\textwidth}|}
\hline
\begin{center}$\begin{aligned}
&\boxed{\range(\Thet{1}\! \compose \Thet{2})}  \subseteq \\
&\range(\Thet{1}) \cup \range(\Thet{2})    
\end{aligned}$\end{center}
& &  \\  \hline
\end{tabular},
\end{center}
taking $\Theta_1$ to be $\theta_0$ and $\Thet{2}$ to be $\{\e{1} \mapsto  \e{2}\}$, we obtain the goal
\begin{center}
\begin{tabular}{|m{0.10\textwidth}|m{0.52\textwidth}||m{0.18\textwidth}|}
 \hline 
  &\begin{center}$\begin{aligned}
  &\range(\thet{0}) \cup  \range(\{\e{1} \mapsto  \e{2}\})) \subseteq \\&\range(\thet{0}) \cup  \{e_1\} \cup\vars(e_2)
   \end{aligned}$\end{center}
& 
\begin{center}$\thet{0} \compose \{\e{1} \mapsto  \e{2}\}$ \end{center}\\
\hline
\end{tabular}.
\end{center}
But by the \emph{equality replacement rule} applied to the \hyperlink{prop-repl-range}{\emph{range property}} of replacement,
\begin{center}
  \begin{tabular}{|m{0.46\textwidth}|m{0.16\textwidth}||m{0.18\textwidth}|}
\hline
\begin{center}$\begin{aligned}
 &\boxed{\isvar(X)}\,\, \uimplies  \\
 &\,\range(\{X \mapsto E\}) =\vars(E)
 \end{aligned}$\end{center}
& &  \\  \hline
\end{tabular},
\end{center}
taking $X$ to be $e_1$ and $E$ to be $e_2$, 
and our \hyperlink{ass:is-var}{\emph{is-var}} case assumption,  that $e_1$  is a variable (that is, $\boxed{\isvar(e_1)}$), the goal becomes
\begin{center}
\begin{tabular}{|m{0.10\textwidth}|m{0.52\textwidth}||m{0.18\textwidth}|}
 \hline 
  &\begin{center}$\begin{aligned}
  &
{\range}(\thet{0}) \cup\vars(\e{2}) \subseteq \\&
{\range}(\thet{0}) \cup  \{e_1\} \cup \vars(e_2)
   \end{aligned}$\end{center}
& 
\begin{center}$\thet{0} \compose \{\e{1} \mapsto  \e{2}\}$ \end{center}\\
\hline
\end{tabular},
\end{center}
which reduces, via set-theoretic reasoning, to
\begin{center}
\begin{tabular}{|m{0.10\textwidth}|m{0.52\textwidth}||m{0.18\textwidth}|}
 \hline 
  &\begin{center}$true$\end{center}
& 
\begin{center}$\thet{0} \compose \{\e{1} \mapsto  \e{2}\}$ \end{center}\\
\hline
\end{tabular}.
\end{center}
As we mentioned, for simplicity we dropped the output entries corresponding to the case assumptions. In the context of the full derivation, the above output entry is just one branch in a larger conditional term.

\subsection*{Introduction of the Recursive Calls}

We have said that recursive calls are introduced by application of the well-founded induction principle. The induction hypothesis is added to the initial tableau as an assertion at the beginning of the derivation.  For the derivation of the unification algorithm, we began with the initial tableau

\begin{center}  
\begin{tabular}{|m{0.20\textwidth}|m{0.40\textwidth}||m{0.20\textwidth}|}
 \hline
 \begin{center} Assertions\end{center} & \begin{center} Goals \end{center}& \begin{center} $\unify(\thet{0}, \e{1}, \e{2})$ \end{center} \\
 \hline\hline  
  & 
  \begin{center}
      $\begin{aligned}[t] &
  \idem(\thet{0}) \uimplies \\  &\,{\mgiu}(\thet{0}, \e{1}, \e{2}, \Theta) \\ \end{aligned}$  
  \end{center}
  & \vspace{5pt}{\begin{center}$\Theta$\end{center}} \\

\hline
\end{tabular}.
\end{center}

   We form the three inputs into a triple (3-tuple), which we treat as a single entity.
  We conduct the proof under the induction hypothesis that recursive calls $\unify(\Thet{0}', \E{1}',\E{2}')$ to the program $\unify$ we are in the process of constructing will satisfy the input-output condition  \begin{align*}\begin{aligned}[t] &\idem(\Thet{0}') \uimplies \\ &\,{\mgiu}(\Thet{0}', \E{1}', \E{2}' , \unify(\Thet{0}', \E{1}', E{2}'))\end{aligned}\end{align*}
for any input triple $\langle \Thet{0}', \E{1}', \E{2}' \rangle$ such that $\langle\Thet{0}', \E{1}', \E{2}'\rangle \, \expsub{\prec}{U} \, \langle\thet{0}, \e{1}, \e{2}\rangle.$ (The symbols $\Thet{0}'$, $\E{1}'$, and $\E{2}'$ are variables.)  Here, $\expsub{\prec}{U}$ is the well-founded relation, previously unspecified, for the unification algorithm.   We add this induction hypothesis to the initial tableau as an assertion:

\noindent

  \begin{center}
  \begin{tabular}{|m{0.40\textwidth}|m{0.20\textwidth}||m{0.20\textwidth}|}
\hline
\begin{center}
{$\begin{aligned}
&\langle\Thet{0}', \E{1}', \E{2}'\rangle \, \expsub{\prec}{U} \, \langle\thet{0}, \e{1}, \e{2}\rangle \uimplies \\
&\left[\begin{aligned}&\idem(\Thet{0}') \uimplies \\
 &\begin{aligned}
    {\,\mgiu}(&\Thet{0}', \E{1}', \E{2}', \\
    &\unify(\Thet{0}', \E{1}', \E{2}')) 
\end{aligned}
\end{aligned}\right]\end{aligned}$} \hspace{1cm} 
\end{center}& &  \\  \hline
\end{tabular}.
\end{center}

\subsubsection{Introduction of the Well-Founded Relation.}\hypertarget{par:wfrelunif} In a \emph{Deus ex Machina} step, we chose the well-founded relation $\expsub{\prec}{\var{U}}$, called the \emph{unification \textup{(}\!well-founded\textup{)} relation}, to be a lexicographic combination of two relations, the \hypertarget{def:range-vars}{\emph{range-vars} relation}
$\expsub{\prec}{\range-\vars}$, defined on substitutions and expressions by

  \begin{center}
  \begin{tabular}{|m{0.60\textwidth}|m{0.10\textwidth}||m{0.10\textwidth}|}
\hline
\begin{center}
{$\begin{aligned}
    &\langle \Thet{0}', \E{1}', \E{2}' \rangle 
\,\expsub{\prec}{\range-\vars}
    \langle \Thet{0}, \E{1}, \E{2} \rangle \uiff \\
    &\left[\begin{aligned}
     &\range(\Thet{0}') \cup \vars(\langle \E{1}', \E{2}'\rangle) \subset \\
     &\range(\Thet{0}) \cup \vars(\langle \E{1}, \E{2}\rangle)
     \end{aligned}\right]
\end{aligned}$} \hspace{1cm} 
\end{center}& &  \\  \hline
\end{tabular},
\end{center}
and the \hypertarget{def:size-first}{\emph{size-first} relation} $\expsub{\prec}{\size1}$, defined by
  \begin{center}
  \begin{tabular}{|m{0.60\textwidth}|m{0.10\textwidth}||m{0.10\textwidth}|}
\hline
\begin{center}
{$\begin{aligned}
    &\langle \Thet{0}', \E{1}', \E{2}' \rangle 
\,\expsub{\prec}{\size1}
    \langle \Thet{0}, \E{1}, \E{2} \rangle \uiff \\
    &\size(\E{1}') < \size(\E{1})
\end{aligned}$} \hspace{1cm} 
\end{center}& &  \\  \hline
\end{tabular}.
\end{center}
We take the \emph{unification well-founded relation} $\expsub{\prec}{U}$ to be the lexicographic combination 
$\expsub{\prec}{\lex(\range-\vars, \size1)}$ of these relations; we shall call this the \hypertarget{def:u-rel}{\emph{u-relation}}. It satisfies the
\hyperlink{prop:lex-refl}{\emph{reflexive lexicographic property}},  
 \begin{center}
    $
  \begin{aligned}
    & \begin{aligned}
 &\langle \Thet{0}', \E{1}', \E{2}' \rangle 
 \, \expsub{\prec}{\var{U}} \,
 \langle \Thet{0}, \E{1}, \E{2} \rangle \\
 \end{aligned}  \uiff \\
    &\left[\begin{aligned}
 &\left\{\begin{aligned}
&\range(\Thet{0}')\cup\vars(\langle  \E{1}', \E{2}' \rangle) \subset \\
  &\range(\Thet{0})\cup\vars(\langle \E{1}, \E{2} \rangle) 
  \end{aligned}\right\} \uor \\ 
&\left[\begin{aligned}
 &\left\{\begin{aligned}
 &\range(\Thet{0}')\cup\vars(\langle  \E{1}', \E{2}' \rangle) \subseteq \\
  &\range(\Thet{0})\cup\vars(\langle \E{1}, \E{2} \rangle) 
  \end{aligned}\right\} \uand  \\
  &\size(\E{1}') < \size(\E{1})
  \end{aligned} \right] 
  \end{aligned} \right]\!.
  \end{aligned}
    $
 \end{center}
By propositional reasoning, using the implication
\[
(
((\mathcal{P} \uor \mathcal{Q}) \uimplies \mathcal{R})
\uimplies
(
(\mathcal{P} \uimplies \mathcal{R})
\uand\,
(\mathcal{Q} \uimplies \mathcal{R})
)
),
\]
taking $\mathcal{P}$ to be \[\left\{\begin{aligned}
&\range(\Thet{0}')\cup\vars(\langle  \E{1}', \E{2}' \rangle) \subset \\
  &\range(\Thet{0})\cup\vars(\langle \E{1}, \E{2} \rangle) 
  \end{aligned}\right\},  \]
  $\mathcal{Q}$ to be
  \[\left[\begin{aligned}
 &\left\{\begin{aligned}
 &\range(\Thet{0}')\cup\vars(\langle  \E{1}', \E{2}' \rangle) \subseteq \\
  &\range(\Thet{0})\cup\vars(\langle \E{1}, \E{2} \rangle) 
  \end{aligned}\right\} \uand  \\
  &\size(\E{1}') < \size(\E{1})
  \end{aligned} \right], \]
and  $\mathcal{R}$ to be
\[\begin{aligned}
 &\langle \Thet{0}', \E{1}', \E{2}' \rangle 
 \, \expsub{\prec}{\var{U}} \,
 \langle \Thet{0}, \E{1}, \E{2} \rangle \\
 \end{aligned},\]
we have that either of the two disjuncts, $\mathcal{P}$ or $\mathcal{Q}$, on the right side implies that the \hyperlink{def:u-rel}{\emph{u-relation}}, $\mathcal{R}$,  holds;  so we have the \hypertarget{prop:range-vars}{\emph{range-vars property} of the u-relation},
  \begin{center}
  \begin{tabular}{|m{0.60\textwidth}|m{0.10\textwidth}||m{0.10\textwidth}|}
\hline
 \begin{center}
    $
  \begin{aligned}
   &\left\{\begin{aligned}
&\range(\Thet{0}')\cup\vars(\langle  \E{1}', \E{2}' \rangle) \subset \\
  &\range(\Thet{0})\cup\vars(\langle \E{1}, \E{2} \rangle) 
  \end{aligned}\right\}     
    & \begin{aligned}
 &\\
 \end{aligned}\, \\
&\uimplies  \boxed{\langle \Thet{0}', \E{1}', \E{2}' \rangle 
\,  \expsub{\prec}{\var{U}} \,
 \langle \Thet{0}, \E{1}, \E{2} \rangle}                     
  \end{aligned} 
    $
 \end{center}
  & &  \\  \hline
\end{tabular},
\end{center}
and the \hypertarget{prop:size-first}{\emph{size-first property}} of the u-relation, 
  \begin{center}
  \begin{tabular}{|m{0.60\textwidth}|m{0.10\textwidth}||m{0.10\textwidth}|}
\hline
 \begin{center}
    $
    \begin{aligned}
      &\left[\begin{aligned}
&\left[\begin{aligned}
 &\left\{\begin{aligned}
 &\range(\Thet{0}')\cup\vars(\langle  \E{1}', \E{2}' \rangle) \subseteq \\
  &\range(\Thet{0})\cup\vars(\langle \E{1}, \E{2} \rangle) 
  \end{aligned}\right\} \uand  \\
  &\size(\E{1}') < \size(\E{1})
  \end{aligned} \right] 
  \end{aligned} \right]
      \\
 & \uimplies \,  \begin{aligned}
 &\boxed{\langle \Thet{0}', \E{1}', \E{2}' \rangle 
 \, \expsub{\prec}{\var{U}}\, 
 \langle \Thet{0}, \E{1}, \E{2} \rangle}
 \end{aligned} 
  \end{aligned} 
    $
 \end{center}
  & &  \\  \hline
\end{tabular}.
\end{center}

By the \emph{resolution rule}, applied to the \hyperlink{prop:range-vars}{\emph{range-vars property} of the u-relation} and the induction hypothesis

  \begin{center}
  \begin{tabular}{|m{0.60\textwidth}|m{0.10\textwidth}||m{0.10\textwidth}|}
\hline
\begin{center}
{$\begin{aligned}
&\boxed{\langle\Thet{0}', \E{1}', \E{2}'\rangle \, \expsub{\prec}{U} \, \langle\thet{0}, \e{1}, \e{2}\rangle} \uimplies \\
&\left[\begin{aligned}&\idem(\Thet{0}') \uimplies \\
 &\begin{aligned}
    {\,\mgiu}(&\Thet{0}', \E{1}', \E{2}', \\
    &\unify(\Thet{0}', \E{1}', \E{2}')) 
\end{aligned}
\end{aligned}\right]\end{aligned}$} \hspace{1cm} 
\end{center}& &  \\  \hline
\end{tabular}
\end{center}
(taking $\Thet{0}$ to be $\thet{0}$ and $\E{1}$ and $\E{2}$ to be $\e{1}$ and $\e{2}$, respectively),  we obtain the assertion
 \begin{center}
  \begin{tabular}{|m{0.60\textwidth}|m{0.10\textwidth}||m{0.10\textwidth}|}
\hline
\begin{center}
{$\begin{aligned}
 &\begin{aligned}
& \left\{\begin{aligned}
&\range(\Thet{0}')\cup\vars(\langle  \E{1}', \E{2}' \rangle) \subset \\
  &\range(\Thet{0})\cup\vars(\langle \e{1}, \e{2} \rangle) 
  \end{aligned}\right\} 
\end{aligned} \\
\uimplies  &\left[\begin{aligned}&\idem(\Thet{0}') \uimplies \\
  &\begin{aligned}
    {\,\mgiu}(&\Thet{0}', \E{1}', \E{2}', \\
    &\unify(\Thet{0}', \E{1}', \E{2}')) 
\end{aligned}
\end{aligned}\right] \end{aligned}  $}  \hspace{1cm} 
\end{center}& &  \\  \hline
\end{tabular},
\end{center}
to be referred to as the \hypertarget{prop:range-vars-hyp}{\emph{range-vars induction hypothesis}}. This says that the induction hypothesis holds if the set of variables in the environment substitution or in the input expressions is strictly reduced.

Similarly, by the \emph{resolution rule} applied to the \hyperlink{prop:size-first}{\emph{size-first property}} of the u-relation and the induction hypothesis (taking $\Thet{0}$ to be $\thet{0}$ and $\E{1}$ and $\E{2}$ to be $\e{1}$ and $\e{2}$, respectively), we obtain the new assertion

  \begin{center}
  \begin{tabular}{|m{0.60\textwidth}|m{0.10\textwidth}||m{0.10\textwidth}|}
\hline
\begin{center}
{$\begin{aligned}
 &\begin{aligned}
&\begin{aligned}
&\left[\begin{aligned}
 &\left\{\begin{aligned}
 &\range(\Thet{0}')\cup\vars(\langle \E{1}', \E{2}' \rangle) \subseteq \\
  &\range(\Thet{0})\cup\vars(\langle \e{1}, \e{2} \rangle) 
  \end{aligned}\right\} \uand  \\
  &\size(\E{1}') < \size(\e{1})
  \end{aligned} \right] 
  \end{aligned} 
\end{aligned} \\
 \uimplies &\left[\begin{aligned}&\idem(\Thet{0}') \uimplies \\
 &\begin{aligned}
    {\,\mgiu}(&\Thet{0}', \E{1}', \E{2}', \\
    &\unify(\Thet{0}', \E{1}', \E{2}')) 
\end{aligned}
\end{aligned}\right] \end{aligned}  $}  \hspace{1cm} 
\end{center}& &  \\  \hline
\end{tabular},
\end{center}
to be referred to as the \hypertarget{hyp:ind-size1}{\emph{size-first induction hypothesis}}. This says that the induction hypothesis holds if the size of the first expression argument is strictly reduced, and if the  set of variables in the environment substitution and in the input expressions is reduced or remains the same.

In the definition of  $\expsub{\prec}{\var{U}}$, we could use an
 analogous \emph{size-second} relation $\expsub{\prec}{\size2}$ instead of $\expsub{\prec}{\size1}$; if we choose to use that in the proof, we obtain a symmetric version of the same algorithm.

 In our implemented derivation, the definition of the u-relation is given to Snark as a lemmas.  In future work, we would like to have Snark discover the combination of well-founded relations itself.
The current proof may be regarded as a work in progress. 

 \subsubsection{Reversing the Expression Arguments.} In this section, we  use the \emph{size-first induction hypothesis} to introduce a recursive call.

We early on developed the \hyperlink{goal:init}{\emph{initial goal}}
\begin{center}
\begin{tabularx}{1.0\textwidth} { 
  | >{\raggedright\arraybackslash}X 
  | >{\raggedright\arraybackslash}X 
  || >{\raggedright\arraybackslash}X | } \hline 
  & \vspace{-.3cm}\begin{center} $\boxed{{\mgiu}(\thet{0}, \e{1}, \e{2} , \Theta)} $  \end{center}  & \vspace{-5pt}\begin{center}$\Theta$\end{center} \\
\hline
\end{tabularx}.
\end{center}

Applying the \emph{equivalence replacement rule} to the \hyperlink{prop:sym-mgiu}{\emph{symmetry property}} of the  most-general idempotent reducing unifier relation, expressed in the valid assertion
 \begin{center}
\begin{tabular}{|m{0.60\textwidth}|m{0.10\textwidth}||m{0.10\textwidth}|}
 \hline
  \[ \begin{aligned}
  &\boxed{{\mgiu}(\Thet{0}, \E{1}, \E{2}, \Theta)} \,\, \uiff \,\, \\
  &\,\,{\mgiu}(\Thet{0}, \E{2}, \E{1}, \Theta)
 \end{aligned} \]
  & 
  &  \\
\hline
\end{tabular}
\end{center}
 (taking $\Thet{0}$ to be $\thet{0}$ and $\E{1}$ and $\E{2}$ to be $\e{1}$ and $\e{2}$, respectively), gives us the new goal
\begin{center}
\begin{tabularx}{1.0\textwidth} { 
  | >{\raggedright\arraybackslash}X 
  | >{\raggedright\arraybackslash}X 
  || >{\raggedright\arraybackslash}X | } \hline 
  & \vspace{-.3cm}\begin{center} $\boxed{{\mgiu}(\thet{0}, \e{2}, \e{1}, \Theta)} $  \end{center}  & \vspace{-5pt}\begin{center}$\Theta$\end{center} \\
\hline
\end{tabularx},
\end{center}
in which the two expression arguments are reversed.

Applying resolution to the \hyperlink{hyp:ind-size1}{\emph{size-first induction hypothesis}},
  \begin{center}
  \begin{tabular}{|m{0.60\textwidth}|m{0.10\textwidth}||m{0.10\textwidth}|}
\hline
\begin{center}
{$\begin{aligned}
 &\begin{aligned}
&\begin{aligned}
&\left[\begin{aligned}
 &\left\{\begin{aligned}
 &\range(\Thet{0}') \cup \vars(\langle \E{1}', \E{2}' \rangle) \subseteq \\
  &\range(\Thet{0})\cup\vars(\langle \e{1}, \e{2}, \rangle) 
  \end{aligned}\right\} \uand  \\
  &\size(\E{1}') < \size(\e{1})
  \end{aligned} \right] 
  \end{aligned} 
\end{aligned} \\
 \uimplies &\left[\begin{aligned}&\idem(\Thet{0}') \uimplies \\
 &\boxed{\begin{aligned}
    \,{\mgiu}(&\Thet{0}', \E{1}', \E{2}', \\
    &\unify(\Thet{0}', \E{1}', \E{2}')) 
\end{aligned}}
\end{aligned}\right] \end{aligned}  $}  \hspace{1cm} 
\end{center}& &  \\  \hline
\end{tabular},
\end{center}
and the above goal, taking $\Thet{0}'$ to be $\thet{0}$, $\E{1}'$ and $\E{2}'$ to be $ \e{2} $ and $ \e{1}$, respectively, and $\Theta$ to be $\unify(\thet{0}, \e{2}, \e{1})$, we obtain the new goal 
\begin{center}
\begin{tabular}{|m{0.18\textwidth}|m{0.36\textwidth}||m{0.26\textwidth}|}
 \hline 
  & 
 \begin{center}
{$\begin{aligned}
 &\begin{aligned}
  &\,\range(\Thet{0}) \cup \vars(\langle \e{2}, \e{1}\rangle ) \\ 
  &\,  \subseteq \range(\Thet{0}) \cup \vars( \langle \e{1}, \e{2}\rangle )  \\
  & \uand \size(\e{2}) < \size(\e{1})
\end{aligned}  \\
& \uand \boxed{\idem(\thet{0})} 
 \end{aligned} $}
\end{center}
& 
\begin{center}$\unify(\thet{0}, \e{2}, \e{1})$ \end{center}\\
\hline
\end{tabular}.
\end{center}
Here, a recursive call in which the expression arguments are reversed has been introduced into the output column.

We simplify this by resolution with the \hyperlink{ass:init}{\emph{initial assertion}} that $\thet{0}$ is idempotent,

\begin{center}
\begin{tabular}{|m{0.18\textwidth}|m{0.36\textwidth}||m{0.26\textwidth}|}
 \hline 
 \begin{center}$\idem(\thet{0} )$\end{center}
  & 
& 
\\
\hline
\end{tabular},
\end{center} 
to obtain
\begin{center}
\begin{tabular}{|m{0.18\textwidth}|m{0.36\textwidth}||m{0.26\textwidth}|}
 \hline 
  & 
 \begin{center}
{$\begin{aligned}
 &\begin{aligned}
  &\,\range(\Thet{0}) \cup \vars(\langle \e{2}, \e{1}\rangle ) \\ 
  &\,  \subseteq \range(\Thet{0}) \cup \vars( \langle \e{1}, \e{2}\rangle )  \\
  & \uand \size(\e{2}) < \size(\e{1})
\end{aligned}  
 \end{aligned} $}
\end{center}
& 
\begin{center}$\unify(\thet{0}, \e{2}, \e{1})$ \end{center}\\
\hline
\end{tabular}.
\end{center} 
We can further simplify this, using the assertion that
 
\begin{center}
\begin{tabular}
{|m{0.60\textwidth}|m{0.10\textwidth}||m{0.10\textwidth}|}
 \hline
  \[ \begin{aligned}
  &\vars(\langle \E{1}, \E{2} \rangle) = 
  \vars(\langle \E{2}, \E{1} \rangle)
 \end{aligned} \]
  & 
  &  \\
\hline
\end{tabular}
\end{center}
\noindent and the reflexivity of the subset relation $\subseteq$, to obtain
\begin{center}
\begin{tabular}{|m{0.18\textwidth}|m{0.36\textwidth}||m{0.26\textwidth}|}
 \hline 
  & 
 \begin{center}
{$\begin{aligned}
 &\begin{aligned}
  & \size(\e{2}) < \size(\e{1})
\end{aligned}  \\
 \end{aligned} $}
\end{center}
& 
\begin{center}$\unify(\thet{0}, \e{2}, \e{1})$ \end{center}\\
\hline
\end{tabular}.
\end{center} 
In other words, in any case in which the second expression is smaller than the first, we can meet the specification for the unification algorithm simply by reversing the arguments.  In particular, we can reverse arguments if the second argument is a variable (of size 0) and the first a constant (of size 1), or if the second argument is an atom and the first is not.  When the second expression is not smaller, if we introduce this recursive call we are in danger of constructing a nonterminating algorithm.

\subsubsection{Introduction of the Instance Recursive Call.} In this section, we use the \emph{range-vars induction hypothesis} to introduce a recursive call.
Again, we start from the \hyperlink{goal:init}{\emph{initial goal}}
\begin{center}
\begin{tabular}
{|m{0.10\textwidth}|m{0.40\textwidth}||m{0.30\textwidth}|}

\hline 
  & \begin{center} $\boxed{{\mgiu}(\thet{0}, \e{1}, \e{2} , \Theta)} $  \end{center}  & \begin{center}$\Theta$\end{center} \\
\hline
\end{tabular}.
\end{center}
We have among our valid assertions the \hyperlink{prop:mgiu-instance}{\emph{instance property}} of most-general idempotent reducing unifiers,

\begin{center}
\begin{tabular}
{|m{0.56\textwidth}|m{0.12\textwidth}||m{0.12\textwidth}|}
 \hline
 \hspace{-20pt}
  {\[\begin{aligned} 
  &\boxed{{\mgiu}(\Thet{0}, \E{1}, \E{2}, \Theta)} \,\, \uiff \,\, \\
  &{\mgiu}(\Thet{0}, \E{1}\!\apply\Thet{0}, \E{2}\!\apply\Thet{0}, \Theta)
 \end{aligned} \]}\hspace{20pt}
  & 
  &  \\
\hline
\end{tabular}.
\end{center}
By the \emph{equivalence replacement rule}, taking $\Thet{0}$ to be $\thet{0}$ and $\E{1}$ and $\E{2}$ to be $\e{1}$ and $\e{2}$, respectively, we can obtain the new goal
\begin{center}
\begin{tabular}
{|m{0.10\textwidth}|m{0.40\textwidth}||m{0.30\textwidth}|}
\hline 
  & \begin{center}$ \boxed{{\mgiu}(\thet{0}, \e{1}\apply\thet{0}, \e{2}\apply\thet{0} , \Theta)}$  \end{center}  & \begin{center}$\Theta$\end{center} \\
\hline
\end{tabular}.
\end{center}
Applying resolution to the \hyperlink{prop:range-vars-hyp}{\emph{range-vars induction hypothesis}}, 
     \begin{center}
  \begin{tabular}{|m{0.56\textwidth}|m{0.12\textwidth}||m{0.12\textwidth}|}
\hline
\begin{center}
{$\begin{aligned}
 &\begin{aligned}
& \left\{\begin{aligned}
&\range(\Thet{0}')\cup\vars(\langle  \E{1}', \E{2}' \rangle) \subset \\
  &\range(\Thet{0})\cup\vars(\langle \E{1}, \E{2} \rangle) 
  \end{aligned}\right\} 
\end{aligned} \\
\uimplies  &\left[\begin{aligned}&\idem(\Thet{0}') \uimplies \\
  &\boxed{\begin{aligned}
    {\mgiu}(&\Thet{0}', \E{1}', \E{2}', \\
    &\unify(\Thet{0}', \E{1}', \E{2}'))
\end{aligned}}
\end{aligned}\right] \end{aligned}  $}  \hspace{1cm} 
\end{center}& &  \\  \hline
\end{tabular},
\end{center}
and the above goal, taking $\Thet{0}'$ to be $\thet{0}$, $\E{1}'$ and $\E{2}'$ to be $ \e{1}\apply\thet{0} $ and $ \e{2}\apply\thet{0}$, respectively, and $\Theta$ to be $\unify(\thet{0}, \e{1}\apply\thet{0}, \e{2}\apply\thet{0})$, we obtain the new goal 
\begin{center}
\begin{tabular}
{|m{0.07\textwidth}|m{0.43\textwidth}||m{0.30\textwidth}|}
 \hline 
  & 
 \begin{center}
{$\begin{aligned}
 &\begin{aligned}
  &\,\range(\Thet{0}) \cup \vars(
  \boxed{\langle \e{1}\apply\thet{0}, \e{2}\apply\thet{0}\rangle} 
  ) \\ 
  &\quad  \subset \range(\Thet{0}) \cup \vars( \langle \e{1}, \e{2}\rangle ) 
\end{aligned}  \\
& \uand \idem(\thet{0}) 
 \end{aligned} $}
\end{center}
& 
\begin{center}$\unify(\thet{0}, \e{1}\apply\,\thet{0}, \e{2}\apply\,\thet{0})$ \end{center}\\
\hline
\end{tabular}.
\end{center}
Here, a recursive call has been introduced into the output column.

We have remarked that proper substitutions \hyperlink{prop:distr-tup}{distribute} over tuples; that is, we have the valid assertion

  \begin{center}
  \begin{tabular}{|m{0.40\textwidth}|m{0.10\textwidth}||m{0.30\textwidth}|}
\hline
\begin{center}
{$\begin{aligned}
        & {\isprop}(\Theta) \uimplies \\
        &\langle \E{1}, \E{2} \rangle \apply \Theta =
        \boxed{\langle \E{1}\apply\Theta, \, \E{2}\apply\Theta \rangle}
    \end{aligned}$}
\end{center}  \hspace{1cm} 
&  & \\  \hline
\end{tabular}.
\end{center} 
By the \emph{equality replacement rule}, right to left, taking $\Theta$  to be $\thet{0}$ and $\E{1}$ and $\E{2}$ to be $\e{1}$ and $\e{2}$, respectively, 
we rewrite the above goal as
\begin{center}
\begin{tabular}{|m{0.10\textwidth}|m{0.40\textwidth}||m{0.30\textwidth}|}
 \hline
  & 
 \begin{center}
{$\begin{aligned}
 &\boxed{\begin{aligned}
  &\,\range(\Thet{0}) \cup \vars(\langle \e{1}, \e{2}\rangle \apply\thet{0} ) \\ 
  &\,  \subset \range(\Thet{0}) \cup \vars( \langle \e{1}, \e{2}\rangle ) 
\end{aligned}}  \\
& \uand \idem(\thet{0}) \\
& \uand \isprop(\thet{0})
 \end{aligned} $}
\end{center}
& 
\begin{center}$\unify(\thet{0}, \e{1}\apply\,\thet{0}, \e{2}\apply\,\thet{0})$ \end{center}\\
\hline
\end{tabular}.
\end{center}
By the \hyperlink{prop:vars-range-proper-subset}{\emph{vars-range proper-subset property}}, we have the following valid assertion:
 \begin{center}
  \begin{tabular}{|m{0.40\textwidth}|m{0.10\textwidth}||m{0.30\textwidth}|}
\hline
\begin{center}
$ \begin{aligned}
    &\,
    \boxed{\begin{aligned}
    &\range(\Theta) \cup \vars(\var{E} \apply \Theta) \\
    &\,\subset\, \range(\Theta) \cup \vars(\var{E}) \\    
    \end{aligned}} \,\uimpliedby \\
     &\idem(\Theta) \uand \unot\, (\misses(\Theta, \var{E}))
\end{aligned}$
\end{center} \hspace{1cm} 
&  & \\  \hline
\end{tabular}.
\end{center} 
By the resolution rule, taking $\Theta$ to be $\thet{0}$ and $\var{E}$ to be $\langle \e{1}, \e{2}\rangle$, we obtain the new goal

\begin{center}
\begin{tabular}{|m{0.10\textwidth}|m{0.40\textwidth}||m{0.30\textwidth}|}
 \hline 
  & 
 \begin{center}
{$
\begin{aligned}
 &\boxed{\idem(\thet{0})} \uand  \\
 &\boxed{\isprop(\thet{0})} \uand  \\
 &\unot\,(\misses(\thet{0},\langle \e{1}, \e{2}\rangle)
\end{aligned}
  $}
\end{center}
& 
\vspace{5pt}
\begin{center}$\unify(\thet{0}, \e{1}\apply\,\thet{0}, \e{2}\apply\,\thet{0})$ \end{center}\\
\hline
\end{tabular}.
\end{center}
In other words, when the environment substitution is proper and idempotent and does not miss both the input expressions, a recursive call in which the environment is applied to the inputs will satisfy the specification. 

If the environment $\thet{0}$ were not idempotent, the above recursive call could lead to an infinite computation. For instance, if the environment substitution merely permuted some variables, it would not be idempotent and no reduction in the well-founded relation would be achieved.  But we know by our \hyperlink{ass:init}{\emph{initial assertion}}
\begin{center}
\begin{tabular}{|m{0.18\textwidth}|m{0.32\textwidth}||m{0.30\textwidth}|}
 \hline 
\begin{center}\hspace{-.3cm}$\boxed{\idem(\thet{0})} $\end{center} &  & \\
\hline
\end{tabular}
\end{center}
that the environment substitution is idempotent, and we have been considering the \hyperlink{ass:prop}{\emph{is-prop}} cases, in which the environment is proper, that is, 

\begin{center}
\begin{tabular}{|m{0.18\textwidth}|m{0.32\textwidth}||m{0.30\textwidth}|}
 \hline 
\begin{center}\hspace{0cm}$\boxed{\isprop(\thet{0})} $\end{center} &  & \\
\hline
\end{tabular}.
\end{center}  The case in which $\thet{0}$ is not proper was discussed in the subsection \hyperlink{sec:fail-env}{\emph {Failure Environment}}.
Applying resolution twice in succession to the two assertions and the most recent goal, we obtain
\begin{center}
\begin{tabular}{|m{0.18\textwidth}|m{0.32\textwidth}||m{0.30\textwidth}|}
 \hline 
  & 
 \begin{center}
{$
\begin{aligned}
 &\unot\,(\misses(\thet{0},\langle \e{1}, \e{2}\rangle)
\end{aligned}
  $}
\end{center}
& 
\vspace{5pt}
\begin{center}$\unify(\thet{0}, \e{1}\apply\,\thet{0}, \e{2}\apply\,\thet{0})$ \end{center}\\
\hline
\end{tabular}. 
\end{center}
That is, the recursive call will satisfy the specification so long as the environment substitution does not miss  both input expressions.  

The cases in which $\thet{0}$ does miss both input expressions are handled by other branches of the search space. For instance, the \hyperlink{ass:miss}{\emph{misses}} case assumptions were introduced in the subsection \hyperlink{sec:replacement}{\emph{Introduction of the Replacement}}.

\subsubsection{Introduction of the Recursive Calls for Nonatomic Expressions.}

This is the most complex part of the derivation. \paragraph{Development of the Nonatomic Induction Hypotheses.}\hypertarget{par:nonatomicindhyp}{} The unification algorithms we obtain employs nested recursive calls when both expressions are nonatomic. This is an instance in which Snark discovered a program simpler than the one we  anticipated.

In this section, we assume that we have handled the cases in which $\e{1}$ or $\e{2}$ is atomic, or in which $\thet{0}$ does not miss $\e{1}$  or $\e{2}$, obtaining subprograms  $\mathcal{T_1}$, $\mathcal{T_2}$, and $\mathcal{T_3}$, respectively; that is, we include among our assertions the \hypertarget{ass:non-atom}{\emph{non-atom}} case assumptions, \begin{center}
  \begin{tabular}{|m{0.60\textwidth}|m{0.10\textwidth}||m{0.10\textwidth}|}
\hline
\begin{center}
$\boxed{{\unot(\isatm(\e{1}))}}$ 
\vspace{-15pt}
\end{center}  \hspace{1cm} 
&  & \begin{center}$\mathcal{T_1}$
\vspace{-5pt}\end{center}\\  \hline
\end{tabular}
\end{center} 
and
  \begin{center}
  \begin{tabular}{|m{0.60\textwidth}|m{0.10\textwidth}||m{0.10\textwidth}|}
\hline
\begin{center}
$\boxed{{\unot(\isatm(\e{2}))}} $  \vspace{-5pt}\hspace{1cm} 
\end{center}&  & \begin{center}$\mathcal{T_2}$ \vspace{-5pt}\end{center}\\  \hline
\end{tabular}
\end{center} 
and the \hyperlink{ass:miss}{\emph{misses}} case assumptions, that the environment substitution $\thet{0}$ misses $\e{1}$ and $\e{2}$ , i.e.,
\begin{center}
  \begin{tabular}{|m{0.40\textwidth}|m{0.20\textwidth}||m{0.20\textwidth}|}
\hline
\begin{center}
 $\boxed{\misses(\thet{0},\e{1})}$
\end{center} 
& & \begin{center} $\expsub{\mathcal{T}}{2}$ \end{center} \\  \hline
\end{tabular}
\end{center}
and 
\begin{center}
  \begin{tabular}{|m{0.40\textwidth}|m{0.20\textwidth}||m{0.20\textwidth}|}
\hline
\begin{center}
 $\boxed{\misses(\thet{0},\e{2})}$
\end{center} 
& & \begin{center} $\expsub{\mathcal{T}}{3}$ \end{center} \\  \hline
\end{tabular}.
\end{center}
in which the satisfying subprograms are output entries.  
Applying resolution to these assertions and other rows can add terms or conditional expressions into the output entries of the resulting rows, but we ignore that complication temporarily for this discussion.

We assume that we have among our assertions
the \emph{left property} of the occurrence relation, 
\begin{center}
  \begin{tabular}{|m{0.60\textwidth}|m{0.10\textwidth}||m{0.10\textwidth}|}
\hline
\begin{center}
$
\begin{aligned}[t]
    &\boxed{\unot(\isatm(E))} \uimplies \\
    & \lef(E) \occursin E
    \end{aligned}
$
\end{center}
& &  \\  \hline
\end{tabular}.
\end{center}
By two applications of the resolution rule to this assertion and the \hyperlink{ass:non-atom}{\emph{non-atom}} case assumptions, that the input expressions $\e{1}$ and $\e{2}$ are nonatomic, taking $\var{E}$ to be $\e{1}$ and $\e{2}$, respectively,  we obtain 
\begin{center}
  \begin{tabular}{|m{0.60\textwidth}|m{0.10\textwidth}||m{0.10\textwidth}|}
\hline
\begin{center}
$
\begin{aligned}
    & \boxed{\lef(\e{1}) \occursin \e{1}}
    \end{aligned}
$
\end{center}
& &  \\  \hline
\end{tabular}
\end{center}
\vspace{5pt}
and 
\begin{center}
  \begin{tabular}{|m{0.60\textwidth}|m{0.10\textwidth}||m{0.10\textwidth}|}
\hline
\begin{center}
$
\begin{aligned}
    & \boxed{\lef(\e{2}) \occursin \e{2}}
    \end{aligned}
$
\end{center}
& &  \\  \hline
\end{tabular}.
\end{center}
We have among our valid assertions the property

\vspace{5pt}
\noindent \begin{tabular}{|m{0.60\textwidth}|m{0.10\textwidth}||m{0.10\textwidth}|}
\hline
\begin{center}$
\begin{aligned}
    & \boxed{\varsub{D}{1} \occursin \varsub{E}{1}} \uand 
     \boxed{\varsub{D}{2} \occursin \varsub{E}{2}} \uimplies \\
    & \vars(\langle \varsub{D}{1}, \varsub{D}{2} \rangle) \subseteq
      \vars(\langle \varsub{E}{1}, \varsub{E}{2} \rangle)
    \end{aligned}
$
\end{center}& &  \\  \hline
\end{tabular},
\vspace{10pt}

\noindent 
which follows from the \hyperlink{prop:occ-sub}{\emph{occurs-subset property}}  of $\vars$,
\[E_{1}^* \occurseq E_{2}^* \uimplies \vars(E_{1}^*) \subseteq \vars(E_{2}^*),\]
and the \hyperlink{prop:vars-union}{\emph{vars-union property}} of tuples, that
\[\vars(\langle E_{1}^*, E_{2}^*\rangle) =
\vars(E_{1}^*) \cup \vars(E_{2}^*),\]
and properties of sets. 

Consequently, by two applications of the \emph{resolution rule}, taking $\varsub{D}{1}$ and $\varsub{E}{1}$ to be $\lef(\e{1})$ and $\e{1}$, respectively, and $\varsub{D}{2}$ 
and $\varsub{E}{2}$ to be  $\lef(\e{2})$ and $\e{2}$, respectively, we obtain

\vspace{5pt}
\noindent \begin{tabular}{|m{0.60\textwidth}|m{0.10\textwidth}||m{0.10\textwidth}|}
\hline
\begin{center}$
\begin{aligned}
       & \boxed{\vars(\langle \lef(\e{1}), \lef(\e{2}) \rangle)} \subseteq \\
      &\boxed{\vars(\langle \e{1}, \e{2} \rangle)}
    \end{aligned}
$
\end{center}& &  \\  \hline
\end{tabular}.

 \vspace{10pt}
Similarly, by the resolution rule applied to 
the \emph{left property} of the size function, 
\begin{center}
  \begin{tabular}{|m{0.60\textwidth}|m{0.10\textwidth}||m{0.10\textwidth}|}
\hline
\begin{center}
$
\begin{aligned}
    &\boxed{\unot(\isatm(E))} \uimplies \\
    & \size(\lef(E)) < \size(E)
    \end{aligned}
$
\end{center}
& &  \\  \hline
\end{tabular}
\end{center}
\vspace{5pt}
and the case \hyperlink{ass:non-atom}{\emph{assumption}} that $\e{1}$ is nonatomic, taking $E$  to be $\e{1}$, we obtain the assertion 
\begin{center}
  \begin{tabular}{|m{0.60\textwidth}|m{0.10\textwidth}||m{0.10\textwidth}|}
\hline
\begin{center}
$
\begin{aligned}
    & \boxed{\size(\lef(\e{1})) < \size(\e{1})}
    \end{aligned}
$
\end{center}
& &  \\  \hline
\end{tabular}.
\end{center}

\vspace{10pt}Earlier, in the subsection entitled \hyperlink{par:wfrelunif}{\emph{Introduction of the Well-Founded Relation}}, we developed the \hyperlink{hyp:ind-size1}{\emph{size-first induction hypothesis}},
  \begin{center}
  \begin{tabular}{|m{0.60\textwidth}|m{0.10\textwidth}||m{0.10\textwidth}|}
\hline
\begin{center}
{$\begin{aligned}
 &\begin{aligned}
&\begin{aligned}
&\left[\begin{aligned}
 &\left\{\begin{aligned}
 &\range(\Thet{0}')\cup\boxed{\vars(\langle \E{1}', \E{2}' \rangle)} \subseteq \\
  &\range(\thet{0})\cup\boxed{\vars(\langle \e{1}, \e{2} \rangle)} 
  \end{aligned}\right\} \uand  \\
  &\size(\E{1}') < \size(\e{1})
  \end{aligned} \right] 
  \end{aligned} 
\end{aligned} \\
 \uimplies &\left[\begin{aligned}&\idem(\Thet{0}') \uimplies \\
 &\begin{aligned}
    {\,\mgiu}(&\Thet{0}', \E{1}', \E{2}', \\
    &\unify(\Thet{0}', \E{1}', \E{2}')) 
\end{aligned}
\end{aligned}\right] \end{aligned}  $}  \hspace{1cm} 
\end{center}& &  \\  \hline
\end{tabular},
\end{center}
or equivalently, by set-theoretic reasoning using the property \[((S_1 \union S_2) \subseteq S_3)
\uiff ((S_1 \subseteq S_3)\uand (S_2 \subseteq S_3)),\] 
  \begin{center}
  \begin{tabular}{|m{0.60\textwidth}|m{0.10\textwidth}||m{0.10\textwidth}|}
\hline
\begin{center}
{$\begin{aligned}
 &\begin{aligned}
&\begin{aligned}
&\left[\begin{aligned}
 &\left\{\begin{aligned}
 &\range(\Thet{0}') \subseteq \\
  &\range(\thet{0})\cup \vars(\langle \e{1}, \e{2} \rangle) 
  \end{aligned}\right\} \uand  \\
 &\left\{\begin{aligned}
 &\boxed{\vars(\langle \E{1}', \E{2}' \rangle)} \subseteq \\
  &\range(\thet{0})\cup\boxed{\vars(\langle \e{1}, \e{2} \rangle)} 
  \end{aligned}\right\} \uand  \\ 
  &\size(\E{1}') < \size(\e{1})
  \end{aligned} \right] 
  \end{aligned} 
\end{aligned} \\
 \uimplies &\left[\begin{aligned}&\idem(\Thet{0}') \uimplies \\
 &\begin{aligned}
    {\,\mgiu}(&\Thet{0}', \E{1}', \E{2}', \\
    &\unify(\Thet{0}', \E{1}', \E{2}')) 
\end{aligned}
\end{aligned}\right] \end{aligned}  $}  \hspace{1cm} 
\end{center}& &  \\  \hline
\end{tabular}.
\end{center}

\noindent By set-theoretic reasoning, using our finding that \[\vars(\langle \lef(\e{1}), \lef(\e{2}) \rangle) \subseteq 
      \vars(\langle \e{1}, \e{2} \rangle)\]
      and taking $\E{1}'$ and $\E{2}'$ to be $\lef(\e{1})$ and $\lef(\e{2})$, respectively, we can reduce this to
  \begin{center}
  \begin{tabular}{|m{0.60\textwidth}|m{0.10\textwidth}||m{0.10\textwidth}|}
\hline
\begin{center}
{$\begin{aligned}
 &\begin{aligned}
&\begin{aligned}
&\left[\begin{aligned}
 &\left\{\begin{aligned}
 &\range(\Thet{0}') \subseteq \\
  &\range(\thet{0})\cup\vars(\langle \e{1}, \e{2} \rangle) 
  \end{aligned}\right\} \uand  \\
  &\boxed{\size(\lef(\e{1})) < \size(\e{1})}
  \end{aligned} \right] 
  \end{aligned} 
\end{aligned} \\
 \uimplies &\left[\begin{aligned}&\idem(\Thet{0}') \uimplies \\
 &\begin{aligned}
    {\,\mgiu}(&\Thet{0}', \lef(\e{1}), \lef(\e{2}), \\
    &\unify(\Thet{0}', \lef(\e{1}), \lef(\e{2}))) 
\end{aligned}
\end{aligned}\right] \end{aligned}  $}  \hspace{1cm} 
\end{center}& &  \\  \hline
\end{tabular}.
\end{center}

Furthermore, using our finding that $\size(\lef(\e{1})) < \size(\e{1})$, we can further reduce our assertion to
  \begin{center}
  \begin{tabular}{|m{0.60\textwidth}|m{0.10\textwidth}||m{0.10\textwidth}|}
\hline
\begin{center}
{$\begin{aligned}
 &\begin{aligned}
&\begin{aligned}
&\left[\begin{aligned}
 &\left\{\begin{aligned}
 &\range(\Thet{0}') \subseteq \\
  &\range(\thet{0})\cup\vars(\langle \e{1}, \e{2} \rangle) 
  \end{aligned}\right\} 
  \end{aligned} \right] 
  \end{aligned} 
\end{aligned} \\
 \uimplies &\left[\begin{aligned}&\idem(\Thet{0}') \uimplies \\
 &\begin{aligned}
    {\,\mgiu}(&\Thet{0}', \lef(\e{1}), \lef(\e{2}), \\
    &\unify(\Thet{0}', \lef(\e{1}), \lef(\e{2}))) 
\end{aligned}
\end{aligned}\right] \end{aligned}  $}  \hspace{1cm} 
\end{center}& &  \\  \hline
\end{tabular}.
\end{center}
We shall refer to this as the \hypertarget{ass:left-int-ind}{\emph{left interim inductive hypotheses}}.

We can perform the same reasoning using the function $\rig$ rather than $\lef$. We then obtain the symmetric assertion
  \begin{center}
  \begin{tabular}{|m{0.60\textwidth}|m{0.10\textwidth}||m{0.10\textwidth}|}
\hline
\begin{center}
{$\begin{aligned}
 &\begin{aligned}
&\begin{aligned}
&\left[\begin{aligned}
 &\left\{\begin{aligned}
 &\range(\Thet{0}') \subseteq \\
  &\range(\thet{0})\cup\vars(\langle \e{1}, \e{2} \rangle) 
  \end{aligned}\right\} 
  \end{aligned}\right] 
  \end{aligned} 
\end{aligned} \\
 \uimplies &\left[\begin{aligned}&\idem(\Thet{0}') \uimplies \\
 &\begin{aligned}
    {\,\mgiu}(&\Thet{0}', \rig(\e{1}), \rig(\e{2}), \\
    &\unify(\Thet{0}', \rig(\e{1}), \rig(\e{2}))) 
\end{aligned}
\end{aligned}\right] \end{aligned}  $}  \hspace{1cm} 
\end{center}& &  \\  \hline
\end{tabular}.
\end{center}
We shall refer to this as the \hypertarget{ass:right-int-ind}{\emph{right interim inductive hypothesis}}.

\paragraph{Treatment of the Left Interim Inductive Hypothesis.} We  handle the left and right interim inductive hypotheses differently.  Applying resolution to the \hyperlink{ass:left-int-ind}{\emph{left interim inductive hypothesis}}, 
  \begin{center}
  \begin{tabular}{|m{0.60\textwidth}|m{0.10\textwidth}||m{0.10\textwidth}|}
\hline
\begin{center}
{$\begin{aligned}
 &\begin{aligned}
&\begin{aligned}
&\left[\begin{aligned}
 &\boxed{\begin{aligned}
 &\range(\Thet{0}') \subseteq \\
  &\range(\thet{0})\cup\vars(\langle \e{1}, \e{2} \rangle) 
  \end{aligned}} 
  \end{aligned} \right] 
  \end{aligned} 
\end{aligned} \\
 \uimplies &\left[\begin{aligned}&\idem(\Thet{0}') \uimplies \\
 &\begin{aligned}
    {\,\mgiu}(&\Thet{0}', \lef(\e{1}), \lef(\e{2}), \\
    &\unify(\Thet{0}', \lef(\e{1}), \lef(\e{2}))) 
\end{aligned}
\end{aligned}\right] \end{aligned}  $}  \hspace{1cm} 
\end{center}& &  \\  \hline
\end{tabular},
\end{center}
and the set theoretic property

\vspace{5pt}
\noindent \begin{tabular}{|m{0.60\textwidth}|m{0.10\textwidth}||m{0.10\textwidth}|}
\hline
\begin{center}$
\begin{aligned}
     \boxed{\varsub{S}{1} \subseteq \varsub{S}{1} \cup \varsub{S}{2}}
\end{aligned}
$
\end{center}& &  \\  \hline
\end{tabular},
\vspace{10pt}

\noindent taking $\Thet{0}'$ to be $\thet{0}$ and $\varsub{S}{1}$ and $\varsub{S}{2}$  to be $\range(\thet{0})$ and $\vars(\langle \e{1}, \e{2} \rangle)$, respectively,  we obtain
  \begin{center}
  \begin{tabular}{|m{0.60\textwidth}|m{0.10\textwidth}||m{0.10\textwidth}|}
\hline
\begin{center}
{$\begin{aligned}
 & &\begin{aligned}&\boxed{\idem(\thet{0})} \uimplies \\
 &\begin{aligned}
    {\,\mgiu}(&\thet{0}, \lef(\e{1}), \lef(\e{2}), \\
    &\unify(\thet{0}, \lef(\e{1}), \lef(\e{2}))) 
\end{aligned}
\end{aligned} \end{aligned}  $}  \hspace{1cm} 
\end{center}& &  \\  \hline
\end{tabular}.
\end{center}

We process this further, by resolution with the \hyperlink{ass:init}{\emph{initial assertion}} that $\thet{0}$ is idempotent,
\begin{center}
\begin{tabular}{|m{0.60\textwidth}|m{0.10\textwidth}||m{0.10\textwidth}|}
 \hline 
\begin{center}$\boxed{\idem(\thet{0})}$\end{center} &  & \\
\hline
\end{tabular},
\end{center}
yielding
  \begin{center}
  \begin{tabular}{|m{0.60\textwidth}|m{0.10\textwidth}||m{0.10\textwidth}|}
\hline
{\begin{center}
{$\begin{aligned}
    {\,\mgiu}(&\thet{0}, \lef(\e{1}), \lef(\e{2}), \\
    &\unify(\thet{0}, \lef(\e{1}), \lef(\e{2}))) 
\end{aligned}  $}  \hspace{1cm} 
\end{center}}& &  \\  \hline
\end{tabular},
\end{center}
or, abbreviating  $\unify(\thet{0}, \lef(\e{1}), \lef(\e{2})))$ as  $\thet{l}$,
  \begin{center}
  \begin{tabular}{|m{0.60\textwidth}|m{0.10\textwidth}||m{0.10\textwidth}|}
\hline
\begin{center}
   ${\,\boxed{\mgiu(\thet{0}, \lef(\e{1}), \lef(\e{2}), \thet{l})}}$
   \end{center}  
& &  \\  \hline
\end{tabular}.
\end{center}
We call this the \hypertarget{hyp:L-ind}{\emph{left induction hypothesis}} and \[\thet{l} = \unify(\thet{0}, \lef(\e{1}), \lef(\e{2})))\] the \emph{left recursive call}.  Thus we have established that \hypertarget{def:L-rec}{the \emph{left recursive call} is a most-general idempotent reducing unifier.}

\paragraph{Expansion and Splitting of the Left Induction Hypothesis.}
We use consequences of the left  inductive hypothesis in dealing with the right induction hypothesis. 

We established the {\hyperlink{prop:idem-mgiu}{\emph{idempotence property}} of the most-general idempotent reducing unifier relation, that a substitution that satisfies the most-general idempotent reducing unifier relation is indeed idempotent:
\begin{center}
  \begin{tabular}{|m{0.58\textwidth}|m{0.10\textwidth}||m{0.10\textwidth}|}
\hline
\begin{center}
{$
\begin{aligned}
&\begin{aligned}
&\boxed{{\mgiu}(\Thet{0}, \E{1}, \E{2}, \Theta)} \uimplies \\
&\begin{aligned}
\idem(\Theta)
\end{aligned}
\end{aligned}
\end{aligned} $}  \hspace{0cm} 
\end{center}& &  \\  \hline
\end{tabular}.
\end{center}
Applying resolution to the  \hyperlink{hyp:L-ind}{\emph{left induction hypothesis}} and this assertion, taking $\Thet{0}$ to be $\thet{0}$, $\E{1}$ and $\E{2}$ to be $\lef(\e{1})$ and $\lef(\e{2}),$ respectively, and $\Theta$ to be $\thet{l}$, we obtain 
\begin{center}
  \begin{tabular}{|m{0.60\textwidth}|m{0.10\textwidth}||m{0.10\textwidth}|}
\hline
\begin{center}
{$
\begin{aligned}
\idem(\thet{l})
\end{aligned} $}  \hspace{0cm} 
\end{center}& &  \\  \hline
\end{tabular}.
\end{center}
That is, \hypertarget{ass:left-idem}{the \emph{left recursive call} is idempotent}. 

We next expand  the \emph{left induction hypothesis} into four separate components, corresponding to the four conjuncts in the definition of the most-general idempotent reducing unifier.  

Our \emph{left induction hypothesis} is
  \begin{center}
  \begin{tabular}{|m{0.60\textwidth}|m{0.10\textwidth}||m{0.10\textwidth}|}
\hline
\begin{center}
$\boxed{\begin{aligned}
    {\,\mgiu}(&\thet{0}, \lef(\e{1}), \lef(\e{2}), \thet{l}) 
\end{aligned}}  $  \hspace{1cm} 
\end{center}& &  \\  \hline
\end{tabular}.
\end{center}
where $\thet{l}$ is the abbreviation for the \hyperlink{def:L-rec}{\emph{left recursive call}}, $\unify(\thet{0}, \lef(\e{1}), \lef(\e{2}))$. We can expand this according to the definition of the most-general idempotent reducing unifier relation \hyperlink{def:mgiu}{$\mgiu$}, using the valid assertion
\begin{center}
  \begin{tabular}{|m{0.60\textwidth}|m{0.10\textwidth}||m{0.10\textwidth}|}
\hline
 \begin{center} 
${\begin{aligned} 
 &\boxed{\mgiu(\Thet{0}, \E{1}, \E{2}, \Theta)} \uiff \\
 &\quad\begin{aligned}
&\E{1} \apply \Theta = \E{2} \apply \Theta \, \uand 
 \\
 \,\, & \Thet{0} \moregen \Theta \, \uand
\\
  \,\, & {\mgi}(\Thet{0}, \E{1}, \E{2}, \Theta) \, \uand \\
  &{\reduce}(\Theta_0, \vars(E_1, E_2)\!\apply\! \Theta_0, \Theta)
\end{aligned}.\end{aligned}}$\hspace{1cm} 
\end{center}& &  \\  \hline
\end{tabular}.
\end{center}
Applying the \emph{equivalence replacement rule}, taking $\E{1}$ and $\E{2}$ to be $\lef(\e{1})$ and  $\lef(\e{2})$, respectively, $\Thet{0}$ to be $\thet{0}$, and $\Theta$ to be $\allowbreak \thet{l},$ we obtain
  \begin{center}
  \begin{tabular}{|m{0.60\textwidth}|m{0.10\textwidth}||m{0.10\textwidth}|}
\hline
\begin{center}
{$\begin{aligned}
 &\begin{aligned}
 &\begin{aligned}
&\lef(\e{1}) \apply \thet{l} = 
 \lef(\e{2}) \apply \thet{l} \, \uand 
 \\
 \,\, &\, \thet{0} \moregen  \thet{l} \, \uand
\\
  \,\, &\, {\mgi}(\thet{0}, \E{1}, \E{2}, \theta_l) \, \uand \\
  &\,{\reduce}(\theta_0, \vars(E_1, E_2)\!\apply\! \theta_0, \theta_l)
\end{aligned}
\end{aligned}\end{aligned}   $} 
\end{center}& &  \\  \hline
\end{tabular}.
\end{center}
This is decomposed, by \emph{and-splitting}, to four separate assertions:
the \hypertarget{hyp:L-un-ind}{\emph{left unify induction hypothesis}} (that the {\emph{left recursive call}} yields a unifier):
  \begin{center}
  \begin{tabular}{|m{0.60\textwidth}|m{0.10\textwidth}||m{0.10\textwidth}|}
\hline
\begin{center}
{$\begin{aligned}
 &\begin{aligned}
 &\begin{aligned}
&\lef(\e{1}) \apply \thet{l} = 
\lef(\e{2}) \apply \thet{l} 
\end{aligned}
\end{aligned}\end{aligned}   $} 
\end{center}& &  \\  \hline
\end{tabular};
\end{center}
 the \hypertarget{hyp:L-ext-ind}{\emph{left extension induction hypothesis}} (that the \emph{left recursive call} yields an extension of the environment):
  \begin{center}
  \begin{tabular}{|m{0.60\textwidth}|m{0.10\textwidth}||m{0.10\textwidth}|}
\hline
\begin{center}
{$\begin{aligned}
 &\begin{aligned}
 &\begin{aligned}
 \,\, &\, \thet{0} \moregen \thet{l} 
\end{aligned}
\end{aligned}\end{aligned}   $} 
\end{center}& &  \\  \hline
\end{tabular};
\end{center}
the \hypertarget{hyp:L-mgi-ind}{\emph{left mgi induction hypothesis}} (that the \emph{left recursive call} yields a most-general idempotent substitution with respect to $\lef(\e{1})$ and $\lef(\e{2})$):
\begin{center}
  \begin{tabular}{|m{0.60\textwidth}|m{0.10\textwidth}||m{0.10\textwidth}|}
\hline
\begin{center}
{$
\begin{aligned}
 \mgi(\thet{0}, \lef(\e{1}), \lef(\e{2}), \thet{l})
\end{aligned} $}  \hspace{0cm} 
\end{center}& &  \\  \hline
\end{tabular};
\end{center}
and the \hypertarget{hyp:L-red-ind}{\emph{left reduce induction hypothesis}}, that the \emph{left recursive call} yields a reduction with respect to $\lef(\e{1})$ and $\lef(\e{2})$):
\begin{center}
  \begin{tabular}{|m{0.60\textwidth}|m{0.10\textwidth}||m{0.10\textwidth}|}
\hline
\begin{center}
{$
\begin{aligned}
 {\reduce}(\thet{0}, \vars(\lef(\e{1}), \lef(\e{2}))\!\apply\! \theta_0, \theta_l)
\end{aligned} $}  \hspace{0cm} 
\end{center}& &  \\  \hline
\end{tabular};
\end{center}
or, by our \hyperlink{ass:miss-inp}{\emph{missed inputs case assumption}}, $\misses(\thet{0}, \langle \e{1}, \e{2} \rangle)$, and reasoning about expressions,
\begin{center}
  \begin{tabular}{|m{0.60\textwidth}|m{0.10\textwidth}||m{0.10\textwidth}|}
\hline
\begin{center}
{$
\begin{aligned}
 {\reduce}(\thet{0}, \vars(\lef(\e{1}), \lef(\e{2})), \theta_l)
\end{aligned} $}  \hspace{0cm} 
\end{center}& &  \\  \hline
\end{tabular}.
\end{center}
In other words, expanding the assertion by the \hyperlink{rule:tab-equiv-AG}{\emph{equivalence replacement rule}} and the \hyperlink{prop:red-def}{\emph{definition of reduction}},
\begin{center}
  \begin{tabular}{|m{0.60\textwidth}|m{0.10\textwidth}||m{0.10\textwidth}|}
\hline
\begin{center}$\begin{aligned}
 &\boxed{{\reduce}(\Thet{0}, V, \Theta)} \,\, \uiff  \\
 &\range(\Theta) \subseteq \range(\Theta_0) \cup  V
 \end{aligned}$\end{center}
& &  \\  \hline
\end{tabular},
\end{center}
taking $\Thet{0}$ to be $\thet{0}$, $V$ to be $\vars(\lef(\e{1}), \lef(\e{2}))$, and $\Theta$ to be $\theta_l$, we have \hypertarget{ass:left-range-vars}{the \emph{left range-vars assertion}},
\begin{center}
  \begin{tabular}{|m{0.58\textwidth}|m{0.10\textwidth}||m{0.10\textwidth}|}
\hline
\begin{center}
{$
\begin{aligned}
&\begin{aligned}
&\begin{aligned}
&{\range}(\theta_l) \subseteq  \\
&\range(\theta_0) \union \vars(\lef(\e{1}), \lef(\e{2}))
\end{aligned}
\end{aligned}
\end{aligned} $}  \hspace{0cm} 
\end{center}& &  \\  \hline
\end{tabular}.
\end{center}

\paragraph{Treatment of the Right Interim Inductive Hypothesis.} 
We now turn our attention to the
 \hyperlink{ass:right-int-ind}{\emph{right interim inductive hypothesis}},
  \begin{center}
  \begin{tabular}{|m{0.60\textwidth}|m{0.10\textwidth}||m{0.10\textwidth}|}
\hline
\begin{center}
{$\begin{aligned}
 &\begin{aligned}
&\begin{aligned}
&{\begin{aligned}
 &\begin{aligned}
 &\range(\Thet{0}') \subseteq \\
  &\range(\thet{0})\cup \boxed{\vars(\langle \e{1}, \e{2} \rangle)}
  \end{aligned}
  \end{aligned}} 
  \end{aligned} 
\end{aligned} \\
 \uimplies &\left[\begin{aligned}&\idem(\Thet{0}') \uimplies \\
 &\begin{aligned}
    {\,\mgiu}(&\Thet{0}', \rig(\e{1}), \rig(\e{2}), \\
    &\unify(\Thet{0}', \rig(\e{1}), \rig(\e{2}))) 
\end{aligned}
\end{aligned}\right] \end{aligned}  $}  \hspace{1cm} 
\end{center}& &  \\  \hline
\end{tabular}.
\end{center}
Expanding the antecedent by the \hyperlink{prop:vars-left-right}{\emph{left-right-property}} of $\vars$,
\[\vars(E) = \vars(\lef(E)) \union \vars(\rig(E)),\]  the \hyperlink{prop:vars-union}{\emph{union property}} of $\vars$, 
\[\vars(E_1, \dots, E_n) =  
\vars(E_1) \union  \ldots \union \vars(E_n),\]
and properties of sets,  we obtain
 \begin{center}
  \begin{tabular}{|m{0.60\textwidth}|m{0.10\textwidth}||m{0.10\textwidth}|}
\hline
\begin{center}
{$\begin{aligned}
 &\begin{aligned}
&\begin{aligned}
&{\begin{aligned}
 &\begin{aligned}
 &\range(\Thet{0}') \subseteq \\
  &\range(\thet{0})\,\union \\
  &\vars(\lef(\e{1})) \union \vars(\lef(\e{2}))\, \union \\
  &\vars(\rig(\e{1})) \union \vars(\rig(\e{2}))
  \end{aligned}
  \end{aligned}} 
  \end{aligned} 
\end{aligned} \\
 \uimplies &\left[\begin{aligned}&\idem(\Thet{0}') \uimplies \\
 &\begin{aligned}
    {\,\mgiu}(&\Thet{0}', \rig(\e{1}), \rig(\e{2}), \\
    &\unify(\Thet{0}', \rig(\e{1}), \rig(\e{2}))) 
\end{aligned}
\end{aligned}\right] \end{aligned}  $}  \hspace{1cm} 
\end{center}& &  \\  \hline
\end{tabular},
\end{center}
or, by properties of sets,
 \begin{center}
  \begin{tabular}{|m{0.60\textwidth}|m{0.10\textwidth}||m{0.10\textwidth}|}
\hline
\begin{center}
{$\begin{aligned}
 &\begin{aligned}
&\begin{aligned}
&{\begin{aligned}
 &\boxed{\begin{aligned}
 &\range(\Thet{0}') \subseteq \\
 &{\begin{aligned}
  &\range(\thet{0})\union
  \vars(\lef(\e{1})) \union \vars(\lef(\e{2})) 
  \end{aligned}}
  \end{aligned}}
  \end{aligned}} 
  \end{aligned} 
\end{aligned} \\
 \uimplies &\left[\begin{aligned}&\idem(\Thet{0}') \uimplies \\
 &\begin{aligned}
    {\,\mgiu}(&\Thet{0}', \rig(\e{1}), \rig(\e{2}), \\
    &\unify(\Thet{0}', \rig(\e{1}), \rig(\e{2}))) 
\end{aligned}
\end{aligned}\right] \end{aligned}  $}  \hspace{1cm} 
\end{center}& &  \\  \hline
\end{tabular}.
\end{center}
By resolution applied to the above assertion and the
\hyperlink{ass:left-range-vars}{\emph{left range-vars assertion}},
\begin{center}
  \begin{tabular}{|m{0.60\textwidth}|m{0.10\textwidth}||m{0.10\textwidth}|}
\hline
\begin{center}
{$
\boxed{\begin{aligned}
&\begin{aligned}
&\begin{aligned}
&{\range}(\theta_l) \subseteq  \\
&\range(\theta_0) \union \vars(\lef(\e{1}), \lef(\e{2}))
\end{aligned}
\end{aligned}
\end{aligned}} $}  \hspace{0cm} 
\end{center}& &  \\  \hline
\end{tabular},
\end{center}
taking $\Theta_0'$ to be $\thet{l}$, we get
 \begin{center}
  \begin{tabular}{|m{0.60\textwidth}|m{0.10\textwidth}||m{0.10\textwidth}|}
\hline
\begin{center}
{$ \begin{aligned}&\boxed{\idem(\thet{l})} \uimplies \\
                &\begin{aligned}
                   {\,\mgiu}(&\thet{l}, \rig(\e{1}), \rig(\e{2}), \\
                             &\unify(\thet{l}, \rig(\e{1}), \rig(\e{2}))) 
                  \end{aligned}
\end{aligned}  $}  \hspace{1cm} 
\end{center}& &  \\  \hline
\end{tabular}.
\end{center}
Next, resolving away the  \hyperlink{ass:init}{\emph{initial assertion}},
\begin{center}
  \begin{tabular}{|m{0.60\textwidth}|m{0.10\textwidth}||m{0.10\textwidth}|}
\hline
\begin{center}
{$
\begin{aligned}
\boxed{\idem(\thet{l})}
\end{aligned} $}  \hspace{0cm} 
\end{center}& &  \\  \hline
\end{tabular},
\end{center}
gives us
  \begin{center}
  \begin{tabular}{|m{0.60\textwidth}|m{0.10\textwidth}||m{0.10\textwidth}|}
\hline
\begin{center}
{$\begin{aligned}
 &\begin{aligned}
 &\begin{aligned}
    {\,\mgiu}(&\thet{l}, \rig(\e{1}), \rig(\e{2}), \\
    &\unify(\thet{l}, \rig(\e{1}), \rig(\e{2}))) 
\end{aligned}
\end{aligned} \end{aligned}  $}  \hspace{1cm} 
\end{center}& &  \\  \hline
\end{tabular},
\end{center}
or, abbreviating $\unify(\thet{l}, \rig(\e{1}), \rig(\e{2}))$ as $\thet{r}$,
  \begin{center}
  \begin{tabular}{|m{0.60\textwidth}|m{0.10\textwidth}||m{0.10\textwidth}|}
\hline
\begin{center}
{$\begin{aligned}
 &\begin{aligned}
 &\begin{aligned}
    {\,\mgiu}(&\thet{l}, \rig(\e{1}), \rig(\e{2}), \thet{r}) 
\end{aligned}
\end{aligned} \end{aligned}  $}  \hspace{1cm} 
\end{center}& &  \\  \hline
\end{tabular}.
\end{center}
We call this the \hypertarget{hyp:R-ind}{\emph{right induction hypothesis}} and 
\[\thet{r}
\begin{aligned}[t]
&= \unify(\thet{l}, \rig(\e{1}), \rig(\e{2})) \\
 &\begin{aligned}[t]
 \,\, = \unify(\unify(\thet{0}, \lef(\e{1}), \lef(\e{2})), \rig(\e{1}), \rig(\e{2}))
\end{aligned}
\end{aligned}
\]
the \hypertarget{def:nest}{\emph{nested recursive call}}.

The \hyperlink{def:L-rec}{\emph{left recursive call}} $\thet{l}$ is the unification of the left components of the input expressions in the context of the input environment $\thet{0}$; the \hyperlink{def:nest}{\emph{nested recursive call}} $\thet{r}$  is the unification of the right components of the input expressions in the context of the environment resulting from the \emph{left recursive call} $\thet{l}$. The difference between the treatment of the left and right components of the inputs is reflected in the algorithm we obtain.  If we reverse the roles of left and right in the derivation, we obtain a symmetric version of the same algorithm, in which the {\emph{nested recursive call}} is
\[\begin{aligned}
\unify(
\begin{aligned}
   \unify(\thet{0}, 
  \rig(\e{1}),  
 \rig(\e{2})),
   \end{aligned} 
\lef(\e{1}), 
\lef(\e{2}))
 \end{aligned}. \]
Which algorithm Snark finds first depends on its search settings, and in particular whether $\lef$ or $\rig$ is greater in its symbol ordering.

\paragraph{Expansion and Splitting of the Right Induction Hypotheses.}
As we did for the \hyperlink{hyp:L-ind} {\emph{left induction hypothesis}}, we now expand the \hyperlink{hyp:R-ind}{\emph{right induction hypothesis}} into four separate components, corresponding to the four conjuncts in the definition of the most-general idempotent reducing unifier.

We can apply the definition of a \hyperlink{def:mgiu}{most-general idempotent reducing unifier},
\begin{center}

  \begin{tabular}{|m{0.60\textwidth}|m{0.10\textwidth}||m{0.10\textwidth}|}
\hline
 \begin{center} 
${\begin{aligned} 
 &\boxed{\mgiu(\Thet{0}, \E{1}, \E{2}, \Theta)} \uiff \\
 &\quad\begin{aligned}
&\E{1} \apply \Theta = \E{2} \apply \Theta \, \uand 
 \\
 \,\, & \Thet{0} \moregen \Theta \, \uand
 \\
  \,\, & {\mgi}(\Thet{0}, \E{1}, \E{2}, \Theta)
\end{aligned}\end{aligned}}$\hspace{1cm} 
\end{center}& &  \\  \hline
\end{tabular},
\end{center}
to the \hyperlink{hyp:R-ind}{\emph{right induction hypothesis}}, 
 \begin{center}
  \begin{tabular}{|m{0.60\textwidth}|m{0.10\textwidth}||m{0.10\textwidth}|}
  \hline
  \begin{center}
{$\mgiu(\thet{l},\rig(\e{1}),\rig(\e{2}), \theta_r) $}
 \end{center}  
& & \\
\hline
\end{tabular},
\end{center}
taking $\Thet{0}$ to be $\thet{l},$ $\E{1}$ and $\E{2}$ to be $\rig(\e{1})$ and $\rig(\e{2})$, respectively, and $\Theta$ to be  $\theta_r$, where $\theta_r$ is our abbreviation for the \hyperlink{def:nest}{{\emph{nested recursive call}}}
\[\begin{aligned}
\unify(
\begin{aligned}
   \unify(\thet{0}, 
  \lef(\e{1}),  
 \lef(\e{2})),
   \end{aligned} 
\rig(\e{1}), 
\rig(\e{2}))
 \end{aligned}. \]
 We obtain 
\begin{center}
  \begin{tabular}{|m{0.60\textwidth}|m{0.10\textwidth}||m{0.10\textwidth}|}
\hline
\begin{center}
{$
\begin{aligned}
&\begin{aligned}
&\begin{aligned}
&\rig(\e{1}) \apply \thet{r} = \rig(\e{2}) \apply \thet{r}  \uand  \\
 \,\, &\, \thet{l} \moregen \thet{r}  \uand
\\
  \,\, &\, {\mgi}(\thet{l}, \rig(\e{1}), \rig(\e{2}), \thet{r}) \uand
\\
  \,\, &\, {\reduce}(\thet{l}, \vars(\rig(\e{1}), \rig(\e{2}))\!\apply\!\theta_l, \thet{r})
\end{aligned}
\end{aligned}
\end{aligned} $}  \hspace{0cm} 
\end{center}& &  \\  \hline
\end{tabular}.
\end{center}

As in the left-side case, this is decomposed, by \emph{and-splitting}, into four separate assertions: the \hypertarget{hyp:R-un}{\emph{right unify induction hypothesis}} (that the right recursive call yields a unifier):
\begin{center}
  \begin{tabular}{|m{0.60\textwidth}|m{0.10\textwidth}||m{0.10\textwidth}|}
\hline
\begin{center}
{$
\begin{aligned}
&\begin{aligned}
&\begin{aligned}
&\rig(\e{1}) \apply \thet{r} = 
 \rig(\e{2}) \apply \thet{r} 
\end{aligned}
\end{aligned}
\end{aligned} $}  \hspace{0cm} 
\end{center}& &  \\  \hline
\end{tabular};
\end{center}
 the \hypertarget{hyp:R-ext-ind}{\emph{right extension induction hypothesis}} (that the right recursive call yields an extension of the environment):
\begin{center}
  \begin{tabular}{|m{0.60\textwidth}|m{0.10\textwidth}||m{0.10\textwidth}|}
\hline
\begin{center}
{$
\begin{aligned}
&\begin{aligned}
&\begin{aligned}
 \,\, & \thet{l} \moregen \thet{r} 
\end{aligned}
\end{aligned}
\end{aligned} $}  \hspace{0cm} 
\end{center}& &  \\  \hline
\end{tabular};
\end{center}
the \hypertarget{hyp:R-mgi}{\emph{right mgi induction hypothesis}} (that the right recursive call yields a most-general idempotent substitution):

\begin{center}
  \begin{tabular}{|m{0.60\textwidth}|m{0.10\textwidth}||m{0.10\textwidth}|}
\hline
\begin{center}
{$
\begin{aligned}
&\mgi(\thet{l}, \rig(\e{1}), \rig(\e{2}), \thet{r})
\end{aligned}
 $}  \hspace{0cm} 
\end{center}& &  \\  \hline
\end{tabular};
\end{center}
and the \hypertarget{hyp:R-red-ind}{\emph{right reduce induction hypothesis}} (that the right recursive call yields a reduction):

\begin{center}
  \begin{tabular}{|m{0.66\textwidth}|m{0.07\textwidth}||m{0.07\textwidth}|}
\hline
\begin{center}
{$
\begin{aligned}
&{\reduce}(\thet{l}, vars(\rig(\e{1}), \rig(\e{2}))\!\apply\!\theta_l, \thet{r})
\end{aligned}
 $}  \hspace{0cm} 
\end{center}& &  \\  \hline
\end{tabular}.
\end{center}
The next section shows how the new induction hypotheses allow the introduction of nested recursive calls into the output column.
\paragraph{Development of the Nested Recursive Calls.}

In the subsection \hyperlink{sec:spec-first}{\emph{The Specification and First Steps}}, we developed the \hyperlink{goal-exp-init}{\emph{expanded initial goal}}

\begin{center}
\begin{tabular}{|m{0.10\textwidth}|m{0.50\textwidth}||m{0.20\textwidth}|}
 \hline 
  & 
  \begin{center}
${\begin{aligned} 
&\begin{aligned}
&\boxed{\e{1} \apply \Theta = \e{2} \apply \Theta} \, \uand 
 \\
 \,\, & \thet{0} \moregen \Theta \, \uand
\\
  \,\, & {\mgi}(\thet{0}, \e{1}, \e{2}, \Theta)\, \uand
\\
  \,\, & {\reduce}(\thet{0},\vars(\e{1}, \e{2})\!\apply\thet{0}, \Theta)
\end{aligned}\end{aligned}}$\hspace{1cm} 
\end{center} & \begin{center}$\Theta$ \end{center} \\
\hline
\end{tabular},
\end{center}

According to the \hyperlink{prop:unify-left-right}{\emph{unify-left-right property}}  substitution unifies two nonatomic expressions if it unifies its left and right components; this is expressed by the valid assertion
 \begin{center}
  \begin{tabular}{|m{0.66\textwidth}|m{0.07\textwidth}||m{0.07\textwidth}|}
\hline
\begin{center}
$
\begin{aligned}
 &\begin{aligned}
&\left[\begin{aligned}&\unot(\isatm(\E{1}))) \uand \\ 
&\unot(\isatm(\E{2})))\end{aligned}\right] 
\end{aligned} \uimplies \\
&\left[\begin{aligned}
&{\left[\begin{aligned} 
      &\lef(\E{1}) \apply \Theta = \lef(\E{2}) \apply \Theta \uand \,\,\\
      & \rig(\E{1}) \apply \Theta = \rig(\E{2}) \apply \Theta 
     \end{aligned}\right]} \\
    & \uimplies \boxed{\E{1} \apply \Theta = \E{2} \apply \Theta} 
     \end{aligned}\right] 
     \end{aligned}
     $
\end{center}
& &  \\  \hline
\end{tabular}.
\end{center}
Applying resolution to this assertion and the above goal, renaming $\Theta$ in the assertion to be $\Thet{r}$ for pedalogical reasons,
taking $\E{1}$ and $\E{2}$ to be $\e{1}$ and $\e{2}$, respectively, and $\Theta$ in the goal to be $\Theta_r$, and then applying resolution to the result and the two  \hyperlink{ass:non-atom}{\emph{nonatomic case assumptions}}, namely,  $\unot(\isatm(\e{1})))$ and  
$\unot(\isatm(\e{2})))$, we obtain
\begin{center}
\begin{tabular}{|m{0.10\textwidth}|m{0.50\textwidth}||m{0.20\textwidth}|}
 \hline 
  & 
  \begin{center}
{$\begin{aligned}
 &  \lef(\e{1}) \apply \Thet{r} = \lef(\e{2}) \apply \Thet{r} \uand \\
 &\boxed{\rig(\e{1}) \apply \Thet{r} = \rig(\e{2}) \apply \Thet{r}} \, \uand \\
  & \,\thet{0} \moregen \Thet{r} \, \uand \\
  & {\mgi}(\thet{0}, \e{1}, \e{2}, \Thet{r})\, \uand
\\
  \,\, &{\reduce}(\thet{0},\vars(\e{1}, \e{2}), \Thet{r})
\end{aligned}$}
\end{center}
& 
\begin{center}$\Thet{r}$ \end{center}\\
\hline
\end{tabular}.
\end{center}
We now apply the resolution rule to this goal and the \hyperlink{hyp:R-un}{\emph{right unify induction hypothesis}},
\begin{center}
  \begin{tabular}{|m{0.66\textwidth}|m{0.07\textwidth}||m{0.07\textwidth}|}
\hline
\begin{center}
$
\begin{aligned}
  \boxed{\begin{aligned}
    \rig(\e{1}) \apply \thet{r} = 
      \rig(\e{2}) \apply \thet{r} 
   \end{aligned}}
\end{aligned} 
$  \hspace{0cm} 
\end{center}& &  \\  \hline
\end{tabular},
\end{center}
taking  $\Thet{r}$ to be $\thet{r},$ to obtain
\begin{center}
\begin{tabular}{|m{0.10\textwidth}|m{0.50\textwidth}||m{0.20\textwidth}|}
 \hline 
  & 
   \begin{center}
{$\begin{aligned}
 &\begin{aligned}    
 &\boxed{
\begin{aligned}
    &\lef(\e{1}) \apply \thet{r} = 
      \lef(\e{2}) \apply \thet{r} \hspace{-10pt}\end{aligned}}
\end{aligned} \hspace{-7pt} \uand  \\
& \,\thet{0} \moregen \thet{r} \, \uand \\ & 
  \,  {\mgi}(\thet{0}, \e{1}, \e{2}, \thet{r})\, \uand
\\
  \,\, &\reduce(\thet{0},\vars(\e{1}, \e{2})\!\apply\thet{0}, \thet{r})
\end{aligned}$}
\end{center}
& 

\begin{center}
$
\hspace{-10pt}
\begin{aligned}
\thet{r}
\end{aligned}
$
\end{center}
\\[10pt]
\hline
\end{tabular}.
\end{center}
 This is the stage at which the \hyperlink{def:nest}{{\emph{nested recursive call}}} $\thet{r}$, our abbreviation for 
\[\begin{aligned}
\unify(
\begin{aligned}
   \unify(\thet{0}, 
  \lef(\e{1}),  
 \lef(\e{2})),
   \end{aligned} 
\rig(\e{1}), 
\rig(\e{2}))
 \end{aligned}, \]
 is introduced into the output entry.

In the boxed conjunct, we need to show that $\thet{r}$ unifies $\lef(\e{1})$ and $\lef(\e{2})$; it suffices to show that $\thet{l}$ unifies $\lef(\e{1})$ and $\lef(\e{2})$, because $\thet{r}$ is an extension of $\thet{l}$. We here spell out this argument in more detail.

We have \hyperlink{prop:unif-ext}{mentioned} that any extension of a unifier is also a unifier;  that is, we have the valid assertion
\begin{center}
  \begin{tabular}{|m{0.66\textwidth}|m{0.07\textwidth}||m{0.07\textwidth}|}
\hline
\begin{center}
$\begin{aligned}
    &\Thet{1} \moregen \Thet{2} \uimplies \\
    &\begin{aligned}
         & \quad \E{1} \apply \Thet{1} = \E{2} \apply \Thet{1} \uimplies \\
         & \boxed{\quad \E{1} \apply \Thet{2} = \E{2} \apply \Thet{2}}
    \end{aligned}
\end{aligned}$
\end{center}& &  \\  \hline
\end{tabular}.
\end{center}
Applying the resolution rule to these rows, taking $\E{1}$ and $\E{2}$ to be
$\lef(\e{1})$ and $\lef(\e{2})$, respectively, and $\Thet{2}$ to be $\thet{r},$ and renaming $\Thet{1}$ to $\Thet{l}$ for pedagogical reasons,  we obtain
\begin{center}
\begin{tabular}{|m{0.10\textwidth}|m{0.50\textwidth}||m{0.20\textwidth}|}
 \hline 
  & 
   \begin{center}
{$\begin{aligned}
 &\Thet{l} \moregen \thet{r}  \uand  \\
  &\boxed{\lef(\e{1}) \apply \Thet{l} = 
          \lef(\e{2}) \apply \Thet{l}} \,\, \uand  \\
&\,
   \,\thet{0} \moregen \thet{r} \,  \uand \\
  &\,  \,{\mgi}(\thet{0}, \e{1}, \e{2}, \thet{r})\,  \uand \\ 
  \,\, &\reduce(\thet{0},\vars(\e{1}, \e{2}), \thet{r})
\end{aligned}$}
\end{center}
& 
\begin{center}
$
\thet{r}
$
\end{center}
\\
\hline
\end{tabular}.
\end{center}
Applying resolution to this row and the \hyperlink{hyp:L-un-ind}{\emph{left unify induction hypothesis}},
  \begin{center}
  \begin{tabular}{|m{0.66\textwidth}|m{0.07\textwidth}||m{0.07\textwidth}|}
\hline
\begin{center}
{$\begin{aligned}
 &\begin{aligned}
 &\boxed{\begin{aligned}
&\lef(\e{1}) \apply \thet{l} = 
\lef(\e{2}) \apply \thet{l} 
\end{aligned}}
\end{aligned}\end{aligned}   $} 
\end{center}& &  \\  \hline
\end{tabular},
\end{center}
taking $\Thet{l}$ to be $\thet{l}$, we obtain
\begin{center}
\begin{tabular}{|m{0.10\textwidth}|m{0.50\textwidth}||m{0.20\textwidth}|}
 \hline 
  & 
   \begin{center}
{$\begin{aligned}
 & \boxed{\begin{aligned}
    &\thet{l} \moregen  \thet{r} \hspace {-10pt}
    \end{aligned}}\, \uand  \\
&\,   \,  \thet{0} \moregen \thet{r} \uand  \\
  &\,    {\mgi}(\thet{0}, \e{1}, \e{2}, \thet{r})\, \uand
\\
  \,\, &\reduce(\thet{0},\vars(\e{1}, \e{2}), \thet{r})
\end{aligned}$}
\end{center}
& 
\begin{center}

$ \begin{aligned}
   \thet{r}
 \end{aligned}    
$
\end{center}
\\
\hline
\end{tabular}.
\end{center}
The boxed subexpression is just the \hyperlink{hyp:L-un-ind}{\emph{right extension induction hypothesis}}, 
\begin{center}
  \begin{tabular}{|m{0.66\textwidth}|m{0.07\textwidth}||m{0.07\textwidth}|}
\hline
\begin{center}
{$
\begin{aligned}
&\begin{aligned}
&\begin{aligned}
 \,\, & \boxed{\thet{l} \moregen \thet{r}} 
\end{aligned}
\end{aligned}
\end{aligned} $}  \hspace{0cm} 
\end{center}& &  \\  \hline
\end{tabular}.
\end{center}
By resolution, we obtain
\begin{center}
\begin{tabular}{|m{0.10\textwidth}|m{0.50\textwidth}||m{0.20\textwidth}|}
 \hline 
  & 
   \begin{center}
{$\begin{aligned}
&  \boxed{\thet{0} \moregen \thet{r}} \uand   \\
  &\,\,{\mgi}(\thet{0}, \e{1}, \e{2}, \thet{r})\, \uand
\\
  \,\, &\reduce(\thet{0},\vars(\e{1}, \e{2})\!\apply\thet{0}, \thet{r})
\end{aligned}$}
\end{center}
& 
\begin{center}

$ \begin{aligned}
   \thet{r}
 \end{aligned}    
$
\end{center}
\\
\hline
\end{tabular}.
\end{center}

We earlier established the \hyperlink{prop:moregen-trans}{\emph{transitivity}} of the \emph{more-general} relation $\moregen$:
\begin{center}
  \begin{tabular}{|m{0.66\textwidth}|m{0.07\textwidth}||m{0.07\textwidth}|}
\hline
\begin{center}
$\begin{aligned}
&\left[\begin{aligned}
    &\Thet{1} \moregen \Thet{2} \, \uand \\
    &\Thet{2} \moregen \Thet{3}
    \end{aligned}\right]
    \uimplies \\
&\boxed{\Thet{1} \moregen \Thet{3}}
   \end{aligned}$  
\end{center}& &  \\  \hline
\end{tabular}.
\end{center}
Applying the resolution rule to these rows, taking $\Thet{1}$ to $\thet{0}$ and $\Thet{3}$ to be $\thet{r}$, and renaming $\Thet{2}$ to be $\Thet{l}$, we obtain 

\begin{center}
\begin{tabular}{|m{0.10\textwidth}|m{0.50\textwidth}||m{0.20\textwidth}|}
 \hline 
  & 
   \begin{center}
{$\begin{aligned}
    &\boxed{\thet{0} \moregen  \Thet{l}}
     \uand \\
 &\boxed{\Thet{l} \moregen \thet{r}}  \uand  \\
  & \,{\mgi}(\thet{0}, \e{1}, \e{2}, \thet{r})\, \uand
\\
  \,\, &\reduce(\thet{0},\vars(\e{1}, \e{2}), \Thet{r})
\end{aligned}$}
\end{center}
& 
\begin{center}

$ \begin{aligned}
   \thet{r}
 \end{aligned}    
$
\end{center}
\\
\hline
\end{tabular}.
\end{center}
Applying resolution twice in succession, to this goal, the \hyperlink{hyp:L-ext-ind}{\emph{left extension induction hypothesis}}:
  \begin{center}
  \begin{tabular}{|m{0.66\textwidth}|m{0.07\textwidth}||m{0.07\textwidth}|}
\hline
\begin{center}
{$\begin{aligned}
 &\begin{aligned}
 &\begin{aligned}
 \,\, &\, \boxed{\thet{0} \moregen \thet{l}} 
\end{aligned}
\end{aligned}\end{aligned}   $} 
\end{center}& &  \\  \hline
\end{tabular}
\end{center}
(taking $\Thet{l}$ to be $\thet{l}$) and the \hyperlink{hyp:L-un-ind}{\emph{right extension induction hypothesis}}:
\begin{center}
  \begin{tabular}{|m{0.66\textwidth}|m{0.07\textwidth}||m{0.07\textwidth}|}
\hline
\begin{center}
{$
\begin{aligned}
&\begin{aligned}
&\begin{aligned}
 \,\, &\, \boxed{\thet{l} \moregen \thet{r}} 
\end{aligned}
\end{aligned}
\end{aligned} $}  \hspace{0cm} 
\end{center}& &  \\  \hline
\end{tabular},
\end{center}
we are left with
\begin{center}
\begin{tabular}{|m{0.10\textwidth}|m{0.50\textwidth}||m{0.20\textwidth}|}
 \hline 
  & 
   \begin{center}
{$\begin{aligned}
  & \,\boxed{\mgi(\thet{0}, \e{1}, \e{2}, \thet{r})}\, \uand
\\
  \,\, &{\reduce}(\thet{0},\vars(\e{1}, \e{2}), \thet{r})
\end{aligned}$}
\end{center}
& 
\begin{center}

$ \begin{aligned}
   \thet{r}
 \end{aligned}    
$
\end{center}
\\
\hline
\end{tabular}.
\end{center}
In other words, it suffices to show that $\thet{r}$ is most-general idempotent and a reduction.

Earlier we established the  \hyperlink{prop:mgi-trans-LR}{\emph{transitivity property} of \emph{mgi}}},
\begin{center}
  \begin{tabular}{|m{0.66\textwidth}|m{0.07\textwidth}||m{0.07\textwidth}|}
\hline
\begin{center}
$
\begin{aligned}
    &\unot(\isatm(\E{1})) \uand \unot(\isatm(\E{2})) \uimplies\\
&\left[    \begin{aligned}
   &\left[\begin{aligned}
  &\mgi (\Thet{0}, \lef(\E{1}), \lef(\E{2}), \Thet{1})
  \, \uand \\
   &\mgi (\Thet{1}, \rig(\E{1}), \rig(\E{2}), \Thet{2}) 
   \end{aligned}\right] \uimplies \\
  &
  \boxed{\begin{aligned}
  \mgi(\Thet{0}, &\lef(\E{1}) {\cons} \rig(\E{1}), \\
     &\,\lef(\E{2}) {\cons} \rig(\E{2}), \Thet{2})
   \end{aligned}}  
   \end{aligned} \right]
     \end{aligned}
     $
\end{center}
& &  \\  \hline
\end{tabular}.
\end{center}
Applying resolution to this assertion and the above goal, taking $\E{1}$ and $\E{2}$  to be  $\e{1}$ and $\e{2}$, respectively, $\Thet{0}$ and 
$\Thet{2}$ to be  $\thet{0}$ and  $\thet{r}$, respectively, renaming $\Thet{1}$ to be $\Thet{l}$, and then applying resolution with the case \hyperlink{ass:non-atom}{\emph{assumptions}} that $\e{1}$ and $\e{2}$ are nonatomic, we obtain 

\begin{center}
\begin{tabular}{|m{0.10\textwidth}|m{0.50\textwidth}||m{0.20\textwidth}|}
 \hline 
  & 
   \begin{center}
{$\begin{aligned}
  &\boxed{\mgi (\thet{0}, \lef(\e{1}), \lef(\e{2}), \Thet{l})}
  \, \uand \\
   &\boxed{\mgi (\Thet{l}, \rig(\e{1}), \rig(\e{2}), \thet{r})}\,\, \uand
\\
  \,\, &\reduce(\thet{0},\vars(\e{1}, \e{2})\!\apply\thet{0}, \thet{r})
\end{aligned}$}
\end{center}
& 
\begin{center}

$ \begin{aligned}
   \thet{r}
 \end{aligned}    
$
\end{center}
\\
\hline
\end{tabular}.
\end{center}
Earlier we have derived the \hyperlink{hyp:L-mgi-ind}{\emph{left mgi induction hypothesis}}

\begin{center}
  \begin{tabular}{|m{0.66\textwidth}|m{0.07\textwidth}||m{0.07\textwidth}|}
\hline
\begin{center}
{$
\begin{aligned}
 \boxed{\mgi(\thet{0}, \lef(\e{1}), \lef(\e{2}), \thet{l})}
\end{aligned} $}  \hspace{0cm} 
\end{center}& &  \\  \hline
\end{tabular},
\end{center}
and the \emph{right mgi induction hypothesis}
\begin{center}
  \begin{tabular}{|m{0.66\textwidth}|m{0.07\textwidth}||m{0.07\textwidth}|}
\hline
\begin{center}
{$
\begin{aligned}
\boxed{\mgi(\thet{l}, \rig(\e{1}), \rig(\e{2}), \thet{r})}
\end{aligned}
 $}  \hspace{0cm} 
\end{center}& &  \\  \hline
\end{tabular}.
\end{center}
  By two applications of the resolution rule, to the above goal and the two induction hypotheses in succession, taking $\Thet{l}$ to be $\thet{l}$, we obtain the goal
  \begin{center}
\begin{tabular}{|m{0.10\textwidth}|m{0.50\textwidth}||m{0.20\textwidth}|}
 \hline 
  & 
   \begin{center}
{$\begin{aligned}
  &\boxed{\reduce(\thet{0},\vars(\e{1}, \e{2}), \thet{r})}
\end{aligned}$}
\end{center}
& 
\begin{center}

$ \begin{aligned}
   \thet{r}
 \end{aligned}    
$
\end{center}
\\
\hline
\end{tabular}.
\end{center}
  
 \paragraph{The Nested Recursive Calls are a Reduction.} 
 It remains only to show that $\theta_r$ is a reduction. By the \hyperlink{prop:red-def}{definition of $\reduce$},
\begin{center}
  \begin{tabular}{|m{0.66\textwidth}|m{0.07\textwidth}||m{0.07\textwidth}|}
\hline
\begin{center}
 $\begin{aligned}
 &\boxed{\reduce(\Thet{0}, V, \Theta)} \,\, \uiff \\ 
 &
\range(\Theta) \subseteq \range(\Theta_0) \cup  V,
 \end{aligned}$.
 \end{center}& &  \\  \hline
\end{tabular}
\end{center}
applied to the most recent goal, taking $\Thet{0}$ to be  $\theta_0$, $V$ to be $\vars(\e{1}, \e{2})$, and $\Theta$ to be $\thet{r}$,  
we obtain
  \begin{center}
\begin{tabular}{|m{0.10\textwidth}|m{0.50\textwidth}||m{0.20\textwidth}|}
 \hline 
  & 
   \begin{center}
{$\begin{aligned}
  &
\range(\thet{r}) \subseteq 
\range(\thet{0}) \cup  \vars(\e{1}, \e{2})\!\apply\thet{0}
\end{aligned}$}
\end{center}
& 
\begin{center}

$ \begin{aligned}
   \thet{r}
 \end{aligned}    
$
\end{center}
\\
\hline
\end{tabular};
\end{center}
that is, by our case assumption that \hyperlink{ass:misses}{the environment substitution $\thet{0}$ misses $\e{1}$ and $\e{2}$},

  \begin{center}
\begin{tabular}{|m{0.10\textwidth}|m{0.50\textwidth}||m{0.20\textwidth}|}
 \hline 
  & 
   \begin{center}
{$\begin{aligned}
  &
\range(\thet{r}) \subseteq 
\range(\thet{0}) \cup  \vars(\e{1}, \e{2})
\end{aligned}$}
\end{center}
& 
\begin{center}

$ \begin{aligned}
   \thet{r}
 \end{aligned}    
$
\end{center}
\\
\hline
\end{tabular}.
\end{center}
Let us refer to this as our \hypertarget{goal:red-nest}{\emph{nested reduction goal.}}

 Recall that we have established the \hyperlink{hyp:R-red-ind}{\emph{right reduce induction hypothesis}},
\begin{center}
  \begin{tabular}{|m{0.66\textwidth}|m{0.07\textwidth}||m{0.07\textwidth}|}
\hline
\begin{center}
{$
\begin{aligned}
&\reduce(\thet{l}, \vars(\rig(\e{1}), \rig(\e{2}))\!\apply\!\theta_l, \thet{r})
\end{aligned}
 $}  \hspace{0cm} 
\end{center}& &  \\  \hline
\end{tabular},
\end{center}
that is, by the \hyperlink{prop:red-def}{\emph{definition of reduction}},
\begin{center}
  \begin{tabular}{|m{0.66\textwidth}|m{0.07\textwidth}||m{0.07\textwidth}|}
\hline
\begin{center}
{$\begin{aligned}
&
\range(\thet{r}) \subseteq \\ 
&\vars(\rig(\e{1}), \rig(\e{2}))\!\apply\!\theta_l\!\apply\!\theta_l \union 
\range(\theta_l) \
\end{aligned} $}  \hspace{0cm} 
\end{center}& &  \\  \hline
\end{tabular},
\end{center}
or, since we have found that \hyperlink{ass:left-idem}{$\theta_l$ is idempotent,}
\begin{center}
  \begin{tabular}{|m{0.66\textwidth}|m{0.07\textwidth}||m{0.07\textwidth}|}
\hline
\begin{center}
{$\begin{aligned}
&
\range(\thet{r}) \subseteq \\ 
&\range(\theta_l)  \union 
\vars(\rig(\e{1}), \rig(\e{2}))\!\apply\!\theta_l
\end{aligned} $}  \hspace{0cm} 
\end{center}& &  \\  \hline
\end{tabular},
\end{center}
or, by the \hyperlink{prop:vars-range-subset}{\emph{vars-range subset property}} and properties of sets,
\begin{center}
  \begin{tabular}{|m{0.66\textwidth}|m{0.07\textwidth}||m{0.07\textwidth}|}
\hline
\begin{center}
$\boxed{\begin{aligned}
&
\range(\thet{r}) \subseteq \\ 
&\range(\theta_l) \union 
\vars(\rig(\e{1}), \rig(\e{2}))  \
\end{aligned}} $  \hspace{0cm} 
\end{center}& &  \\  \hline
\end{tabular}.
\end{center}

Similarly, recall that we have established the \hyperlink{hyp:L-red-ind}{\emph{left reduce induction hypothesis}}, 
\begin{center}
  \begin{tabular}{|m{0.66\textwidth}|m{0.07\textwidth}||m{0.07\textwidth}|}
\hline
\begin{center}
{$
\begin{aligned}
 \reduce(\thet{0}, \vars(\lef(\e{1}), \lef(\e{2})), \theta_l)
\end{aligned} $}  \hspace{0cm} 
\end{center}& &  \\  \hline
\end{tabular}
\end{center}
By a similar chain of reasoning, using $\lef$ instead of $\rig$ and $\theta_0$ instead of $\theta_l$, we can infer
\begin{center}
  \begin{tabular}{|m{0.66\textwidth}|m{0.07\textwidth}||m{0.07\textwidth}|}
\hline
\begin{center}
$\boxed{\begin{aligned}
&
\range(\thet{l}) \subseteq \\ 
&\range(\theta_0) \union \vars(\lef(\e{1}), \lef(\e{2})) 
\end{aligned}}$  \hspace{0cm} 
\end{center}& &  \\  \hline
\end{tabular}.
\end{center}
Combining the boxed formulas from two recent assertions with the transitivity of the subset relation and a property of equality, we obtain
\begin{center}
  \begin{tabular}{|m{0.66\textwidth}|m{0.07\textwidth}||m{0.07\textwidth}|}
\hline
\begin{center}
$\begin{aligned}
&\range(\thet{r}) \subseteq \\
&\range(\theta_0) \, \union \\
&\begin{aligned}[t]
&\boxed{\begin{aligned}
&\vars(\rig(\e{1}), \rig(\e{2}))\,  \union  \\
&\vars(\lef(\e{1}), \lef(\e{2}))
\end{aligned}}
\\
&
\end{aligned}
\end{aligned}  $  \hspace{0cm} 
\end{center}& &  \\  \hline
\end{tabular}.
\end{center}
By our \hyperlink{ass:non-atom}{\emph{non-atom} case assumptions} that $\e{1}$ and $\e{2}$ are nonatomic, we can deduce 
\[\begin{aligned}
vars(e_1, e_2) &= 
\begin{aligned}[t]
  &\vars(\lef(e_1)) \union \vars(\rig(e_1))\, \union \\ 
  &\vars(\lef(e_2)) \union \vars(\rig(e_2)) 
\end{aligned} \\
&= \boxed{\begin{aligned}[t] 
&\vars(\rig(\e{1}), \rig(\e{2})) \,
\union \\ &\vars(\lef(\e{1}), \lef(\e{2}))`
\end{aligned}}
\end{aligned}\]
Hence we can infer the assertion
\begin{center}
  \begin{tabular}{|m{0.66\textwidth}|m{0.07\textwidth}||m{0.07\textwidth}|}
\hline
\begin{center}
$\boxed{\begin{aligned}
&
\range(\thet{r}) \subseteq  
\range(\theta_0) \union \vars(\e{1},\e{2}) 
 \\
\end{aligned}} $  \hspace{0cm} 
\end{center}& &  \\  \hline
\end{tabular}.
\end{center}
But this resolves against our \hyperlink{goal:red-nest}{nested reduction goal,}
  \begin{center}
\begin{tabular}{|m{0.10\textwidth}|m{0.50\textwidth}||m{0.20\textwidth}|}
 \hline 
  & 
   \begin{center}
{$\begin{aligned}
  &
\boxed{\range(\thet{r}) \subseteq 
\range(\thet{0}) \cup  \vars(\e{1}, \e{2})}
\end{aligned}$}
\end{center}
& 
\begin{center}

$ \begin{aligned}
   \thet{r}
 \end{aligned}    
$
\end{center}
\\
\hline
\end{tabular},
\end{center}
yielding
  \begin{center}
\begin{tabular}{|m{0.10\textwidth}|m{0.50\textwidth}||m{0.20\textwidth}|}
 \hline 
  & 
   \begin{center}
{$\begin{aligned}
  &\true
\end{aligned}$}
\end{center}
& 
\begin{center}

$ \begin{aligned}
   \thet{r}
 \end{aligned}    
$
\end{center}
\\
\hline
\end{tabular}.
\end{center}
Here $\thet{r}$ is our abbreviation for the \hyperlink{def:nest}{{\emph{nested recursive call}}}
\[\begin{aligned}
\unify(
\begin{aligned}
   \unify(\thet{0}, 
  \lef(\e{1}),  
 \lef(\e{2})),
   \end{aligned} 
\rig(\e{1}), 
\rig(\e{2}))
 \end{aligned}. \]

 Initially, we omitted the output entries corresponding to the case assumptions, which make up the other terms of the conditional expression that comprises the entire unification algorithm. 

 We have presented representative branches of the derivation of one of the algorithms constructed by Snark, rewritten for readability.   The \href{https://www.ai.sri.com/~waldinge/The-Full-Unification-Proof.pdf}{Snark proof} appears online. The final algorithm appears in the next section.

 \subsection*{The Final Program.}
\hypertarget{sec:program}{}

\vspace{-\baselineskip}
\vspace{10pt}

Here is the program extracted from the proof.  We introduce by hand annotations, bold and in braces (\textbf{$\mathbf{\{ \ldots \}}$}), to indicate some conditions that hold when control passes through the corresponding point in the program. This is merely to make the program easier to read; it has no effect on its evaluation.


\newpage
\vspace{-1cm}
\[
\begin{aligned}
&\unify(\thet{0}, \e{1}, \e{2}) = \\
&{\ucond 
  {\isprop(\thet{0})}
  {\ucond 
  {\e{1} \occursin \e{2}}
  {\fail}
  {\ucond 
    {\e{1} = \e{2}} 
    {\thet{0}} 
    {\ucond 
      {\iscnst(\e{1})} 
      {\ucond 
       {\iscnst(\e{2})} 
       {\fail} 
       {\ucond 
  {\isvar(\e{2})} 
  {\unify(\thet{0}, \e{2}, \e{1})} 
  {\begin{aligned}[t]
        &\textbf{$\mathbf{\{\unot (\isatm(\e{2}) \}}$} \\
        &{\fail}
        \end{aligned}
        } }  
      }
      {\ucond
      {\isvar(\e{1})}
      {\ucond
{\misses(\thet{0}, \e{2})}
{\ucond
  {\misses(\thet{0}, \e{1})}
  {\thet{0} \compose \repl{\e{1}} {\e{2}}}
  {\unify(\thet{0},\, \e{1}\apply \thet{0},\, \e{2}\apply\thet{0} )}
}
{\unify(\thet{0},\, \e{1}\apply\thet{0},\, \e{2}\apply\thet{0} )}
}
      {\begin{aligned}[t]
&\textbf{$\mathbf{\{\unot (\isatm(\e{1}))\}}$} \\
&{\ucond
{\iscnst(\e{2})}
{\fail}
{\ucond
{\isvar(\e{2})}
{\unify(\thet{0}, \e{2}, \e{1})}
{\unify(\!\!\!\begin{aligned}[t]
&\unify(\thet{0},\lef(\e{1}),\lef(\e{2}) \\ 
&\rig(\e{1}),\rig(\e{2}))
\end{aligned})
}
}
}
\end{aligned}}
      }}}
    }
    {\fail} }
    \end{aligned}
\]

\subsection*{The Snark Proof.} 
\hypertarget{proof}{}
The proof we have presented is based on the \href{https://www.ai.sri.com/~waldinge/The-Full-Unification-Proof.pdf}{Snark proof}, but it has been altered for readability.
  The original Snark proof is a refutation, so the conjecture is negated and Snark derives a contradiction.  All the rows are in clausal form, and in Snark syntax, which is based on Lisp syntax. 
 The proof was found in some 10 seconds, but this number is not really meaningful.   Presumably, more modern theorem provers would be faster, but if we extended the theory with additional axioms to enable the construction of other algorithms, all proof times could be longer.
 
 No changes were made in Snark for finding the derivation, but some modifications were made to the answer extraction mechanism, to allow it to delete orphaned output entries. We were able to alter its strategy by providing weights on the symbols in the theory's vocabulary and an ordering between them.  The weights determined the which row is worked on next, and the ordering determined which part of a given row was the focus of attention. The ordering on symbols we employed was the \emph{recursive path ordering} \citep{dersh}. 

   The \hyperlink{sec:program}{final program} was subjected to automatic simplification; e.g., redundant tests were eliminated. Snark was able to find many proofs for the theorem, each giving a different unification algorithm. By altering the symbol ordering, we were able to influence the algorithm Snark discovered.  For example, in the algorithm we presented, we test if the environment is a proper substitution, $\isprop(\thet{0})$, before we perform the occurs check, $\e{1} \occursin \e{2}$. This is because in the symbol ordering we have declared, the properness symbol $\isprop$ is less than the occurrence symbol $\occursin$.  When we reverse this ordering, we get an algorithm in which we perform the occur check before we test if the environment is proper. Some orderings give us longer proof searches, or no proof at all; some orderings give better programs than others.

\section*{Discussion} As we have remarked, the proof from which the program is extracted establishes the correctness of the derived program and also provides its rationale, an explanation of its working.  Of course, the correctness of the program depends on the correctness of the specification and the assertions we provide; if these have errors, all bets are off. In an earlier version of this paper, referees detected that the proof depended on a false assertion, casting into doubt the correctness of the algorithm.

The body of knowledge required for the derivation is surprisingly large, and assembling it was labor intensive.  One may ask what the point is of devoting so much effort toward constructing a program that is already known.  Furthermore, the automatic construction of the unification algorithm itself required a unification algorithm.

For one thing, the derivation is useful as a case study;  many of the problems that arise, such as the discovery of a well-founded relation or the generalization of the specification appear in other program synthesis problems.

It has been found that theory-specific unification algorithms (e.g. associative-commutative unification  \citep{sti:acu, liv-siek:acu}, higher-order unification \citep {huet}) can allow a theorem prover in that theory to find proofs more quickly than if the properties of the relevant relation and function symbols are represented by axioms.  If one could construct the appropriate unification algorithm automatically, it would be possible to discard those axioms mid-proof and replace the unification algorithm of the theorem prover with the newly synthesized one, completing the proof with the new algorithm.   For some theories there is no single most-general unifier;  one must find a set or stream of unifiers. This must be reflected in the specification, and the theorem prover must be prepared to incorporate such an algorithm.

Also, the same knowledge may allow the derivation of many algorithms, including other unification algorithms and, presumably, other algorithms with the same subject domain, such as pattern matchers and anti-unification \citep{plo:au} and disunification  \citep{comon:disu} algorithms. We may look forward to a day in which program synthesis systems improve and specialize themselves in the course of solving a particular programming problem.

The derivation was challenging for Snark to discover; perhaps more modern theorem provers would have an easier time of it.    For instance, in current work, the contemporary theorem prover  Vampire \citep{hoz} is being extended and applied to program synthesis. The rules of Vampire are similar to those of Snark, but it performs many more inferences per second.  Vampire employs the Green [\citeyear{ccg}] answer literal in place of Snark's output columns.  As we do, it introduces a new assertion to play the role of the induction hypothesis in an induction proof.

Several decisions we made in the design of the theory of expressions were taken to reduce the search space.  We specified the more-general idempotence relation so that it would imply idempotence of the unifier---this meant that we did not have to include idempotence as a separate condition.  We decided to treat the failure indicator ($\fail$) as a special substitution which yielded the black hole constant ($\blk$) when applied to any expression.  This meant that we did not have to treat ununifiable expressions as a special case. All of this led to a shorter specification and a correspondingly shorter proof, which was easier to discover.

We did not attempt to discover the proof from first principles---rather we included lemmas that described the properties of the relations in our specification. (In the same way, a number theorist does not attempt to prove a complex theorem using the basic axioms, such as the Peano postulates.)  In proving the main theorem, Snark did not use the \emph{equality replacement rule}, which coincides with the paramodulation rule in resolution theorem provers. We found that we got better performance when we dispensed with the rule and instead included equality substitutivity properties for the principal function and relation symbols. At this point, all the lemmas we used have been proved, either by Snark or by hand. In proving the lemmas, we did use the \emph{equality replacement rule}.

\subsubsection{Notes on Large Language Models.}  Recently, impressive results have been obtained using \emph{large language models}  \citep{wik:llm}, which  apply neural net technology to very large collections of information, including text, programs, and images.  Neural-net-based systems are now superior to humans at chess and go. They do a decent job of translating languages.  Some of these systems will produce code in a variety of subject domains and programming languages, and are being used to assist professional programmers as a time saver.  The systems are particularly effective in recovering known algorithms that have been documented in the literature.  

Unfortunately, large language models have been known to pick up erroneous information from their data sources, and confabulate information when it seems opportune.  They can invent plausible references to articles that do not exist, which appear in journals that also do not exist.  One system was unsure about whether “seven" was greater than “eight"; another insisted that 6 was a prime, even after it was told that 2 times 3 is 6.  In our own experiments, a system was unsure whether, say, December 31, 1999 came before or after January 1, 2000.  While the capabilities of these systems are constantly improving, we are not at the point at which we can trust their results. Nor do they always explain or justify their findings.

This brings up the question of whether we can combine the virtues of large language models and deductive methods.
One obstacle for a theorem proving approach is collecting and encoding as axioms in logic the knowledge we require to construct a program; perhaps some of this effort could be relegated to a large language model, with the resulting theory tested for inconsistency and otherwise validated by a theorem prover as well as human scrutiny. The task of formulating a specification in logic is an obstacle to all of us; perhaps an LLM could perform this, with the resulting formula translated back into natural language for confirmation by the user.  Also,  determining the settings a theorem prover requires to be effective is something of a black art; in Snark we had to determine experimentally the weights we assigned to symbols and the ordering we applied between them.  A large language model might be effective at this; it doesn't seem very different from playing a game.  

This is an active area of research; for instance, the system LeanDojo \citep{leandojo} has been successful in applying a large language model to guide the search in the interactive theorem prover Lean \citep{lean}. 
If these systems reach the point at which they can find constructive proofs of the theorems obtained from specifications, like our unification proof, we can use them to develop provably correct programs.

\section*{Acknowledgments}

This paper is dedicated to the memory of Zohar Manna, Mark Stickel, and Cordell Green, without whom this work would not have been possible.  We would like to thank the SRI Artificial Intelligence Center for years of support, and for comments and suggestions.  We have had valuable comments from members of the SRI Computer Science Lab, the SRI Crazy Idea Seminar,  the Kestrel Institute, and the Stanford SLUGS group.  The paper was influenced by the Flex Project of the Kestrel Institute.  Karthik Nukala read early versions of this paper and made valuable suggestions. We have also had useful discussions with and encouragement and/or criticism from Franz Baader, Maria-Paola Bonacina, Alessandro Coglio, Steve Eker, Michael Genesereth, Cordell Green, Beorn Johnson, Laura Kovacs, Yanhong Annie Liu, Neil Murray, Ray Perrault, Kyle Richardson, Natarajan Shankar, Eric Smith, Asuman Suenbuel, Geoff Sutcliffe, Andrei Voronkov, Eva Maria Wagner, David S. Warren, Stephen Westfold, Dave Wolf, and the referees for the \emph{Journal of Symbolic Computation.}  Julie Thomas edited the entire draft.

\end{document}